\documentclass{aa}

\usepackage{graphicx}
\usepackage{natbib}
\usepackage{threeparttablex}
\usepackage{hyperref}
\usepackage{txfonts}
\begin{document}

\title{Investigating the hot molecular core, G10.47+0.03: A pit of nitrogen-bearing complex organic molecules}
     \author{
           Suman Kumar Mondal\inst{2}
      \and Wasim Iqbal\inst{1}\thanks{Corresponding authors}
      \and Prasanta Gorai\inst{2,3}
      \and Bratati Bhat\inst{2}
      \and Valentine  Wakelam\inst{4}
      \and Ankan Das\inst{5\star}
          }
\institute{
     South-Western Institute for Astronomy Research (SWIFAR), Yunnan University (YNU), Kunming 650500, People’s Republic of China
\and Indian Centre for Space Physics, 43 Chalantika, Garia Station Road, Kolkata 700084, India
\and Department of Space, Earth \& Environment, Chalmers University of Technology, SE-412 96 Gothenburg, Sweden
\and Laboratoire d’astrophysique de Bordeaux, CNRS, Univ. Bordeaux, B18N, allée Geoffroy Saint-Hilaire, F-33615 Pessac, France
\and Institute of Astronomy Space and Earth Science, AJ 316, Salt lake Sector II, Kolkata 700091, West Bengal, India \\\\
\email{ankan.das@gmail.com \& wasim@ynu.edu.cn}
         }
%
%\titlerunning{The rotational spectrum and ISM search for glycinamide}
\authorrunning{Mondal et. al.}
%
%
%\abstract{}{}{}{}{}
%5{}token are mandatory
%
%
\abstract
{
Recent observations have shown that Nitrogen-bearing complex organic species are present in large quantities in star-forming regions. Thus, investigating the N-bearing species in a hot molecular core,  such as G10.47+0.03, is crucial to understanding the molecular complexity in star-forming regions. They also allow us to investigate the chemical and physical processes that determine the many phases during the structural and chemical evolution of the source in star-forming regions.
} 
{
The aim of this study is to investigate the spatial distribution and the chemical evolution states of N-bearing complex organic molecules in the hot core G10.47+0.03.
}
{
We used the  Atacama Large Millimeter/submillimeter
Array (ALMA) archival data of the hot molecular core G10.47+0.03. The extracted spectra were analyzed assuming local thermodynamic equilibrium (LTE). LTE methods are used to estimate the column density of observed species. Furthermore, robust methods such as  Markov chain Monte Carlo (MCMC) and rotational diagram methods are implemented for molecules for which multiple transitions were identified to constrain the temperature and column density. Finally, we used the Nautilus gas-grain code to simulate the nitrogen chemistry in the hot molecular core. We carried out both 0D and 1D simulations of the source. We compared the simulated abundances with observational results.
}
{ 
We report various transitions of  nitrogen-bearing species (NH$_2$CN, HC$_3$N, HC$_5$N, C$_2$H$_3$CN, C$_2$H$_5$CN, and H$_2$NCH$_2$CN) together with some of their isotopologues and isomers. Besides this, we also report the identification of CH$_3$CCH and one of its isotopologues. We present detailed chemical simulation results to investigate the possible N-bearing chemistry in the source.
} 
{
In this study, various transitions of nitrogen-bearing molecules are identified and discussed. The emissions originating from vinyl cyanide, ethyl cyanide, cyanoacetylene, and cyanamide are compact, which could be explained by our astrochemical modeling. Our 0D model shows that the chemistry of certain N-bearing molecules can be very sensitive to initial local conditions such as density or dust temperature. In our 1D model, simulated higher abundances of species such as HCN, HC$_3$N, and HC$_5$N toward the inner shells of the source confirm the observational findings.
}
\keywords{Astrochemistry–line:  identification–ISM: individual(G10.47+0.03)–ISM: molecules,  ISM: abundances, ISM: evolution}
\maketitle
%%%%%%%%%%%%%%%%%%%%%%%%%%%%%%%%%%%%%%%%%%%%%%%%%%%%%%%%%%%%%%%%%%%%%%%%%%%%%%%%%%%%%%
%
%
\section{Introduction}
Numerous molecules (about 250) are identified in the interstellar medium (ISM) or circumstellar shells  \citep[CDMS\footnote{\url{https://cdms.astro.uni-koeln.de/classic/molecules}},][]{mcgu18}. Among them, 96 molecules contain at least one nitrogen atom. Of these, 33 have cyanide (-CN) and 12 
isocyanide (-NC). CN was the first nitrogen-bearing molecule that was observed in 1940 \citep{mcke14}. It is the key precursor of cyanopolynes, some hydrocarbons, and bio-molecules. Isomerization can play a key role in the conversion of cyanide to isocyanide and vice versa. 
The study of nitrogen-bearing complex molecules with  CN bonds is of prime importance in prebiotic chemistry. Protein synthesis occurs through a peptide bond formation; for example, CN bonds formed between -COOH and -NH$_2$ groups from different amino acids \citep{gold10}. 
Nitrogen-bearing molecules play a crucial role in the synthesis of biomolecules \citep{balu09}. Complex organic nitrogen-bearing molecules were detected in various regions of our galaxy, such as star-forming
regions, comets, and atmospheres of planets and the Moon \citep{john77,irvi84,kawa92,jimn77,ziur99}.
Some  nitrogen-containing ions such as $\rm{HC_3NH^+, NH_4^+, H_2NCO^+, NCCNH^+}$, $\rm{CN^-, C_3N^-, C_5N^-}$, and $\rm{HCNH^+}$  have been identified in the star forming regions \citep{kawa94,cern13,marc18,agun15,agun10,thad08,cern08,ziur86}.  

The study of complex organics in star-forming region can shed some light on the region's evolutionary history. Dense, hot, and chemically enriched inner regions of high-mass star-forming regions are hot molecular cores (HMCs). %These HMCs are chemically rich. 
Many simple and complex species were identified in these regions  \citep{dish98,bell19,jorg20}. 
The detection of complex organic molecules is an important tool for explaining the physical and chemical structure of high-mass star-forming regions. G10.47+0.03 (hereafter G10) is a massive hot molecular core
located at a distance of 8.6 kpc \citep {sann14}. The luminosity of this source is $\sim$ 7$\times$10$^{5}$ L$_{\odot}$ 
\citep {cesa10}. 

Interstellar dust is used as a catalyst for the synthesis of complex organic molecules. For example, vinyl cyanide (C$_2$H$_3$CN) was primarily detected toward Sgr B2 in 1975 \citep{gard75}. Later, this molecule was identified in various regions of the ISM, such as cold cores, Orion KL, the circumstellar envelope of the carbon-rich AGB star CW Leo, which is also known as IRC +10216, and so on \citep{agun07,lope14}.
The observed abundance of vinyl cyanide in IRC+10216 is two times lower than in Orion KL. \cite{agun07} found a good agreement between their modeling and observational results for IRC+10216. 
Ethyl cyanide was first detected in the hot core OMC-1 and Sgr B2 \citep{john77}. Later it was also observed in different hot cores (G327.3-0.6, Orion 
IRC2, Orion KL, Sgr B2(N), Sgr B2 (M)) and in hot corinos (IRAS 16293-2422). Along with the main isotopic species, singly and doubly 
$^{13}$C-substituted ethyl cyanide was identified in the ISM \citep{marg16}. Cyanoallene was observed with the 100m Green Bank Telescope in the cold core of TMC by \cite{lova06}. Cyanamide is a potential prebiotic molecule \citep{iban71} that was first detected several decades ago in the hot core,  Sgr B2 \citep{turn75}.
%, no gas-grain modeling result is available due to the lack of its formation route. 
To date, the 
most complex cyanide molecule observed in the ISM is the  propyl cyanide (n-$\rm{C_3H_7CN}$) and its isomer, first branch chain molecule, isopropyl 
cyanide (i-$\rm{C_3H_7CN}$) \citep{bell09,bell14}.

\cite{rolf11} observed G10 in three different frequency ranges (200 GHz, 345 GHz, and 690 GHz) with the submillimeter array (SMA). They identified many oxygen-bearing, nitrogen-bearing, and sulfur-bearing species in this source. Furthermore, many complex organic molecules and some highly excited lines of HC$_3$N, HCN, and HC$^{15}$N observed in this source reveal that it is an active site for chemical enrichment \citep {rolf11a,rolf11}. Here, we analyze the  Atacama Large Millimeter/submillimeter
Array (ALMA) archival data of G10 in the ALMA band 4 frequency region. This paper focuses on nitrogen-bearing molecules, their distributions, complex chemistry, and fractional abundances.
Various nitrogen-related molecules such as HCN, CH$_3$CN, CH$_3$NH$_2$, CH$_2$NH, C$_2$H$_3$CN, C$_2$H$_5$CN, HC$_3$N, NH$_2$CN, CH$_3$NCO, NH$_2$CHO, and HNCO have been observed \citep{rolf11,font07,masa19,suzu16,gora20,wyro99} in G10. In this work, we detect some new nitrogen-bearing molecules such as HC$_5$N, H$_2$NCH$_2$CN, some isotopologues of vinyl cyanide, ethyl cyanide, and HC$_3$N. 

This paper is organized as follows. First, in Section 2, observational details are given. Astrochemical modeling and its results  are discussed in Section 3 and Section 4, respectively. Finally, in Section 5, we conclude.

%%%%%%%%%%%%%%%%%%%%%%%%%%%%%%%%%%%%%%%%%%%%%%%%%%%%%%%%%%%%%%%%

\section{Observational data and analysis}
Here, ALMA Cycle 4 data of G10 is used for the analysis. The phase center of the observation is $\alpha$ (J2000)= 18$^h$08$^m$38.232$^s$, 
$\delta$(J2000)= -19$^0$51$'$50.4$^{''}$. The systematic velocity of the source is 68 km.s$^{-1}$. The flux calibrator was 
J1733-1304, the phase calibrator was J1832-2039, and the bandpass calibrator was J1924-2914. This observation was performed with the four spectral bands, covering the sky frequencies of 
$129.50 - 131.44$ GHz, $147.50 - 149.43$ GHz, $153.00 - 154.93$ GHz,  and $158.49 - 160.43$ GHz and corresponding  angular resolution at four different frequencies are $1.5 ^{''}$ (12900 AU), $1.72^{''}$ (14792 AU), $1.66^{''}$ (14276 AU), and $1.76^{''}$ (15136 AU), considering the distance to the source is 8.6 kpc.
%Each spectral window is divided into two data cubes; continuum and line emission.
Summaries of the observational details (frequency range, channel width, baseline, etc.) are already presented in \cite{gora20}. For data analysis and imaging we used CASA (version 4.7.2) \citep{mcmu07}, and for the line identification and analysis we used the CASSIS software \footnote{\url{http://cassis.irap.omp.eu}  \citep{vast15}} (developed by IRAP-UPS/CNRS). 

%%%%%%%%%%%%%%%%%%%%%%%%%%%%%%%%%%%%%%%%%%%%%%%%%%%%%%%%%%%%%%

\subsection{Images}
\cite{cesa10} resolved the hot core G10 and found three distinct HII regions inside the core, which are A (Ultra Compact HII region), B1, and B2 (Hyper compact HII regions). \cite{rolf11} also observed this source with the submillimeter Array (SMA). They observed the continuum at three different frequencies (201/211 GHz, 345/355 GHz, 681/691 GHz). 
The beam sizes were insufficient to resolve the continuum emission at 201/211 and 681/691 GHz. However, at 345/355 GHz, the beam size was sufficient ($\sim 0.3^{''}$) to resolve this source. Here, continuum maps of G10 were observed at four different wavelengths, 1.88 mm (159.45 GHz), 1.94 mm (153.96 GHz), 2.02 mm (148.51 GHz), and 2.3 mm (130.50 GHz), respectively.
%Angular resolution of our observation is $\sim$ 2$^{''}$.
Summaries of continuum images (synthesized beam, peak flux, integrated flux, etc.) are discussed in detail in \cite{gora20}. 
\cite{gora20} calculated the H$_2$ column density for each continuum map and obtained an average value of $\sim$ 1.35$\times$10$^{25}$cm$^{-2}$. \cite{gora20} also estimated the dust opacity of each continuum map and found all are optically thin ($\tau_\nu$ $\sim$0.135).

%%%%%%%%%%%%%%%%%%%%%%%%%%%%%%%%%%%%%%%%%%%%%%%%%%%%%%
\subsection{Analysis}
All lines are extracted from the 2.0$^{''}$ (17200 AU) diameter region centered at G10  ($\alpha$(J2000)=18$^h$08$^m$38.232$^s$, $\delta$(J2000)=-19$^0$51$'$50.4$^{''}$). Line identification has been carried out with CASSIS together with the spectroscopic database, `Cologne Database for Molecular Spectroscopy' \citep[CDMS,][]{mull01,mull05}\footnote{\url{(https://www.astro.uni-koeln.de/cdms)}} , and the Jet Propulsion Laboratory \citep[JPL,][]{pick98}\footnote{\url{(http://spec.jpl.nasa.gov/)}} database. The observed transitions of various nitrogen-bearing molecules together with their quantum numbers, upper state energies (E$_u$), V$_{LSR}$, line parameters such as peak brightness temperature (I$_{max}$), line width (full width at half maximum; FWHM), and the integrated intensity ($\rm{\int T_{mb}dV}$) are noted in Table \ref{tab:dataobs}. The line parameters (I$_{max}$, FWHM, $\rm{\int T_{mb}dV}$) are obtained using a single Gaussian fit to the observed spectral profile of each unblended transition. To see the spatial distribution of the observed species, moment zero maps of each transition are shown in Figures \ref{fig:mm_vc}, \ref{fig:mm-ec}, and \ref{fig:mm-oth}, and \ref{fig:mm-othh}. The emitting regions of all the transitions are calculated using a two-dimensional Gaussian fitting of their moment maps (see Table \ref{tab:mm}).

%In general, a good fit is obtained except for a few lines where these are blended with other species.
%such as molecular transitions (quantum numbers) 
%along with their  rest frequency ($\nu$), full-width at half maximum ($dV$), integrated intensity ($\rm{\int T_{mb}dV}$), 
%upper state energy (E$_u$), V$_{LSR}$, and the intensity of transition (S$\mu^{2}$) are presented in Table \ref{tab:dataobs}. 

%%%%%%%%%%%%%%%%%%%%%%%%%%%%%%%%%%%%%%%%%%%%%

%%%%%%%%%%%%%%%%%%%%%%%%%%%%%%%%%%%%%%%%%%%%%%

\subsection{Rotational diagram analysis \label{sec:RD}} % 
Among all the observed nitrogen-bearing species toward G10, we only find multiple unblended transitions (more than 2) for four molecules (see Table \ref{tab:dataobs}), which are vinyl cyanide (C$_2$H$_3$CN),  $^{13}$C isotopologue of vinyl cyanide ($^{13}$CH$_2$CHCN), ethyl cyanide (C$_2$H$_5$CN), and $^{13}$C isotopologue of ethyl cyanide ($^{13}$CH$_3$CH$_2$CN). In addition to these four nitrogen-bearing molecules, multiple transitions are observed for methyl acetylene (CH$_3$CCH). Thus, a rotational diagram analysis was carried out for these five species to measure their excitation temperatures and the total column densities.

%Since it is recommended by the CASSIS manual to use the obtained column density and temperature from the rotational diagram analysis to generate the synthetic spectrum and compare it with the observations.

\begin{figure*}[t]
\includegraphics[width=9cm]{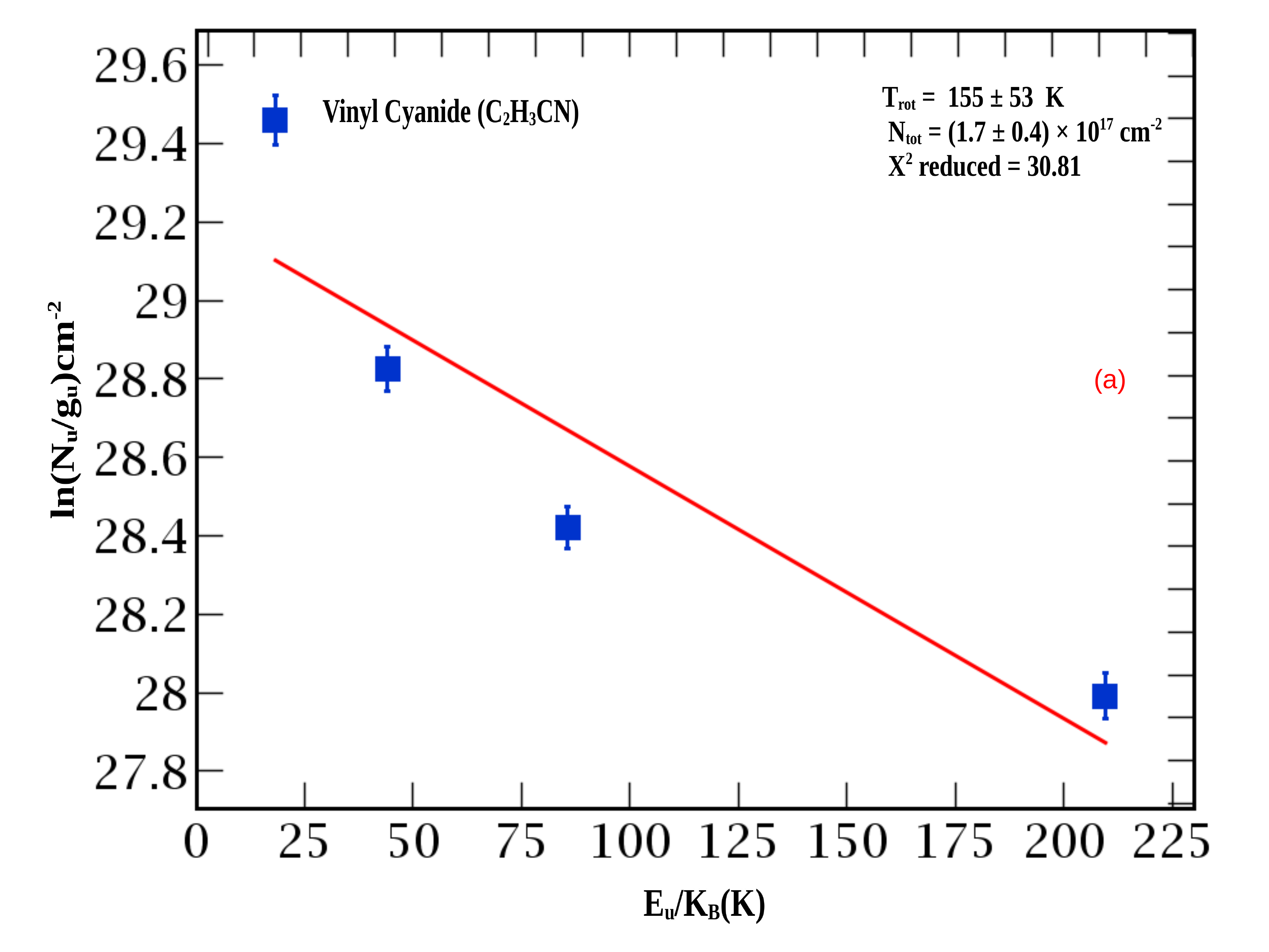}
\includegraphics[width=9cm]{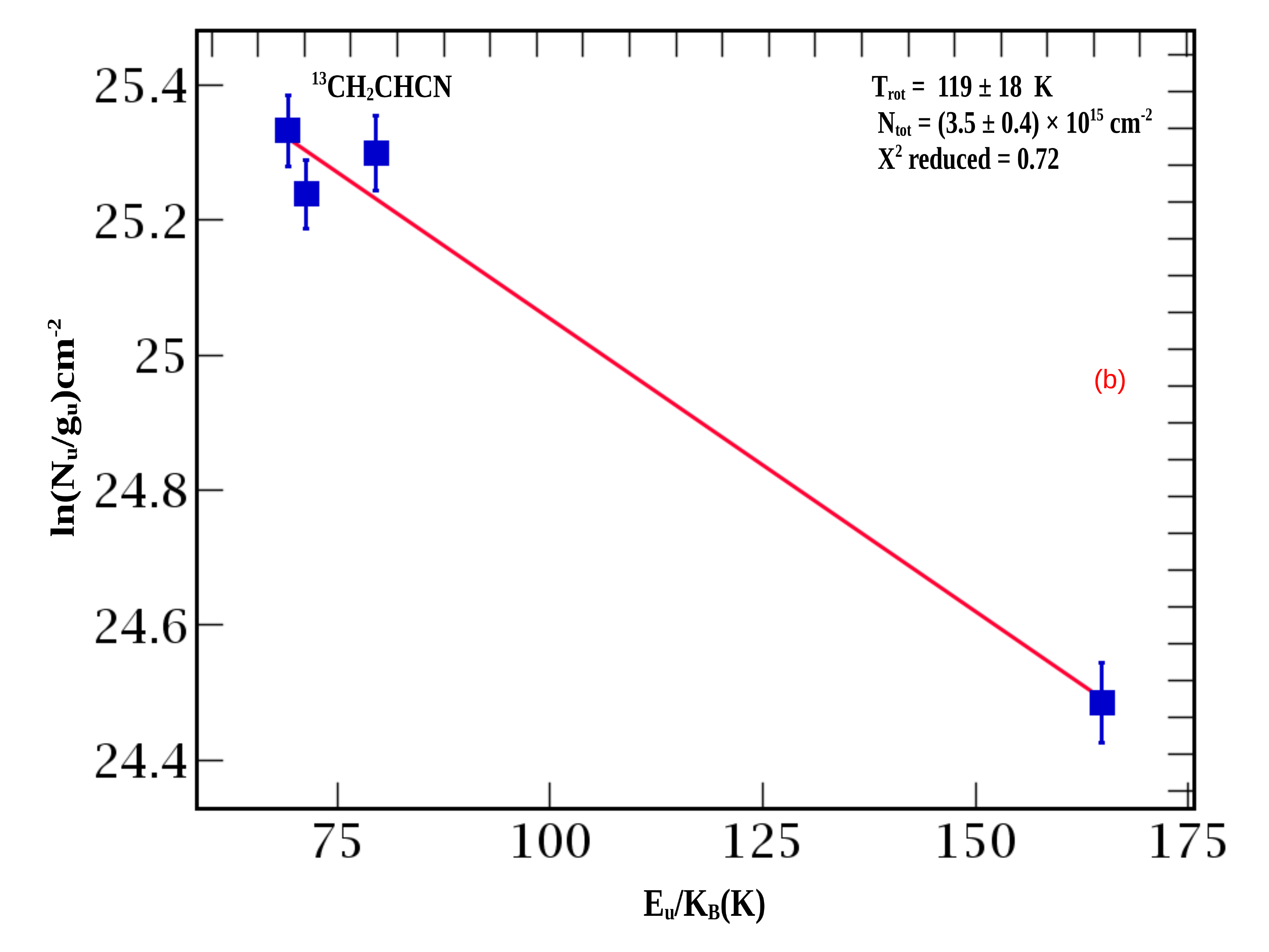}
\caption{Rotational diagram of (a) vinyl cyanide (C$_2$H$_3$CN) and (b) its one istotopologue ($^{13}$CH$_2$CHCN). The best fitted rotational temperature and column density are given in each panel. Filled blue squares are the data points and the blue line represents the error bar. The error bar of each transition is calculated from rms noise (varies 117 mK to 306 mK) and calibration error (considering 5$\%$). }
\label{fig:rot-dia}
\end{figure*}

For the optically thin transitions, the column density (N$_u^{thin}$) at local upper state can be expressed as
\begin{equation}
\frac{N_u^{thin}}{g_u}=\frac{3k_B\int{T_{mb}dV}}{8\pi^{3}\nu S\mu^{2}},
\label{eqn:clmn}
\end{equation}
where g$_u$ is the degeneracy of the upper state, k$_B$ is the Boltzmann constant, $\rm{\int T_{mb}dV}$ is the integrated intensity,
$\nu$ is the rest frequency, $\mu$ is the electric dipole moment, and S is the transition line strength, \citep{gold99}. Under the  local thermodynamic equilibrium (LTE) condition and for optically thin transitions, the total
column density ($N_{total}$) of a molecular species can be prescribed as
\begin{equation}
\frac{N_u^{thin}}{g_u}=\frac{N_{total}}{Q(T_{rot})}\exp(-E_u/k_BT_{rot}),
\end{equation}
where $T_{rot}$ is the rotational temperature, E$_u$ is the upper state energy, $\rm{Q(T_{rot})}$ is the partition function at rotational
temperature. This can be rearranged as
\begin{equation}
ln\Bigg(\frac{N_u^{thin}}{g_u}\Bigg)=-\Bigg(\frac{1}{T_{rot}}\Bigg)\Bigg(\frac{E_u}{k}\Bigg)+ln\Bigg(\frac{N_{total}}{Q(T_{rot})}\Bigg).
\end{equation}
There is a linear relationship between the upper state energy and column density at the upper level when all lines are optically thin and all lines are probing the same physical conditions. The column density and rotational temperature are extracted from the rotational diagram. Rotational diagrams of all the five species are shown in Figures \ref{fig:rot-dia}, \ref{fig:rot-dia1}, and \ref{fig:rot-ch3cch}. 
 %The MCMC method can be used as an alternative to the Rotation diagram method \citep{vast18} for those species for which multiple transitions (more than 2) of different upper state energies are obtained. Therefore, we also apply MCMC method to estimate  best fit excitation temperatures and column densities of the above mentioned five molecules and compare them with measured values that are obtained with rotational diagram analysis.

%These are used in the Line Analysis module of CASSIS to generate LTE spectra. For this purpose, an average FWHM of the unblended optically thin transitions and a source size of 2$^{''}$ is used.
%Results obtained from the analysis are discussed in section \ref{sec:mol}.

\subsection{Markov chain Monte Carlo model \label{sec:MCMC}}
Though the rotational diagram is used for the species for which multiple transitions are observed, another method, Markov chain Monte Carlo (MCMC), is used to constrain the excitation temperature and column density for the comparison. The MCMC algorithm was initially developed for dealing with the probabilistic behavior of the collection of atomic particles. However, it was challenging to do this analytically. Thus, an iterative process was implemented to solve it through simulation. This stochastic model describes a sequence of possible events in which the probability of each event depends only on the state attained in the previous one. The MCMC method is an interactive process that goes through all line parameters (e.g., molecular column density, excitation temperature, source size, line width) with a random walk and heads into the solution's space. $\chi^2$ minimization gives the final solution. 

%The MCMC method can be used as an alternative to the Rotation diagram method for those species for which multiple transitions (more than 2) of different upper state energies are obtained.

%The $\chi^2$ is defined as,
%$$
%{\rm {\chi_i}^2=\sum_{j=1}^{N_i} \frac{  (I_{obs,ij}-I_{model,ij})^2}{rms^2_i+(cal_i\times I_{obs,ij})^2}},
%$$
%where, ${\rm I_{obs,ij}}$ and ${\rm I_{model,ij}}$  are observed and modeled intensity in the channel j of transition i respectively,
%rms$_i$ is the rms of the spectrum i, and cal$_i$ is the calibration error.

For MCMC fitting, the FWHM of all the transitions of a species are kept constant at their average measured value, v$_{LSR}$ is kept constant at $68$ km/s, and source size is kept constant at 2$^{"}$. Then, the obtained results of these species are compared with those obtained with the rotation diagram method. Obtained results with these two methods are presented in Table \ref{tab:clmn} and discussed in Section \ref{sec:mol}.  %Additionally, the MCMC method is meticulously implemented to determine the best-fitted parameters (i.e., column density, excitation temperature, size) for those species for which multiple transitions are not identified. 
%For example, we identify two transitions each of CH$_2$CDCN, C$_2$H$_5$NC, NH$_2$CN, and CH$_3$$^{13}$CCH.
%one transition each of CH$_2$$^{13}$CHCN, CH$_2$CH$^{13}$CN,CH$_3$CH$_2$$^{13}$CN H$_2$NCH$_2$CN, HC$_3$N, HC$^{13}$CCN, HCC$^{13}$CN, HC$_5$N, HCC$^{13}$CCCN. 
The obtained line parameters of the MCMC fitting of five species are noted in Table \ref{table:mcmc_lte}. MCMC fitted spectra are shown in Figures \ref{fig:c2h3cn-mcmc}, \ref{fig:c2h5cn-mcmc}, and \ref{fig:ch3cch-mcmc}.
%finally, we apply LTE model for all species to measure their column densities.

\begin{figure*}
\centering
\includegraphics[width=18cm]{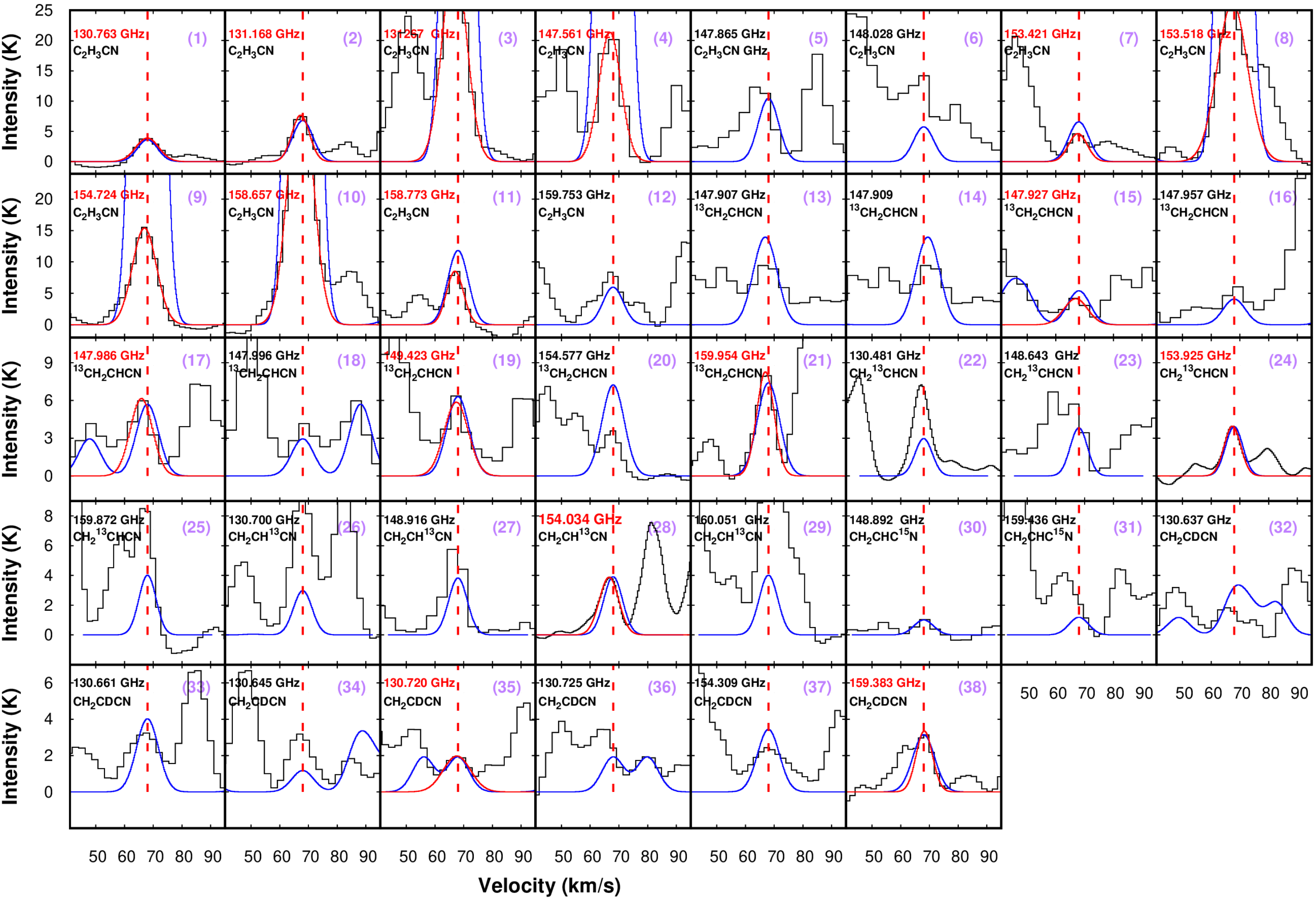}
\caption{Black lines represent the observed spectra of vinyl cyanide and its isotopologues ($^{13}$CH$_2$CHCN, {CH$_2$}$^{13}$CCN, CH$_2$CH$^{13}$CN, CH$_2$CNC$^{15}$N, and CH$_2$CDCN). 
Unblended transitions are noted in red, whereas blended transitions are indicated in black. The solid red lines represent the Gaussian fit of the unblended transitions.
LTE spectra are shown in blue by considering N(H$_2)=1.35 \times 10^{25}$ cm$^{-2}$, excitation temperature $150$ K, source size of 2$^{''}$, and the average FWHM (obtained from the unblended optically thin transitions) of the species. The column density of the species is varied until eye-estimated good fits are obtained.  The red dashed lines show the systematic velocity (V$_{LSR}$) of the source at $\sim$ 68 km s$^{-1}$. In addition, the name of the species and their respective transitions (in GHz) are provided in each panel.}
\label{fig:fit1}
\end{figure*}

%%%%%%%%%%%%%%%%%%%%%%%%%%%%%%%%%%%%%%%%%%%%%%%%%%%%%%
\begin{table*}
{\scriptsize
%\hskip -1.5cm
%\vbox{
%\tiny{
\caption{Estimated temperature and column density of the observed species.}
\label{tab:clmn}
\begin{center}
%\addtolength{\leftskip} {-2cm}
%\addtolength{\rightskip}{-2cm}
\begin{tabular}{|c|c|c|c|c|c|c|c|c|}
%\hline
\hline
species&\multicolumn{4}{|c|}{Temperature}&\multicolumn{4}{|c|}{Column density}\\
&\multicolumn{4}{|c|}{(K)}&\multicolumn{4}{|c|}{(cm$^{-2}$)}\\
\hline
&other observation & LTE/LTE2 & RD & MCMC & other observation & LTE/LTE2(*) & RD & MCMC \\
&                  &     &    &      &                   &     &    &     \\
\hline
\multicolumn{9}{|c|}{}\\
\multicolumn{9}{|c|}{Vinyl cyanide and its isotopologues}\\
\hline
$\rm{C_2H_3CN}$&200$^a$&150&$155\pm53$&209.82$\pm$18.02&$7.0 \times {10^{17}}^a$&$2.9\times 10^{17}/6.0\times 10^{17}(1.07)$&$(1.7\pm 0.4) \times 10^{17}$&$(3.4\pm0.44 )\times10^{17}$\\
&$176\pm35^b$&&&&$2.0\times10^{14}$/$6.1\times10^{16}$&&&\\
&&&&&&&&\\
$\rm{^{13}CH_2CHCN}$&-&150&$119 \pm 18$&125.86$\pm$13.67&-&$6.5 \times 10^{15}/9.60\times10^{15}(1.07)$&$(3.5\pm0.4) \times 10^{15}$&(5.3$\pm0.46)\times10^{15}$\\
&&&&&&&&\\
{CH$_2$}$^{13}$CHCN&-&150&-&-&-&$2.8 \times 10^{15}/5.30\times10^{15}$(0.78)&-&-\\
&&&&&&&&\\
CH$_2$CH$^{13}$CN&-&150&-&-&-&$3.0 \times 10^{15}/5.60\times10^{15}$(0.70)&-&-\\
&&&&&&&&\\
CH$_2$CHC$^{15}$N&-&150&-&-&-&$9.0 \times 10^{14}$&-&-\\
&&&&&&&&\\
$\rm{CH_2CDCN}$&-&150&-&-&-&$3.3\times10^{15}/1.00\times10^{16}$(0.67)&-&-\\
\hline
\multicolumn{9}{|c|}{}\\
\multicolumn{9}{|c|}{\it Ethyl cyanide with its isotopologues and isomer}\\
\hline
$\rm{C_2H_5CN}$&200$^a$&150&$92\pm18$&104.11$\pm$9.98&$9 \times 10^{17a}$&$2.3\times 10^{17}/3.80\times 10^{17}(1.35)$&(1.2$\pm0.3)\times 10^{17}$&$(1.7\pm0.10)\times10^{17}$\\
&$103\pm12^b$&&&&$3.6\times10^{14}$/$1.1\times10^{17}$&&&\\
&&&&&&&&\\
$^{13}$CH$_3$CH$_2$CN&-&150&$220\pm112$&130.79$\pm$23.17&-&$8.5\times 10^{15}/1.5\times10^{16}(1.16)$&(9.7$\pm2.3)\times10^{15}$&$(7.8\pm1.12)\times10^{15}$\\
&&&&&&&&\\
$\rm{{CH_3}^{13}CH_2CN}$&&150&-&&-&$8.0 \times 10^{15}$&-&-\\
&&&&&&&&\\
$\rm{CH_3{CH_2}^{13}CN}$&&150&-&-&&$3.2\times10^{15}/4.70\times10^{15}(1.22)$&-&-\\
%&&&&&&&&\\
%$\rm{CH_3$^{13}$CH_2N}$&&&&&&&&\\
%&&&&&&&&\\
%$\rm{CH_3CH_2$^{13}$CN}$&&&&&&&&\\
&&&&&&&&\\
C$_2$H$_5$C$^{15}$N&-&150&-&-&-&$3.1 \times 10^{15}$&-&-\\
&&&&&&&&\\
$\rm{C_2H_5NC}$&&150&-&-&&$4.0\times10^{15}/6.40\times10^{15}(1.05)$&-&-\\
\hline
\multicolumn{9}{|c|}{}\\
\multicolumn{9}{|c|}{\it Cyanoacetylene and its isotopologues}\\
\hline
$\rm{HC_3N}$&200$^a$/500$^a$&150&&&$5.0 \times 10^{16a}/10^{18a}$&$4.1\times10^{15}/4.0\times10^{15}(2.18)$&&\\
&&&&&&&&\\
$\rm{HC^{13}CCN}$&&150&&&&$1.4\times10^{15}/2.30\times10^{15}(1.
17)$&&\\
&&&&&&&&\\
$\rm{HCC^{13}CN}$&&150&&&&$1.7\times10^{15}/2.80\times10^{15}$(1.16)&&\\
\hline
\multicolumn{9}{|c|}{}\\
\multicolumn{9}{|c|}{\it Cyanodiacetylene and its isotopologues}\\
\hline
$\rm{HC_5N}$&&150.0&&&&$7.7\times10^{14}/1.70\times10^{15}$(1.20)&&\\
&&&&&&&&\\
$\rm{HCC^{13}CCCN}$&&150.0&&&&$1.0\times10^{15}/2.35\times10^{15}(0.95)$&&\\
\hline
\multicolumn{9}{|c|}{}\\
\multicolumn{9}{|c|}{\it Cyanamide}\\
\hline
$\rm{NH_2CN}$&200$^a$&150.0&&&$2.0 \times 10^{16a}$&$4.8\times10^{15}/1.30\times10^{16}$(0.86)&&-\\
\hline
\multicolumn{9}{|c|}{}\\
\multicolumn{9}{|c|}{\it Aminoacetonitrile}\\
\hline
$\rm{H_2NCH_2CN}$&&150.0&&&&$5.5\times 10^{15}/9.30\times10^{15}(1.07)$&&-\\
\hline
%\multicolumn{9}{|c|}{\bf Nitrogen Sulfide}\\
%$\rm{NS-33}$&&&&&&&&\\
%&&&&&&&&\\
%$\rm{NS-34}$&&&&&&&&\\
%\hline
\multicolumn{9}{|c|}{}\\
\multicolumn{9}{|c|}{\it Methylacetylene and its isotopologues}\\
\hline
$\rm{CH_3CCH}$&&150.0&$188.9\pm$16&219.12$\pm$14.80&&$7.0\times10^{16}/8.30\times10^{16}(1.67)$&$(6.0\pm0.5)\times10^{16}$&$(8.7\pm0.75)\times10^{16}$\\
&&&&&&&&\\
$\rm{{CH_3}^{13}CCH}$&&150&&&&$8.0\times10^{15}/9.80\times10^{15}(1.17)$&&-\\
\hline
\multicolumn{9}{|c|}{}\\
\multicolumn{9}{|c|}{\it Other nitrogen-bearing molecules}\\
\hline
$\rm{HCN}$&200$^a$/500$^a$&&&&$1.0\times10^{18}$/$3.0\times10^{18}$&&&\\
$\rm{CH_3NH2}$&64$^c$&&&&$6.3\times10^{15}$&&&\\
$\rm{CH_2NH}$&84/84$^d$&&&&$2.1 \times10^{17}$/$4.7 \times10^{15}$&&&\\
$\rm{CH_3CN}$&408/82$^e$&&&&$5.1 \times10^{17}$/$4.1\times10^{14}$&&&\\
&240/180$^f$&&&&$6.0\times10^{16}$/$3.6\times10^{15}$&&&\\
$\rm{CH_3NCO}$&$248\pm19^b$&&&&$1.2 \times10^{17}$&&&\\
$\rm{HNCO}$&$317\pm25^b$&&&&$1.3\times10^{17}$&&&\\
$\rm{NH_2CHO}$&$439\pm100^b$&&&&$3.8 \times10^{16}$&&&\\
&$200^a$&&&&$2.0\times10^{17}$&&&\\
\hline
\end{tabular}

\begin{tablenotes}
$^a$\citep[consider source size 1.5$^{''}$, H$_2$ column density $5.0\times10^{24}$ cm$^{-2}$  , and an additional component of $0.5^{''}$ and 500 K, which are used for HC$_3$N and HCN]{rolf11}; $^b$\citep[low value column density is beam averaged ($23{''}$) and high value column density is source averaged ($1.3{''}$), consider H$_2$ column density $1.2\times10^{25}$ cm$^{-2}$]{font07}; 
$^c$\citep[considering source size $10{''}$ and H$_2$ column density $2.6\times10^{23}$cm$^{-2}$]{masa19};
$^d$\citep[for high column density, considering source size $1.4{''}$ and H$_2$ column density $6.69\times10^{24} cm^{-2}$ and low column density obtained, considering source size $10{''}$ and  H$_2$ column density $1.3\times10^{23} cm^{-2}$]{suzu16};
$^e$\citep[a high rotational temperature and column density are for compact component $\sim0.6^{''}$ and low for extended components $\sim 4.0^{''}$, consider H$_2$ column density  $6.69\times10^{24} cm^{-2}$]{hern14}; $^f$ \citep[consider two component structure:a dense compact core $\sim 1{''}-2{''}$ and the halo extended about 10 times larger than core]{olmi96b}, $^*$ The average source size (in arcsec) that value is used for LTE modeling (LTE2) is noted in parentheses.
\end{tablenotes}
\end{center}}
\end{table*}
\clearpage
\section{Column densities and spatial distributions of all molecules \label{sec:mol}}
\begin{figure}
\includegraphics[width=9cm]{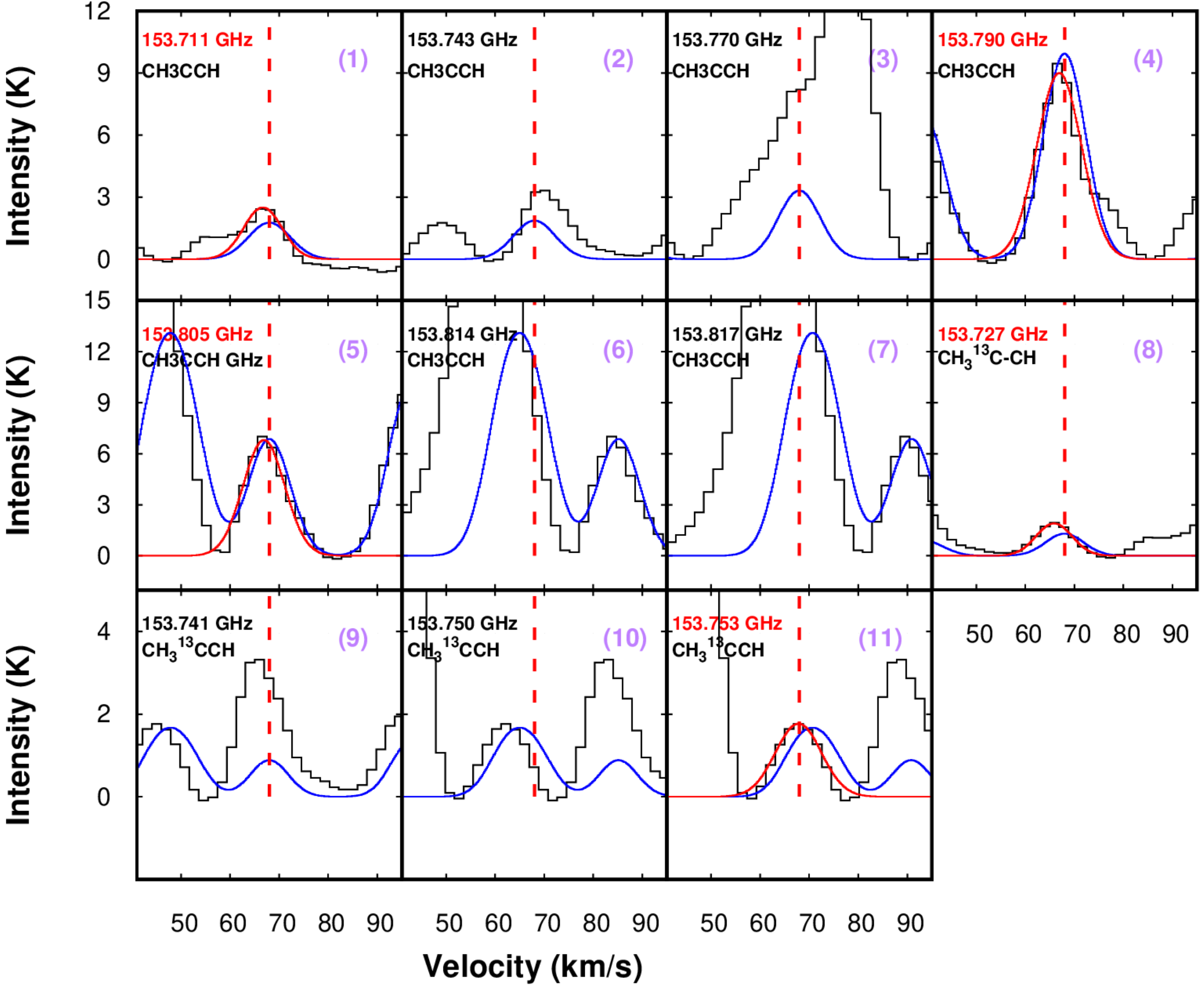}
\caption{Gaussian-fit spectra and LTE spectra (same as Figure \ref{fig:fit1}) of CH$_3$CCH and its isotopologues.}
\label{fig:fit4}
\end{figure}

%\vskip -2cm
The column densities of all species are estimated by generating the synthetic spectra under an LTE approximation. The blended transitions are also identified by carefully examining the synthetic spectra of two nearby transitions. As the input parameters of the LTE model, the source size of $\sim 2^{''}$ (half-maximum
diameter of our beam size), an excitation temperature of 150 K, and molecular hydrogen column density $\rm N(H_2) \sim 1.35 \times 10^{25}$ cm$^{-2}$ \citep{gora20} are considered. The FWHM of a species is chosen by taking an average FWHM of the detected, unblended, optically thin transitions. The column density of a species is varied to obtain a satisfactory fit with the observed data. Figures \ref{fig:fit1}, \ref{fig:fit2}, \ref{fig:fit3}, and \ref{fig:fit4} show the LTE-fit spectra (see captions of these figures for more details). Table \ref{eqn:clmn} provides the best-fit column densities of all species.
We also generated another set of LTE spectra (called LTE2) by keeping all other parameters the same but considering a different source size. For this fitting, the measured emitting regions of an unblended transition and the average value of emitting regions for multiple transitions are used as an input source size. 
 These LTE spectra (LTE2) are shown in Appendices \ref{fig:fit_size1}, \ref{fig:fit_size2}, \ref{fig:fit_size3}, and \ref{fig:fit_size4}, and the obtained column densities are noted in Table \ref{tab:clmn}. 
 We find that the estimated column density slightly increased by using measured emitting regions as source size instead of constant value ($\sim 2^{''}$).
The estimated values of column densities and temperatures with the rotational diagram, MCMC, and LTE methods are summarized in Table \ref{tab:clmn}. All the estimated parameters agree well among themselves. However, the rotational diagram and MCMC method provide more realistic values because the MCMC method is an alternative method to the rotational diagram. Therefore, for comparison and to better constrain the estimated parameters, we employed the MCMC approach. In the LTE model, we have only one free parameter:  column density. However, if all other fixed parameters (e.g., FWHM, source size, excitation temperature) are known with certainty, the simple LTE model also provides an outstanding result. Here, the rotational diagram method uses excitation temperature measurements for five molecules. Therefore, we know the gas's excitation temperature.
Furthermore, the FWHM and source size of each molecule are also known from our measurement. To test the reliability of the simple LTE method, we apply the LTE model to five species for which we made a rotational diagram. Interestingly, the obtained column densities with the LTE method are comparable to that obtained with the rotational diagram (see Table \ref{tab:clmn}. Therefore, unless otherwise stated, for the rest of the paper in discussions or comparison with modeling results, we use measured values of column densities obtained using a rotational diagram. For all other molecules (if the result with the rotational diagram method is unavailable), column densities obtained with the LTE model are used.
We discuss individual species in the following sections. 
\subsection{Vinyl cyanide and its isotopologues}
\subsubsection{C$_2$H$_3$CN}
Altogether, twelve transitions of vinyl cyanide are identified toward G10, which are noted in Table \ref{tab:dataobs} and shown in Figure \ref{fig:fit1}.  Among them, nine are unblended transitions, and three are blended ($147.86523$ GHz, $148.028678$ GHz, and $159.753858$ GHz). A rotational diagram (see Figure \ref{fig:rot-dia}a) is performed by considering only the unblended, optically thin transitions. It yields a rotational temperature of $155\pm 53$ K and column density of $(1.7\pm 0.4)\times 10^{17}$ cm$^{-2}$, respectively (see Table \ref{tab:clmn}). \cite{font07} found a rotational temperature of vinyl cyanide of $176\pm35$ K, which is in good agreement with our measured value.  Comparing our results with \cite{rolf11}, we obtain a similar temperature, while their column density is about two to four times larger than ours. The optically thin unblended transitions of vinyl cyanide have an emitting region of about 0.72$^{''}$-1.11$^{''}$ (see Figure \ref{fig:mm_vc} and Table \ref{tab:mm}).  The $131.168$ GHz and $158.773$ GHz transitions are found to be very compact (0.72-0.77$^{''}$), whereas the other transitions are comparatively little extended (1.04-1.11${''}$).
Though these transitions are marginally resolved, our analysis indicates that vinyl cyanide is possibly emitting from the inner regions of the source.  
\subsubsection{$^{13}$CH$_2$CHCN}
Nine transitions of the $^{13}$CH$_2$CHCN (isotopologue of vinyl cyanide) are observed toward G10 (see Table \ref{tab:dataobs}). Among them, five are blended. The four unblended transitions are optically thin, and a column density of $(3.5 \pm 0.4) \times 10^{15}$ cm$^{-2}$ and rotational temperature of $119 \pm 18$ K are obtained from rotational diagram. The $159.954$ GHz transition of $^{13}$CH$_2$CHCN is found (see panels 10-13 of Figure \ref{fig:mm_vc} and Table \ref{tab:mm}) to be very compact ($\sim 0.73^{''}$), whereas the other transitions are slightly extended ($1.16-1.20^{''}$).\\
\subsubsection{$\rm{{CH_2}^{13}CHCN}$ \& $\rm{ CH_2CH^{13}CN}$}
Another two $^{13}$C isotopologues of vinyl cyanide (CH$_2^{13}$CHCN and CH$_2$CH$^{13}$CN) are tentatively observed toward G10. Only one transition of each of these isotopologues is clearly identified (see Figure \ref{fig:fit1}). 
Column densities of $2.8 \times 10^{15}$ cm$^{-2}$ and $3.0 \times 10^{15}$ cm$^{-2}$ for ${\rm {CH_2}^{13}CHCN}$ and CH$_2$CH$^{13}$CN, respectively, depict a best LTE fit to the observed data (see solid blue lines of panels 22-25, 26-29 of Figure \ref{fig:fit1}). Both the transitions are found to be very compact (0.70-0.78 $^{''}$; see panels 14 and 15 of Figure \ref{fig:mm_vc} and Table \ref{tab:mm}). \\
\subsubsection{CH$_2$CHC$^{15}$N} 
Our simple LTE model found two possible  
transitions of $^{15}$N isotopologue of vinyl cyanide (CH$_2$CHC$^{15}$N)  (see panels 30-31 of Figure \ref{fig:fit1}). However, the transition at 148.898 GHz is partially contaminated with ethylene glycol, and the other transition at 159.436 GHz is blended with other species. So, no clear transitions are detected toward G10. We derive an upper limits column density of $\sim 9.0 \times 10^{14}$ cm$^{-2}$. \\
\subsubsection{CH$_2$CDCN}
Altogether, seven transitions of a deuterated form of vinyl cyanide (CH$_2$CDCN) are observed  toward G10 (Table \ref{tab:dataobs}). Among them, five are blended.  
The 130.645 GHz transition is possibly blended with propynal and C$_2$H$_5$C$^{15}$N. 130.661 GHz is probably blended with its other transitions and CH$_3$NCO. 130.725 GHz is probably blended with CH$_3$OCH$_3$. 130.637 GHz is probably blended with its other transitions, and 154.309 GHz is probably blended with CH$_3$OCH$_3$. Since only two transitions are well detected, we consider that deuterated vinyl cyanide is tentatively detected from our observation.The best LTE fit (see panels 32-38 of Figure \ref{fig:fit1}) yields a column density of $ 3.3 \times 10^{15}$ cm$^{-2}$. 
The emitting region of these two transitions (see Table \ref{tab:mm} and panels 11-12 of Figure \ref{fig:mm_vc}) depict that they are comparatively more compact ($0.64 - 0.69^{''}$) than vinyl cyanide and other isotopologues of vinyl cyanide.
\subsubsection{Fractionation of vinyl cyanide}
The results obtained with the rotational diagram method, MCMC method, and simple LTE estimation reflect the $^{12}$C/$^{13}$C isotopic ratio of vinyl cyanide, which varies from $45-64$ (see Table \ref{tab:clmn}). Despite a high elemental atomic H/D ratio in the ISM \citep[$6.667 \times 10^4$]{lins98}, we see a substantial deuterium enrichment ($\sim 88$ with simple LTE estimation) for vinyl cyanide. However, the $^{12}$C/$^{13}$C isotopic ratio \citep[$\sim 89$]{clay04} is roughly conserved in vinyl cyanide (45-64, with all the methods implemented).
\begin{figure*}[t]
\includegraphics[width=9cm]{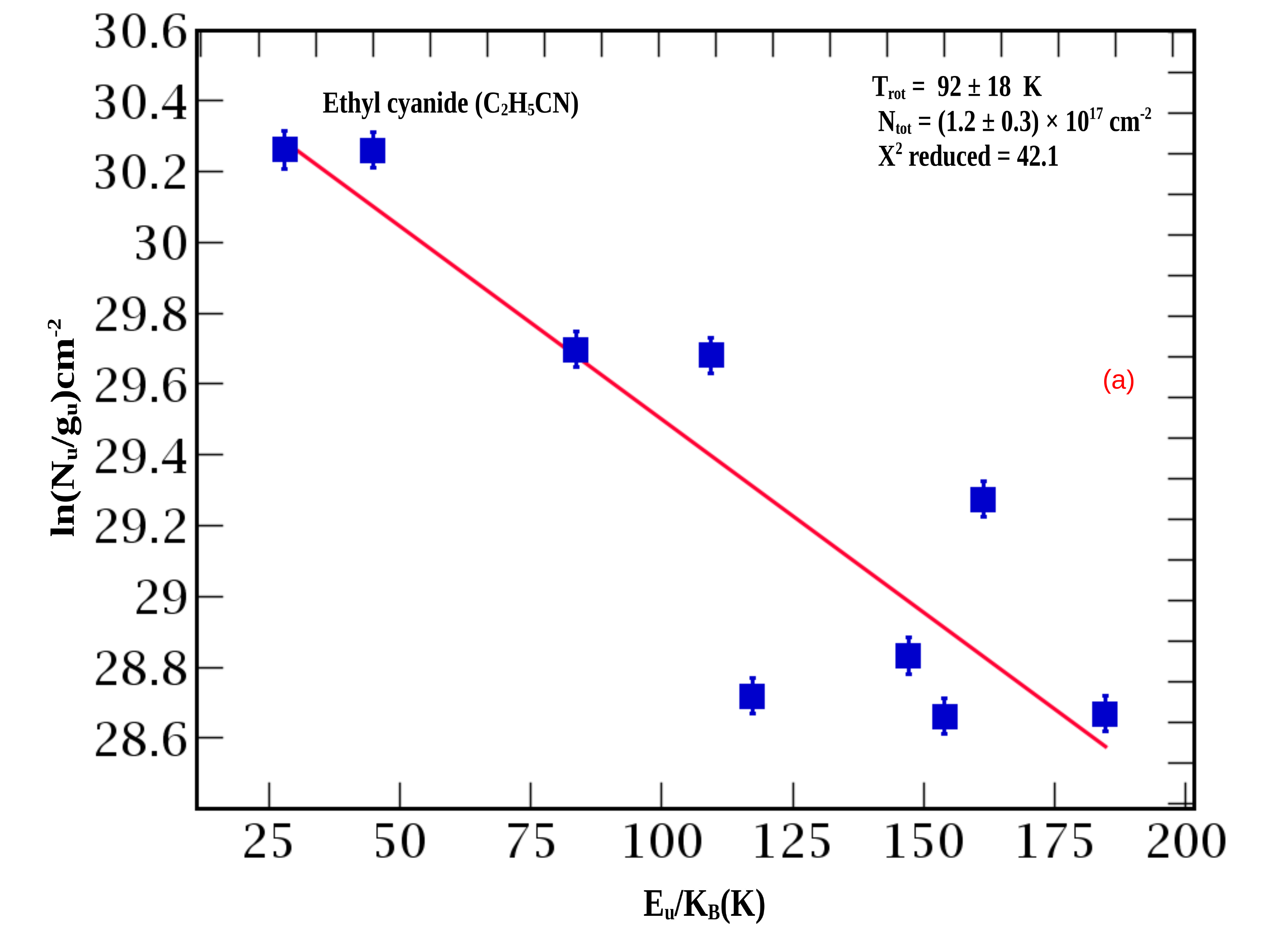}
\includegraphics[width=9cm]{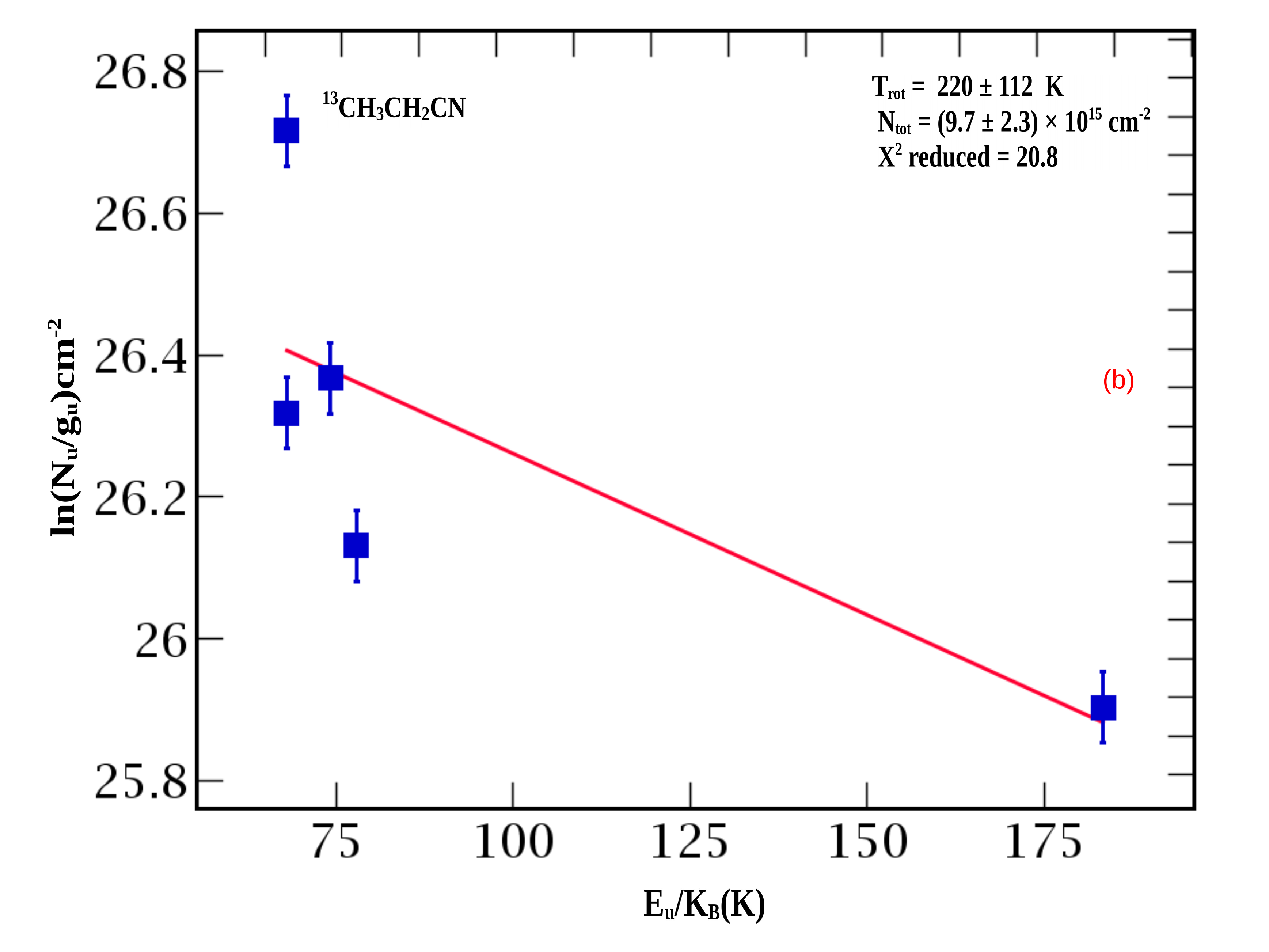}
\caption{Rotational diagram of (a) ethyl cyanide (C$_2$H$_5$CN) and (b) its one istotopologue ($^{13}$CH$_3$CH$_2$CN). The best-fit rotational temperature and column density are given in each panel. Filled blue squares are the data points and blue line represents the error bar.  Error bar of each transition is calculated from rms noise (varying from 70 mK to 155 mK) and calibration error (considering 5$\%$).}
\label{fig:rot-dia1}
\end{figure*} 
\begin{figure*}
\centering
\includegraphics[width=18cm]{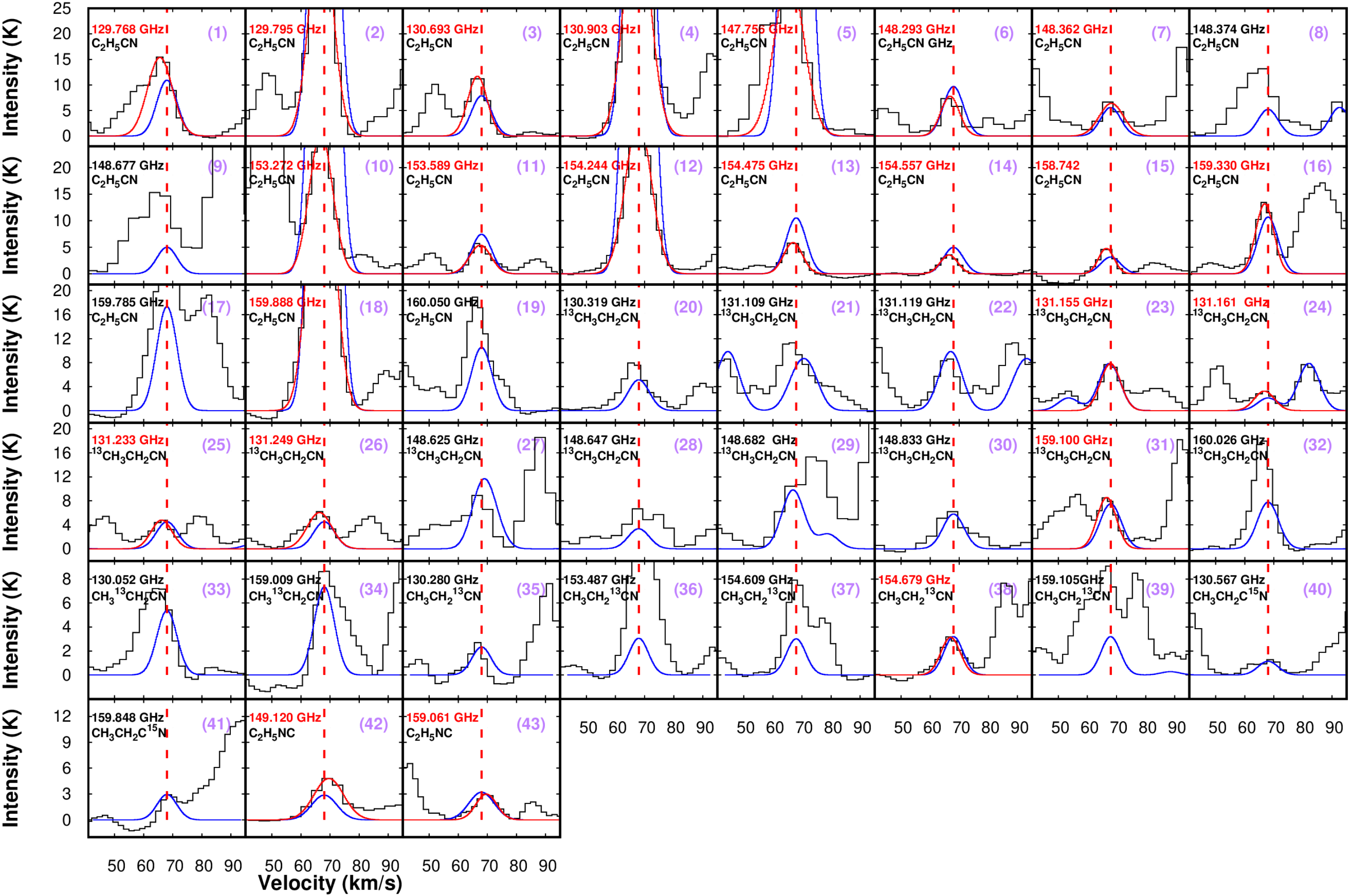}
\caption{Gaussian-fit and modeled LTE spectra (same as Figure \ref{fig:fit1}) of the unblended transitions of ethyl cyanide, its isotopologues ($^{13}$CH$_3$CH$_2$CN, CH$_3$$^{13}$CH$_2$CN, CH$_3$CH$_2$$^{13}$CN, C$_2$H$_5$C$^{15}$N), and one of its isomers (C$_2$H$_5$NC).}
\label{fig:fit2}
\end{figure*} 
\clearpage
\subsection{Ethyl cyanide with its isotopologues and isomer}
\subsubsection{C$_2$H$_5$CN}
In total, nineteen transitions of ethyl cyanide are identified toward G10 (see Table \ref{tab:dataobs}). Among them, four are blended. Within the fifteen unblended transitions, six are optically thick, and nine are optically thin. From rotational diagram analysis (see Figure \ref{fig:rot-dia1}a), we obtained  a column density of $1.2 \times 10^{17}$ cm$^{-2}$ and rotational temperature of $92$ K.
Our derived rotational temperature is similar to the rotational temperature ($103\pm12$ K) of ethyl cyanide derived by \cite{font07}. \cite{rolf11} estimated a higher column density of ethyl cyanide ($9 \times 10^{17}$ cm$^{-2}$) than vinyl cyanide ($7.0 \times 10^{17}$ cm$^{-2}$). However, with all our methods (simple LTE, rotational diagram, and MCMC), a higher column density of vinyl cyanide is obtained compared to ethyl cyanide (see Table \ref{tab:clmn}).  \cite{font07} found vinyl cyanide is $\sim 10$ times more abundant than ethyl cyanide. However, when they considered all the transitions (strong, weak and vibrationally excited lines), they obtained the best fit parameters: column density $\sim1.7 \times 10^{17}$cm$^{-2}$, rotational temperature $\sim140$ K, source size $\sim 1^{''}$. The emitting regions of the optically thin transitions vary from $1.08^{''}-1.48^{''}$ (see panels 1-9 of Figure \ref{fig:mm-ec} and Table \ref{tab:mm}).\\
\subsubsection{$^{13}$CH$_3$CH$_2$CN}
Thirteen transitions of the $^{13}$C isotopologue of ethyl cyanide are observed toward this source (Table \ref{tab:dataobs}). Among them, five transitions are unblended. The rotational diagram yielded a rotational temperature and a column density of $220$ K and $9.7 \times 10^{15}$ cm$^{-2}$, respectively (see Figure \ref{fig:rot-dia1}b). The emitting region of the unblended and optically thin transitions of this isotopologue is found (see Figure \ref{fig:mm-ec}) to be more compact (0.93-1.29$^{''}$) than the transitions of ethyl cyanide.\\
\subsubsection{CH${_3}^{13}$CH$_2$CN}
Two transitions of this $^{13}$C isotopologues are detected toward G10 (Table \ref{tab:dataobs}) but both transitions are found to be blended. A column density upper limit of $8.0 \times 10^{15}$ cm$^{-2}$ is estimated by the LTE approximation.\\
\subsubsection{CH$_3$CH$_2^{13}$CN}
Five transitions of this species are observed toward G10 (Table \ref{tab:dataobs}. Our LTE analysis indicates that three of them are blended.
Since only two transitions are clearly detected, we consider this tentatively detected molecule with a column density of $2.9 \times 10^{15}$ cm$^{-2}$. The emitting region's unblended transitions of CH$_3$CH$_2^{13}$CN vary from 1.35-1.22$^{''}$ (panels 15 and 16 of Figure \ref{fig:mm-ec}). \\
\subsubsection{\rm{C$_2$H$_5$C$^{15}$N}}
Any transitions of a $^{15}$N isotopologue of ethyl cyanide (C$_2$H$_5$C$^{15}$N) are not identified toward G10 with full confidence.  Our LTE model found only two possible transitions (see panels 40-41 of Figure \ref{fig:fit2}), with an upper-limit column density of  $\sim 3.1\times 10^{15}$ cm$^{-2}$.\\
\subsubsection{CH$_3$CH$_2$NC}
 CH$_3$CH$_2$NC is tentatively detected toward G10. Only two unblended transitions of the isomer of ethyl cyanide are observed. Our best-fit LTE spectra of these transitions yield a column density of $4.0 \times 10^{15}$ cm$^{-2}$ (blue lines in panels 38-39 of Figure \ref{fig:fit2}). The emitting region of the transition at 159.061 GHz is found to be very compact ($0.75{''}$), whereas the other transition at 149.12 GHz is found to be comparatively not very extended ($1.34^{''}$).\\
\subsubsection{Fractionation of ethyl cyanide}
Multiple transitions of ethyl cyanide and one of its $^{13}$C isotopologues ($^{13}$CH$_3$CH$_2$CN) are identified. Based on the various methods (simple LTE, rotational diagram, and MCMC), a $^{12}C/^{13}C$ ratio of $12-27$ is estimated. Two transitions of another $^{13}$C isotopologue (CH$_3$CH${_2}^{13}$CN) of ethyl cyanide are identified. With the simple LTE method, a carbon fractionation of $72$ is estimated. From the simple LTE fitting, an isomeric ratio between the cyanide (C$_2$H$_5$CN) and isocyanide (C$_2$H$_5$NC) of $57.5$ is estimated.

\subsection{Cyanopolyynes}
Cyanopolyynes are a group of species with the chemical formula HC$_n$N $(n=3,5,7..)$. In addition, it contains the cyano group (CN).   
Previously, \cite{rolf11} identified  HCN, HNC, H$^{13}$CN, and HC$_3$N in G10. Here, the identifications of cyanoactylene (HC$_3$N) and its isotopologues and cyanodiacetylene (HC$_5$N) and its isotopologue are reported.\\
\subsubsection{Cyanoacetylene and its isotopologue}
HC$_3$N is observed toward G10 with one transition at 154.657 GHz.  
Additionally, transitions of its two $^{13}$C isotopologues, HC$^{13}$CCN and HCC$^{13}$CN are detected at $154.001$ GHz and $154.016$ GHz, respectively. 
 We obtained best-fit synthetic spectra with column densities of $4.1 \times 10^{15}$ cm$^{-2}$, $1.4 \times 10^{15}$ cm$^{-2}$, and $1.7 \times 10^{15}$ cm$^{-2}$, for HC$_3$N, HC$^{13}$CCN, and HCC$^{13}$CN, respectively (see blue lines in panels 1-3 of Figure \ref{fig:fit3}).
The transition of HC$_3$N is comparatively more extended ($\sim 2.18^{''}$) than the other species reported in this article (see Figure \ref{fig:mm-oth}). The isotopologues of HC$_3$N are found to be relatively compact ($1.16^{''}-1.17^{''}$).
\cite{rolf11} obtained a column density of $5.0 \times 10^{16}$ cm$^{-2}$ when they used an excitation temperature of 200 K. Additionally, with the 500 K and 0.5$^{''}$ they estimated a very high column density ($\sim 10^{18}$ cm$^{-2}$) of HC$_3$N. Here, with our simple LTE analysis, we estimate a column density of HC$_3$N is $4.1 \times 10^{15}$ cm$^{-2}$. This mismatch is possibly due to our low spatial resolution observation compared to that of \cite{rolf11}. With the simple LTE calculation, a $\rm{^{12}C/^{13}C}$ fractionation is found to be $2.41-2.93$.\\  

\subsubsection{Cyanodyacetylene}
We clearly detected one transition of HC$_5$N and one transition of its isotopologue (HCC$^{13}$CCCN). Both the transitions yield best LTE fits with the column densities of $7.7 \times 10^{14}$ cm$^{-2}$ and $1.0 \times 10^{15}$ cm$^{-2}$, respectively.  
The spatial distribution (see Figure \ref{fig:mm-oth} and Table \ref{tab:mm}) of HC$_5$N is found to be more compact ($1.2^{''}$) than HC$_3$N ($\sim 2.18^{''}$). Furthermore, the HC$_5$N line we observed has an upper state energy of $\sim 156$ K. In contrast, for HC$_3$N, it is $\sim 67$ K. The observed spatial distribution of HC$_3$N suggests that it traces the envelopes regions, whereas HC$_5$N may originate from the warm inner part of the hot molecular core. The emitting areas (see Table \ref{tab:mm}) of the isotopologue ($^{13}$C) of HC$_5$N depicts that it is more compact ($0.95^{''}$) than their parent isotopologues. The same compactness is also observed for the observed isotopologues ($^{13}$C) of HC$_3$N.
With the simple LTE calculation, a $\rm{^{12}C/^{13}C}$ fractionation of 0.77 is obtained.\\
\subsection{Cyanamide}
Six transitions of cyanamide (NH$_2$CN) are observed toward G10 (Table \ref{tab:dataobs}).  However, two of them are clearly detected. Our simple
LTE model yields a good fit (Figure \ref{fig:fit3}) to the observed spectra with a column density of $4.8 \times 10^{15}$ cm$^{-2}$. The emitting region (see Figure \ref{fig:mm-oth} and Table \ref{tab:mm}) of the two unblended transitions are compact ($\sim 0.91^{''}$).\\
\subsection{Aminoacetonitrile}
\begin{figure*}
\centering
\includegraphics[width=18cm]{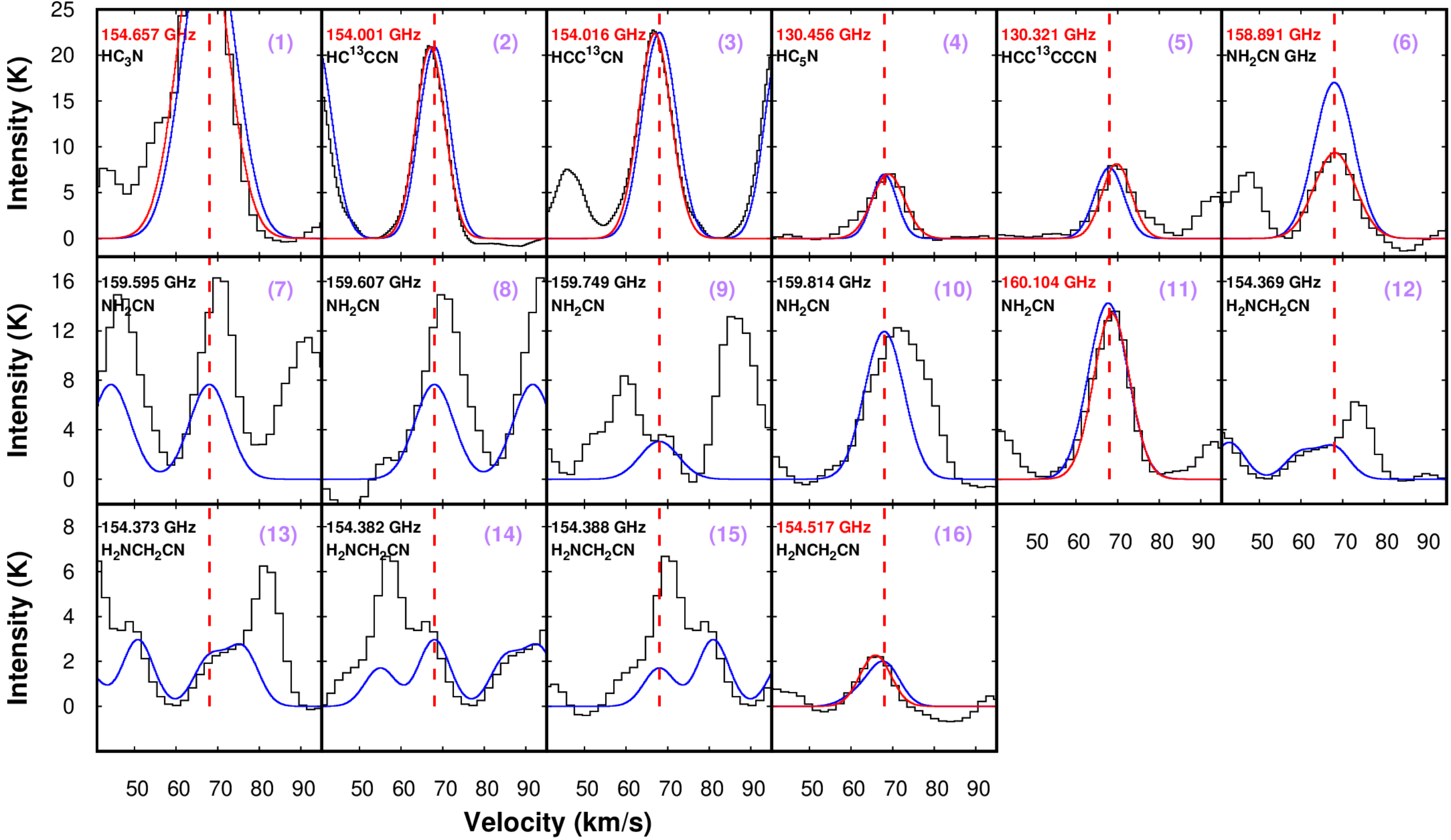}
\caption{Gaussian-fit spectra and LTE spectra (same as Figure \ref{fig:fit1}) of cyanoacetylene, cyanodiacetylene, cyanamide, and  aminoacetonitrile.}
\label{fig:fit3}
\end{figure*}

 Aminoacetonitrile (H$_2$NCH$_2$CN) is a well-known precursor of glycine. It hydrolyses to form glycine. \cite{dang11} studied the NH$_2$CH$_2$CN formation under the interstellar condition by thermal processing of ices containing methanimine, ammonia, and hydrogen cyanide. Five transitions of aminoacetonitrile are observed toward G10 ( \ref{tab:dataobs}), but four of them are blended with other species. Only the transition at 154.51748 GHz with upper state energy of 86 K is successfully identified. Therefore, we consider that this species was tentatively detected in this observation.
 A column density of $5.5 \times 10^{15}$ cm$^{-2}$ yields a good fit to the observed spectra (see panel 16 of Figure \ref{fig:fit3}). The emitting diameter of aminoacetonitrile is 1.07$^{''}$. The moment zero map of this transition is shown in panel 8 of Figure \ref{fig:mm-oth}.\\
\begin{figure}
\includegraphics[width=9cm]{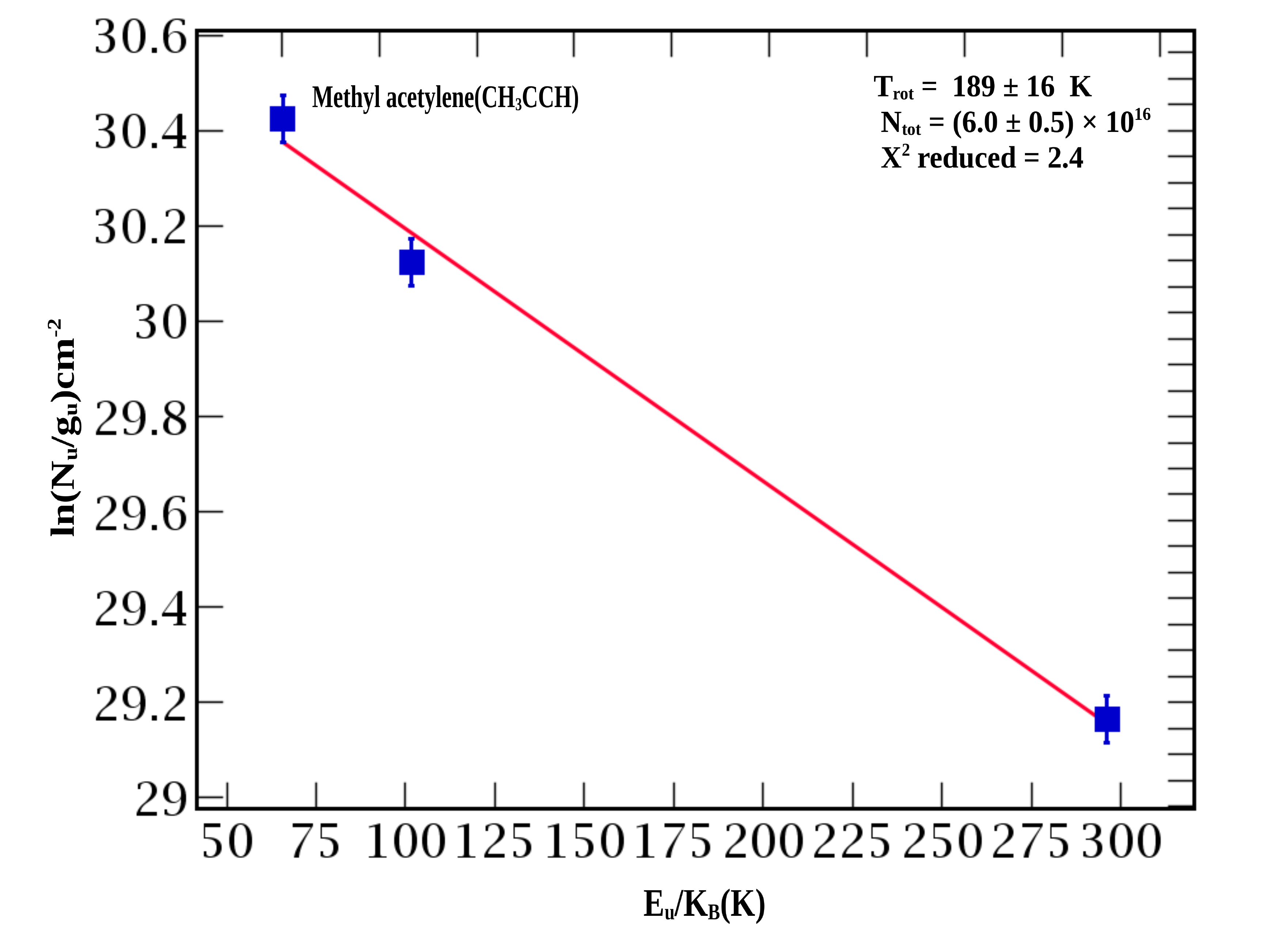}
\caption{Rotational diagram of methyl acetylene (CH$_3$CCH; same as Figures \ref{fig:rot-dia} and \ref{fig:rot-dia1}).  Error bar of each transition is calculated from rms noise (varies 70 mK to 136 mK) and calibration error (considering 5$\%$).}
\label{fig:rot-ch3cch}
\end{figure}
\subsection{Methyl acetylene}
\cite{snyd73} first identified methyl acetylene (CH$_3$CCH) in Sgr B2. Though methyl acetylene is not a nitrogen-bearing species, being a symmetric top molecule, its identification is reported here because of its utility as a tracer of the physical condition of the star-forming region.
Three transitions of methyl acetylene are clearly identified. All the transitions correspond to $J = 9 - 8$. The three transitions have $k_a$= 6, 3, and 2, respectively. This is the first identification of CH$_3$CCH in G10. 
All three unblended transitions are found to be optically thin. Rotational diagram analysis of the CH$_3$CCH (Figure \ref{fig:rot-ch3cch}) depicts a column density of $6.0 \times 10^{16}$ cm$^{-2}$ and a rotational temperature of $189$ K. Panels 1-3 of Figure \ref{fig:mm-othh} shows  moment map of CH$_3$CCH for different $k_a$ components of $J = 9 - 8$ transitions. Interestingly, it shows that the emitting regions of CH$_3$CCH transitions are related to their corresponding $k_a$ components. 
For example, the emitting region of $J = 9 - 8$ transitions with $k_a$ = 2 is 1.8$^{''}$ ($E_u=65.8 K$), and for $k_a=3$ it is $1.9^{''}$ ($E_u=101.7$ K).
Since with our spatial resolution these two transitions are marginally resolved, we did not find significant differences for close-lying energy states. However, for $k_a=6$ ($E_u=296.9$ K), a compact emission ($1.28^{''}$) is obtained. With the increase in the $k_a$ component, the upper state energy increases, which yields a more compact emitting region. 
All three detected transitions have $\Delta k_a=0$. The ratio of lines between the same J-state but other $k_a$-states can be used as temperature probe \citep{mang93}. To estimate the kinetic temperature ($T_K$), we chose some selected line ratios ($\frac{J1_{ka}-J2_{ka}}{J3_{ka'}-J4_{ka'}}$). These ratios are selected based on the following criteria \citep{mang93,das19}: 1) $\Delta J=1$, J1=J3, and J2=J4, 2) $\Delta k_a=0$ for both the transitions,
3) ${k_a'> k_a}$,
4) both the transitions are optically thin,
5) The ratio between the difference in excitation temperature belongs to different k$_a$ values.
 \begin{figure*}[t]
\centering
\includegraphics[height=7cm,width=14cm]{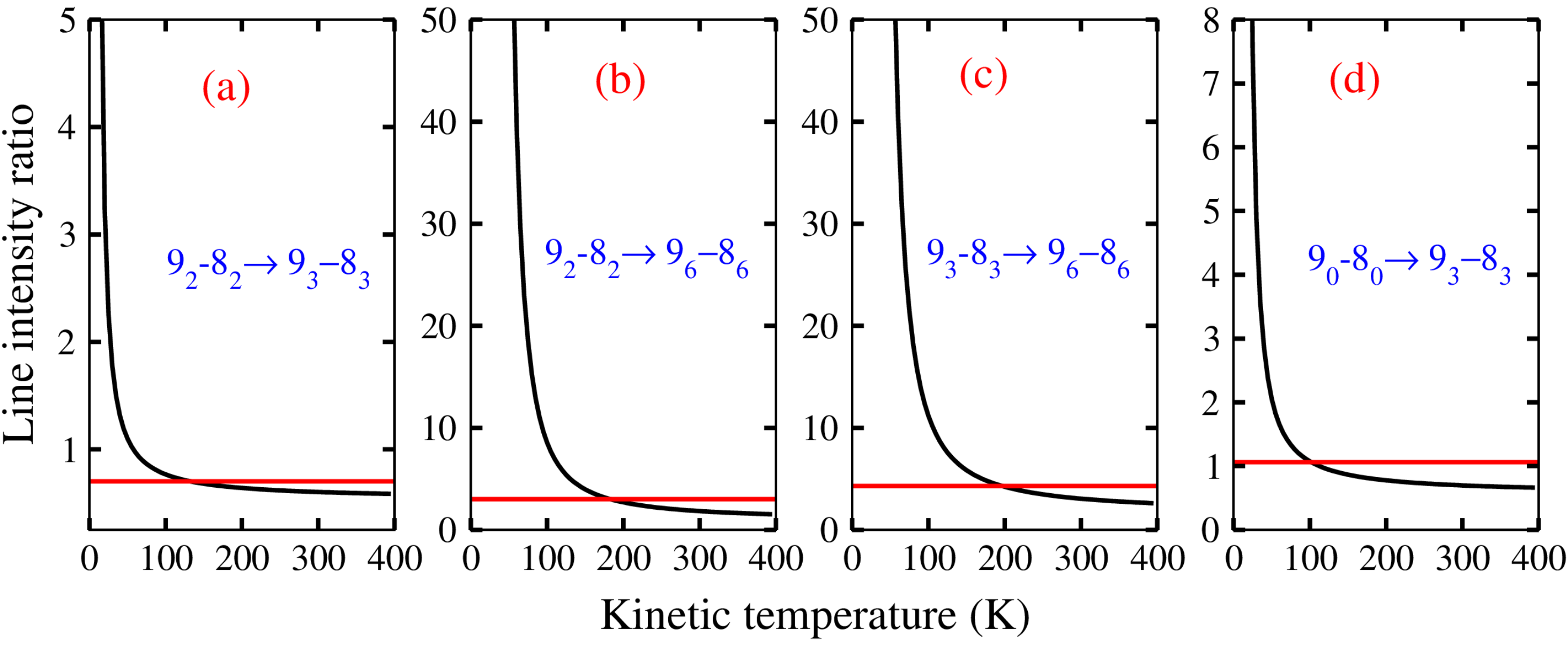}
\caption {Black line represents the expected line intensity ratio variation with the kinetic temperature, whereas the red line represents the observed intensity ratio. The ratios of CH$_3$CCH transitions are plotted in panels (a), (b), and (c), and that of CH$_3^{13}$CCH is given in panel (d).} 
\label{fig:LTE-ratio}
\end{figure*}

With the LTE approximation, the line ratio (R) between the two transitions satisfying the above criteria can be expressed as \citep{mang93} 
\begin{equation}
R= S_R exp(D/T_K),    
\end{equation}
where $D= E(J3,k_a')-E(J1,k_a)$ and $S_R= \frac{S_{J1k_a}}{S_{J3k_a'}}$. \cite{mang93} derived this relation for H$_2$CO, which is also applicable to any slightly asymmetric or symmetric rotor species. 

According to the above selection rules, in between the three observed unblended transitions, $\frac{9_2-8_2}{9_3-8_3}$, $\frac{9_2-8_2}{9_6-8_6}$, and $\frac{9_3-8_3}{9_6-8_6}$ are considered for estimating the excitation temperature. The black solid curve in Figure \ref{fig:LTE-ratio}(a-c) depicts the expected line ratio with the excitation temperature. The observed line ratio of these three transitions is shown with the solid red line. Interestingly, the observed line ratio intersects the curve between 133-193 K, closer to the estimated hot core temperature. 

Additionally, two transitions ($J = 9 - 8$ with $k_a =0, 3$) of  $^{13}$C isotopologue of methyl acetylene (CH${_3}^{13}$CCH) are observed toward G10. %Among them, $k_a=4$ transition is blended. 
Our LTE synthetic spectra fit the observed spectra well,  with a column  density of $8.0 \times 10^{15}$ cm$^{-2}$.
% SM Line parameters obtained by the Gaussian fitting of these two transitions are noted in Table \ref{tab:dataobs}. 
% SM The MCMC fitting of the two unblended transitions yields a column density and temperature of $1.7 \times 10^{16}$ cm$^{-2}$ and $82$ K, respectively. 
%The emitting regions of these two unblended transitions of CH${_3}^{13}$CCH are noted in Table \ref{tab:mm}. 
A similar trend of emitting regions is also obtained here. 
For
$k_a = 0, the$ emitting region is 1.31$^{''}$, whereas it is $1.01^{''}$ for $k_a=3$. 
%From the MCMC fitting reported in Table \ref{table:mcmc_lte}, 
%a column density of $1.7 \times 10^{16}$ cm$^{-2}$ and excitation temperature of $82$ K is obtained. 
With the simple LTE calculation, a carbon fractionation ($\frac{^{12}C}{^{13}C}$ ratio) of $8.8$ is obtained between CH$_3$CCH and CH${_3}^{13}$CCH. By considering the column density obtained with the rotational diagram and MCMC methods (for $\rm{CH_3CCH}$), a $\frac{^{12}C}{^{13}C}$ ratio of 7.5-10.9 is obtained.
%Whereas, with the MCMC fitting a fractionation of $2.7$ is obtained.

A similar process for estimating the kinetic temperature from the transitions obtained with the CH${_3}^{13}$CCH is carried out. Following the selection rule, $\frac{9_0-8_0}{9_3-8_3}$ is used to estimate the kinetic temperature. Figure \ref{fig:LTE-ratio}d depicts that the observed line ratio intersects the expected line ratio at $102$ K.

 We identify several transitions of nitrogen-bearing species along with their various $^{13}$C isotopologues. Only the unblended transitions are used for further analysis. In the case of the $^{13}$C isotopologue of vinyl cyanide, we identify four transitions of $^{13}$CH$_2$CHCN, one of $\rm{C{H_2}^{13}CHCN,}$ and one of CH$_2$CH$^{13}$CN. We obtain two transitions from one of the $^{13}$C isotopologues of ethyl cyanide ($\rm{CH_3C{H_2}^{13}CN}$). Two transitions of the $^{13}$C isotopologue of cyanoacetylene (HC$^{13}$CCN and $\rm{HCC_{13}CN}$) and one transition of the $^{13}$C isotopologue of cyanodiacetylene (HCC$^{13}$CCCN) are identified.
Furthermore, we identify two transitions of a $^{13}$C isotopologue of methyl acetylene ($\rm{C{H_3}^{13}CCH}$). We use the rotational diagram analysis and MCMC method to estimate the column density of these species if more than two transitions were identified. Otherwise, we consider the simple LTE approximation. Despite all these issues, we found that the obtained $\frac{12_C}{13_C}$ ratio of all these species varies from 0.8-72. In the case of methyl acetylene, vinyl, and ethyl cyanide, it is 7.5-72, which is generally consistent with the elemental  $\frac{12_C}{13_C}$ ratio  ($\sim 89$). For the cyanopolyynes (cyanoacetylene and cyanodiacetylene), we obtain a small $^{13}$C enrichment (0.77-2.93). Since not many transitions were obtained for the cyanopolyynes, we restrict our comments based on the upper limit of the column densities obtained.

\begin{table}
\centering{
\tiny{
\caption{Emitting regions of observed molecules.\label{tab:mm}}
\begin{tabular}{|c|c|c|c|}
\hline
Molecules &   frequency (GHz)  &  E$_u$ (K) &  Diameter (arcsec)\\
%          &  GHz         &      K  &   (arcsec)\\
\hline
C$_2$H$_3$CN&130.763576&18.2&1.04$\pm$0.033\\
&131.168737&209.5&0.77$\pm$0.030\\
&131.267473$^b$&47.5&1.07$\pm$0.022\\
&147.561701$^b$&62.5&1.28$\pm$0.059\\
&153.42175&44.13&1.11$\pm$0.015\\
&153.518944$^b$&70.9&1.48$\pm$0.016\\
&154.72454$^b$& 65.4&1.10$\pm$0.016\\
&158.657428$^b$&69.0&1.07$\pm$0.027\\
&158.773785&85.56&0.72$\pm$0.047\\
\hline
$^{13}$CH$_2$CHCN&147.927607&164.8612&1.20$\pm$0.054\\
&147.986686&79.5&1.17$\pm$0.045\\
&149.423715&69.2&1.16$\pm$0.049\\
&159.954635&71.3&0.73$\pm$0.048\\
\hline
CH$_2$$^{13}$CHCN&153.925606&65.0&0.78$\pm$0.02\\
\hline
CH$_2$CH$^{13}$C-N&154.034285&65.1&0.70$\pm$0.029\\
\hline
CH$_2$CDCN&130.720066&74.1&0.69$\pm$0.059\\
&159.38385&83.9&0.64$\pm$0.071\\
\hline
C$_2$H$_5$CN&129.76814&44.9&1.23$\pm$0.023\\
&129.795661$^b$&51.1&1.47$\pm$0.021\\
&130.693882&109.4&1.13$\pm$0.025\\
&130.903902$^b$&50.8&1.63$\pm$0.052\\
&148.293988&147.03&1.32$\pm$0.024\\
&147.756711$^b$&64.5&1.63$\pm$0.052\\
&148.36276&28.1&1.48$\pm$0.019\\
&153.272198$^b$&75.9&1.59$\pm$0.018\\
&153.589524&184.6&1.10$\pm$0.017\\
&154.244297$^b$&68.2&1.67$\pm$0.016\\
&154.47598&117.2&1.22$\pm$0.017\\
&154.55736&154.0&1.14$\pm$0.015\\
&158.742366&161.4&1.08$\pm$0.049\\
&159.3308&83.6&1.22$\pm$0.026\\
&159.888831$^b$&77.6&1.46$\pm$0.027\\
\hline
$^{13}$C-H$_3$CH$_2$CN&131.155447&77.8&1.21$\pm$0.023\\
&131.161428&183.4&0.93$\pm$0.040\\
&131.233679&67.9&1.17$\pm$0.025\\
&131.249984&67.9&1.29$\pm$0.026\\
&159.10002&74.2&1.18$\pm$0.031\\
\hline
CH3CH2$^{13}$C-N%&130.280373&50.5&$1.35\pm0.136$\\
&154.679342&71.8&1.22$\pm$0.022\\
\hline
C$_2H_5$NC&149.120363&62.1&1.34$\pm$0.040\\
&159.061678&69.60&0.75$\pm$0.060\\
%$16_{1,16}\rightarrow15_{1,15}$&62.2&1.36\\
%C$_2$H$_5$NC&$23_{3,21}\rightarrow23_{1,22}$&138.4&0.93\\
% 
%C$_2H_5$CN-15&$15_{5,10}\rightarrow14_{5,9}$&77.93&1.02\\
%C$_2H_5$CN-15&$17_{6,11}\rightarrow16_{6,10}$&103.9&1.09\\
\hline
HC$_3$N&154.657284&66.8&2.18$\pm$0.031\\
HC$^{13}$CCN&154.001217&66.5&1.17$\pm$0.019\\
HCC$^{13}$CN&154.016078&66.5&1.16$\pm$0.017\\
\hline
HC$_5$N&130456.437&156.5&1.20$\pm$0.027\\
HCC$^{13}$C-CCN&130321.351&156.3&0.95$\pm$0.030\\
\hline
NH$2$CN&158.891146&48.8&0.80$\pm$0.109\\
&160.104805&164.9&0.91$\pm$0.026\\
\hline
NH$_2$CH$_2$CN&154.51748&86.4&1.07$\pm$0.018\\
\hline
CH$_3$CCH&153.71152&296.1&1.28$\pm$0.021\\
&153.790769&101.7&1.92$\pm$0.037\\
&153.805457&65.71&1.80$\pm$0.06\\
%9$_6\rightarrow8_6$&296.88&1.29\\
%&9$_5\rightarrow8_5$&217.489&1.498\\
%&9$_3\rightarrow8_3$&101.71&1.86\\
%&9$_2\rightarrow8_2$&65.815&1.61\\
\hline
CH$_3$$^{13}$CCH&153.72701&101.7&1.01$\pm$0.10\\
&153.75337&36.9&1.31$\pm$0.034\\
%9$_3\rightarrow8_3$&101.70&1.01\\
%&9$_2\rightarrow8_2$&65.697&1.509\\
%&9$_0\rightarrow8_0$&36.895&1.22\\
%&9$_6\rightarrow8_6$&153.71153&296.88&7.835$\pm$0.32870&67.604$\pm$0.1235&3.742&44.016$\pm$1.6536\\
\hline
\end{tabular}
\begin{tablenotes}
\item Spatial resolution at four different frequencies are $1.5^{''}$(129.5-131.5 GHz), $1.72^{''}$(147.5-149.4 GHz),$1.66^{''}$(153-154.9 GHz), and $1.76^{''}$(158.5-160.43 GHz).
\end{tablenotes}
}}
\end{table}
% 
%\begin{table}
%\centering{
%\scriptsize{
%\caption{Fractional abundance of  observed molecules.}
%\begin{tabular}{|c|c|c|c|p{1.1cm}|}
%\hline
%Molecules&G10.47+0.03&G31.41+0.31&SgrB2&OrionKL\\
%\hline
%C$_2$H$_3$CN&&&&\\
%% 
%\hline
%C$_2$H$_5$CN&&&&\\
%\hline
%HC$_3$N&&&&\\
%\hline
%HC$^{13}$CCN&&&&\\
%\hline
%HCC$^{13}$CN&&&&\\
%\hline
%HCC$^{13}$C-CCN&&&&\\
%\hline
%H$_2$NCH$_2$CN&&&&\\
%\hline
%NH$_2$CN&&&&\\
%\hline
%NS-33&&&&\\
%%\hline
%CH$_3$CCH&&&&\\
%\hline
%CH$_3$$^{13}$CCH&&&&\\
%\hline
%\end{tabular}}}
%\end{table}
%% 
%% 
%%%%%%%%%%%%%%%%%%%%%%%%%%%%%%%%%%%%%%%%%%%%
\section{Modeling}
\label{sec:model}
Modeling the time evolution of chemical composition of high mass sources such as G10 requires knowledge of the history of their physical structure's evolution over the whole hot core formation stage, including the early diffuse stage, collapse stage, and the warm-up stage. However, the physical evolution during the initial stages of massive hot cores and its timescales are still poorly known, and the difficult task of reconstructing the evolutionary history of the physical structure of massive hot sources such as G10 is beyond the scope of this work. Nevertheless, to better understand the factors that can have a major impact on the chemical history of the hot cores such as G10, we did extensive chemical modeling. 

In the present work, following \cite{Garrod2013}), we adopted a three-stage physical model to simulate the time-dependent chemical evolution of N-bearing species detected in G10. The three stages describing the evolution of physical structure of G10-like sources are 1) an isothermal cloud collapse stage, 2) a relatively short warm-up stage, post cloud collapse, and 3) a hot and dense post-warm-up stage. Here, the three-stage model is only explained briefly as this was already explained in detail in \citet{sil2018,das19,gora20,bhat21}. In the first stage, the cloud collapses iso-thermally from a diffuse state (initial H density, $n_{\rm H}$ = $10^3$ cm$^{-3}$) to a dense state ($n_{\rm H}$ = $10^6$ cm$^{-3}$ or $10^7$ cm$^{-3}$ depending on the model). During the cloud collapse stage, visual extinction ($A_v$) also varies from its minimum to maximum value. The first stage is followed by a warm-up stage during which the gas temperature ($T_g$) and the dust temperature ($T_d$) vary. The third stage is called the post-warm-up stage. 
 We use the above-described three-stage physical model in our 0D and 1D simulations. First, we use 0D simulation for its simplicity to test if our adopted physical model can explain the observed abundances of N-bearing species in G10. 0D simulations do not give us any information about the possible spatial distribution of species abundances. So, we try to find the possible spatial distribution of simulated species in our 1D simulations.  We explain our 0D and 1D simulations in Sections \ref{sec:physicalG10} and \ref{sec:physicalG101D}, respectively. Results for both types of simulations are presented in Sections \ref{model-I} and  \ref{model-II}, respectively.

\subsection{0D physical model}
\label{sec:physicalG10}

We started with the simple 0D simulation. We simulate the source as a single point and do not consider any spatial distribution of physical parameters ($n_{\rm H}$, $A_v$, $T_g$, $T_d$). Following \cite{gora20}, we carried out our first simulation with generally accepted parameters for a source such as G10 (initially $n_{\rm H} = 10^3$ cm$^{-3}$ , which goes up to $10^7$ cm$^{-3}$ during cloud collapse; initial $A_v = 2,$ which goes up to 250; initial $T_g =40$ K, which goes up to 200 during the warm-up stage; and initial  $T_d$ = 20 K, which goes up to 200 K). Since G10 is massive hot core, we selected a collapse time of $10^5$ years and a warm-up time of $5 \times 10^4$ years. 
%With above selected parameter values, we note that the best fit time lies between $1.01 \times 10^6$ years to $1.02 \times 10^6$ years. This time in simulation corresponds to the warm-up time of simulated cloud. This time is shorter than what \cite{gora20} reported. \cite{gora20} obtained a good matching between simulated and observed abundances at $1.12 \times 10^6$ years. \textbf{ We discuss this in later section}  {\color{red} need to highlight our short collapsing time results}. 

Knowing that chemistry is very sensitive to certain physical parameters such as temperature and density, we are required to test a number of 0D models by varying the values of most relevant physical parameters. Hence, first we repeat our simulation with a different peak core density of $10^6$ cm$^{-3}$. We also simulated a number of different models where we used different values of peak gas temperature (400, 300, and 200 K) during the warm-up phase and initial dust temperature (15, 20, and 25 K) during the collapse phase to cover the possible physical evolution history of hot cores such as G10. 
The warm-up phase is a very important phase in chemical evolution history as the dust grain chemistry is very sensitive to dust temperature. This makes the warm-up duration a very critical parameter. Therefore, we used three different values for the total duration ($T_w = 10^4, 5\times 10^4,$ and $10^5$ years) of the warm-up phase in order to properly investigate the dependence of simulated abundances of N-bearing species on the warm-up time of the collapsed cloud. 

Depending on the values of physical parameters used during the time evolution of these three-stages, a number of 0D models are implemented to study the chemical evolution in G10-like dense cores. In Table \ref{tbl:physicalEvo}, we list values of different physical parameters used in different models. More details about these models can be found in Section \ref{model-I}, where we discuss results for the 0D model.

\subsection{1D physical model}
\label{sec:physicalG101D}
\begin{figure}
\includegraphics[width=.48\textwidth]{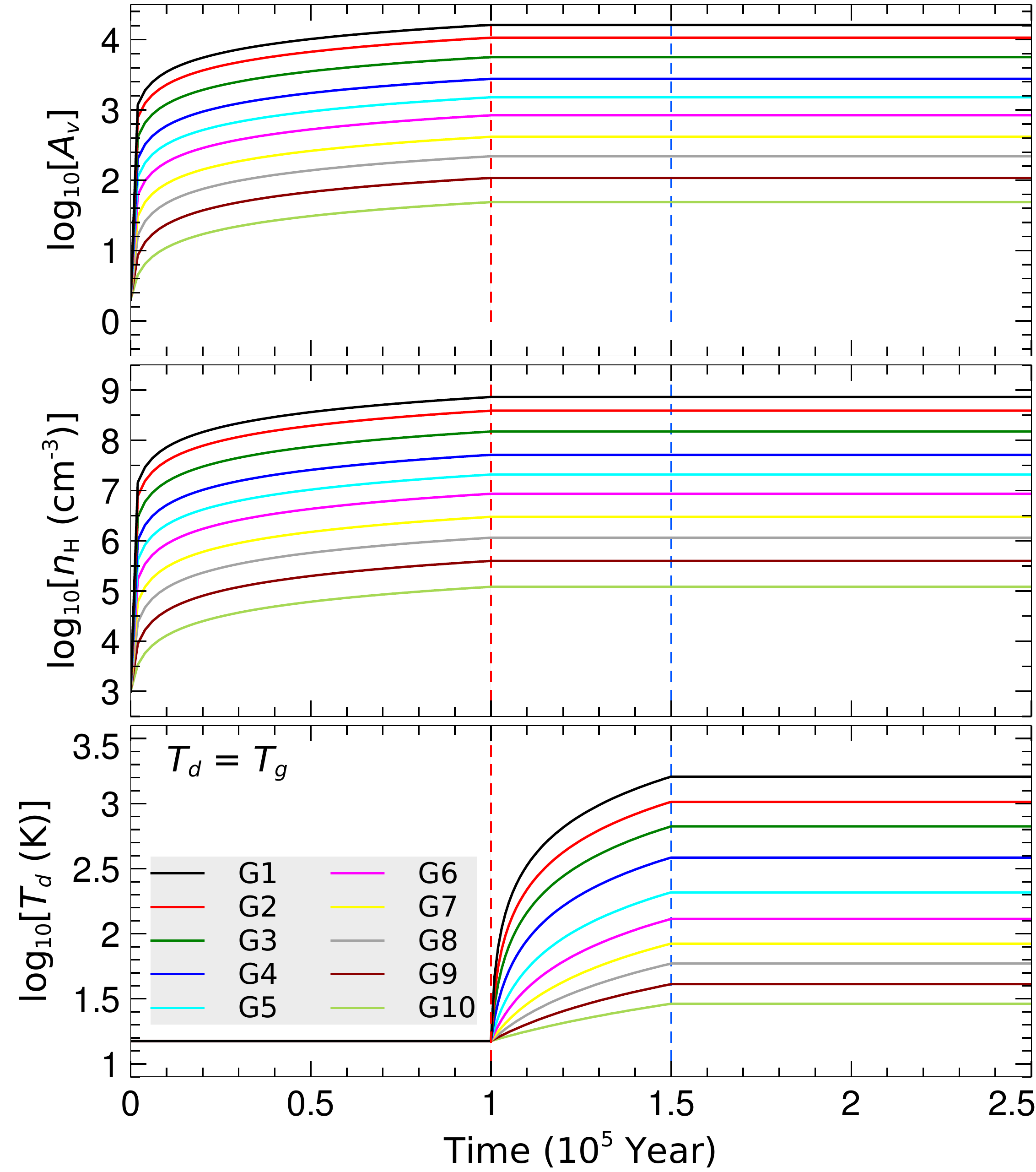}
\caption{Time evolution of $n_{\rm H}$, $A_v$, and $T_d = T_g$ considered from \cite{van13} for ten different grids. Red and blue vertical dashed lines show the end of collapse and warm-up phases, respectively.} 
\label{fig:physical_profiles2}
\end{figure}
 To study the spatial distribution of the chemical abundances of G10, we considered the the density and temperature distribution obtained by SED fitting by \cite{van13}. Very recently, \cite{bhat21} constructed a similar model for G31.041+0.31. They divided the cloud envelope into 23 logarithmic spacing shells. Here, for G10, we divided the cloud envelope into ten logarithmic spacing shells. The time evolutions of the physical parameters ($n_{\rm H}$, $A_v$, $T_d = T_g$) for these ten shells are shown in Figure \ref{fig:physical_profiles2}. 

For each shell, we consider that, at $t=0$ year, all the shells have $n_H=10^3$ cm$^{-3}$. In the first stage, every shell is allowed to evolve up to its spatial density distribution obtained by \cite{van13}. For example, for the fifth shell's (at $\sim 1717$ AU) density may evolve up to $2.08 \times 10^7$ cm$^{-3}$ at the end of $10^5$ years. Beyond that (warm-up and post-warm-up stages), the density of each shell is kept constant at its maximum value. Since dust temperature is a critical parameter, we used two profiles (15 and 20 K) for the initial dust temperature. For simplicity, we kept the gas temperature $T_g$ equal to the dust temperature $T_d$. In the first stage, $T_d = T_g$ was kept constant at initial value (15 or 20 K depending on model). In the warm-up stage, $T_d = T_g$ evolves from its initial value to final value respective of the spatial temperature distribution as reported by \cite{van13} (for example, $207$ K for the fifth shell). In the post-warm-up stage, gas and dust temperatures are kept constant at their highest values respective to each shell. To calculate $A_V$, we used the relation $A_V = A_{V,0} \times (n_{\rm{}H}/n_{\rm{}H,0})^{(2/3)}{}$, as given in Sect. 3.3 of \cite{Garrod2011}, where A$_{V,0}$ is the initial value of visual extinction and $n_{\rm{}H,0}$ is the initial total number density. We used  A$_{V,0} = 2$ and $n_{\rm{}H,0} = 10^3$ cm$^{-3}$.
\subsection{Chemical model}
\label{sec:nautilus}
To explain the observed abundances of chemical species in G10, a three-phase (gas phase, grain surface, and grain mantle) version of the Nautilus gas-grain chemical code \citep{Ruaud2016} is used. This code is based on the rate equation approximation \citep{Hasegawa1992, Hasegawa1993b}. In all our simulations, the surface of the dust grain is defined as the top two monolayers of species. All ice species under the surface are treated as mantles. When a species desorbs from the surface, another species from the mantle comes to the top to form the new surface layer, and, similarly, with the accretion of a new species on the surface, another surface species moves to the mantle to form a new mantle layer. In our model, standard physical processes such as accretion, diffusion, recombination, and thermal desorption of species are considered. The cosmic-ray-induced desorption, UV (direct and indirect) photo-desorption, and chemical desorption are also considered. These physical processes are explained in detail in  \citet{Ruaud2016, Wasim2018,das19,Wakelam2020}. We used a common cosmic-ray ionization rate of $1.3 \times 10^{-17}$ s$^{-1}$. In Table \ref{tbl:initial-abundances}, we list the initial abundances of all the gas and the grain species we used in our all chemical models.
% In Table \ref{tbl:initial-abundances} we list the initial abundances of all the gas and the grain species we used in our all chemical model.
% 
%
%
\begin{table}
   \begin{center}
   \caption{Elemental abundances and initial abundances.}
   \label{tbl:initial-abundances}
   \begin{tabular}{@{}lll}
   \hline
   \hline
   Element                      &       Abundance relative to H    & References  \\
   \hline
   H$_2$                        &       0.5                        &             \\
   He                           &       0.09                       &$^\textrm{a}$ \\
   N                            &       6.2$\times10^{-5}$         &$^\textrm{b}$ \\
   O                            &       2.4$\times10^{-4}$         &$^\textrm{c}$ \\      
   C$^+$                        &       1.7$\times10^{-4}$         &$^\textrm{b}$ \\   
   S$^+$                        &       8.0$\times10^{-9}$         &$^\textrm{d}$ \\      
   Si$^+$                       &       8.0$\times10^{-9}$         &$^\textrm{d}$ \\
   Fe$^+$                       &       3.0$\times10^{-9}$         &$^\textrm{d}$ \\
   Na$^+$                       &       2.0$\times10^{-9}$         &$^\textrm{d}$ \\
   Mg$^+$                       &       7.0$\times10^{-9}$         &$^\textrm{d}$ \\   
   P$^+$                        &       2.0$\times10^{-10}$        &$^\textrm{d}$ \\      
   Cl$^+$                       &       1.0$\times10^{-9}$         &$^\textrm{d}$ \\ 
   ice                          &       0                          &              \\
   all other gas species        &       0                          &              \\
   \hline
   \end{tabular}
   \end{center} 
%    \medskip
   $^\textrm{a}$\citet{Wakelam08}, $^\textrm{b}$\citet{Jenkins09}, 
   $^\textrm{c}$\citet{Hincelin2011}, $^\textrm{d}$\citet{Graedel82}   
\end{table}
\begin{table*}
\begin{center}
\caption{Simulated models and physical parameters that describe the 0D model.}
\label{tbl:physicalEvo}
\begin{tabular}{llllllllllllll}
\hline
& Simulated models        & $t_{c}$ (Years)   & $t_{w}$ (Years)   & $t_{pw}$ (Years)  & $n_{\rm H}$ (cm$^{-3}$) & $A_v$   & $T_g$ (K)    &  $T_d$ (K)   \\
\hline    
      &N7-Tg200-Td20      & $10^5$            & $5\times 10^4$    & $10^5$            & $10^3$ --- $10^7$ & 2 ---  250    &   40 --- 200 &    20 --- 200 \\            
      &N7-Tg300-Td20      & $10^5$            & $5\times 10^4$    & $10^5$            & $10^3$ --- $10^7$ & 2 ---  250    &   40 --- 300 &    20 --- 300 \\
      &N7-Tg400-Td20      & $10^5$            & $5\times 10^4$    & $10^5$            & $10^3$ --- $10^7$ & 2 ---  250    &   40 --- 400 &    20 --- 400 \\
 %    &N7-Tg400-Td20      & $10^6$            & $10^5$            & $10^5$            & $10^3$ --- $10^7$ & 2 ---  250    &   40 --- 400 &    20 --- 400 \\
Set 1 &N7-Tg200-Td15      & $10^5$            & $5\times 10^4$    & $10^5$            & $10^3$ --- $10^7$ & 2 ---  250    &   40 --- 400 &    15 --- 200 \\
%     &N7-Tg400-Td20      & $10^6$            & $10^4$            & $1.9 \times 10^5$ & $10^3$ --- $10^7$ & 2 ---  250    &   40 --- 400 &    20 --- 400 \\
      &N7-Tg200-Td25      & $10^5$            & $5\times 10^4$    & $10^5$            & $10^3$ --- $10^7$ & 2 ---  250    &   40 --- 400 &    25 --- 200 \\
      &N7-Tg200-Td20-slow-wup & $10^5$            & $10^5$            & $5 \times 10^4$   & $10^3$ --- $10^7$ & 2 ---  250    &   40 --- 400 &    20 --- 200 \\
      &N7-Tg200-Td20-fast-wup & $10^5$            & $10^4$            & $1.4 \times 10^5$ & $10^3$ --- $10^7$ & 2 ---  250    &   40 --- 400 &    20 --- 200 \\ 
\hline     
      &N6-Tg200-Td20      & $10^5$            & $5\times 10^4$    & $10^5$            & $10^3$ --- $10^6$ & 2 ---  250    &   40 --- 200 &    20 --- 200 \\            
      &N6-Tg300-Td20      & $10^5$            & $5\times 10^4$    & $10^5$            & $10^3$ --- $10^6$ & 2 ---  250    &   40 --- 300 &    20 --- 300 \\
      &N6-Tg400-Td20      & $10^5$            & $5\times 10^4$    & $10^5$            & $10^3$ --- $10^6$ & 2 ---  250    &   40 --- 400 &    20 --- 400 \\
 %    &N6-Tg400-Td20      & $10^6$            & $10^5$            & $10^5$            & $10^3$ --- $10^6$ & 2 ---  250    &   40 --- 400 &    20 --- 400 \\
Set 2 &N6-Tg200-Td15      & $10^5$            & $5\times 10^4$    & $10^5$            & $10^3$ --- $10^6$ & 2 ---  250    &   40 --- 400 &    15 --- 200 \\
%     &N6-Tg400-Td20      & $10^6$            & $10^4$            & $1.9 \times 10^5$ & $10^3$ --- $10^6$ & 2 ---  250    &   40 --- 400 &    20 --- 400 \\
      &N6-Tg200-Td25      & $10^5$            & $5\times 10^4$    & $10^5$            & $10^3$ --- $10^6$ & 2 ---  250    &   40 --- 400 &    25 --- 200 \\
      &N6-Tg200-Td20-slow-wup & $10^5$            & $10^5$            & $5 \times 10^4$   & $10^3$ --- $10^6$ & 2 ---  250    &   40 --- 400 &    20 --- 200 \\
      &N6-Tg200-Td20-fast-wup & $10^5$            & $10^4$            & $1.4 \times 10^5$ & $10^3$ --- $10^6$ & 2 ---  250    &   40 --- 400 &    20 --- 200 \\ 
%
%      &N6-Tg250-Td15 & $10^5$            & $5\times 10^4$    & $1.5 \times 10^5$ & $10^3$ --- $10^6$ & 2 ---  250 &   40 --- 250 &    15 --- 250 \\                  
%      &N6-Tg300-Td15 & $10^5$            & $5\times 10^4$    & $1.5 \times 10^5$ & $10^3$ --- $10^6$ & 2 ---  250 &   40 --- 300 &    15 --- 300 \\
%      &N6-Tg400-Td15 & $10^5$            & $5\times 10^4$    & $1.5 \times 10^5$ & $10^3$ --- $10^6$ & 2 ---  250 &   40 --- 400 &    15 --- 400 \\
%     &N6-Tg400-Td20 & $10^6$            & $10^5$            & $10^5$            & $10^3$ --- $10^6$ & 2 ---  250 &   40 --- 400 &    20 --- 400 \\
%Set 2 &N6-Tg400-Td20 (N6-mid)& $10^5$            & $5\times 10^4$    & $1.5 \times 10^5$ & $10^3$ --- $10^5$ & 2 ---  250 &   40 --- 400 &    20 --- 400 \\
%     &N6-Tg400-Td20 & $10^6$            & $10^4$            & $1.9 \times 10^5$ & $10^3$ --- $10^6$ & 2 ---  250 &   40 --- 400 &    20 --- 400 \\
%      &N6-Tg400-Td25 & $10^5$            & $5\times 10^4$    & $1.5 \times 10^5$ & $10^3$ --- $10^6$ & 2 ---  250 &   40 --- 400 &    25 --- 400 \\
%      &N6-slow       & $10^5$            & $10^5$            & $10^5$            & $10^3$ --- $10^6$ & 2 ---  250 &   40 --- 400 &    20 --- 400 \\
%      &N6-fast       & $10^5$            & $10^4$            & $1.9 \times 10^5$ & $10^3$ --- $10^6$ & 2 ---  250 &   40 --- 400 &    20 --- 400 \\  
                                                                                                                                                      \\
\hline
\end{tabular}
\end{center}
During cloud collapse period $t_{c}$: $T_g$ and $T_d$ values remain constant at their initial values and $n_{\rm H}$ and $A_v$ vary from initial to final values as given above. During the warm-up period $t_{w}$: $T_d$ = $T_g$, which varies from the initial to final value as given above, while $n_{\rm H}$ and $A_v$ values remain unchanged at their final values. During post-warm-up period $t_{pw}$: all $n_{\rm H}$, $A_v$, $T_g$, and $T_d$ remain constant at their final values. Total time of simulation in each model is $t_c + t_w + t_{pw} = 2.5 \times 10^5$ years.
\end{table*}
\subsection{Chemical network}
\label{sec:reaction}
Our gas-phase chemistry is mostly based on the kida.uva.2014 public network \citep{Wakelam2015}. We updated our gas phase chemistry by adding all relevant reactions from KIDA and UMIST databases. Apart from this we also followed Table 8 and Table 9 of \citet{gora20}, and Table 5 of \citet{sil2018} to complete our gas-phase nitrogen chemistry. Our surface chemistry is mostly based on \citet {Wasim2018}. To complete the surface chemistry, we added all relevant formation and destruction reactions in our chemical network from Table 14 of \cite{bell09} and Table 8 of \citet{gora20}. We also added all surface reactions, related to the formation and destruction of C$_2$H$_3$CN and C$_2$H$_5$CN, given in Table B.3 and Table B.4 of \cite{garrod2017}.   
\section{Chemical simulation results}
\label{sec:result}

We present our results in two sections. First, we present our simulation results for 0D models shown in Table \ref{tbl:physicalEvo}. Results from these simulations were considered to select the most suitable parameters for our 1D simulations. Results for 1D simulations are presented in Section \ref{model-II}. 
To find the best-fit-simulated model, we used a simple method in which we calculate the mean distance of disagreement between the simulated and observed values. This method is explained in detail in \cite{Wasim2018}.

\subsection{Results obtained with 0D model \label{model-I}}

Density is one of the fundamental properties of a core that determines the future state of the core. It is challenging to consider the final density of a collapsed core just after its isothermal collapse without having comprehensive knowledge of its collapse. Hence, in our simulations, we adopt two final density profiles for G10 to study their impact on chemical abundances at the end of the simulation.

\subsubsection{Impact of peak H density}
\label{sec:resultnH}
\begin{figure}
\includegraphics[width=.48\textwidth]{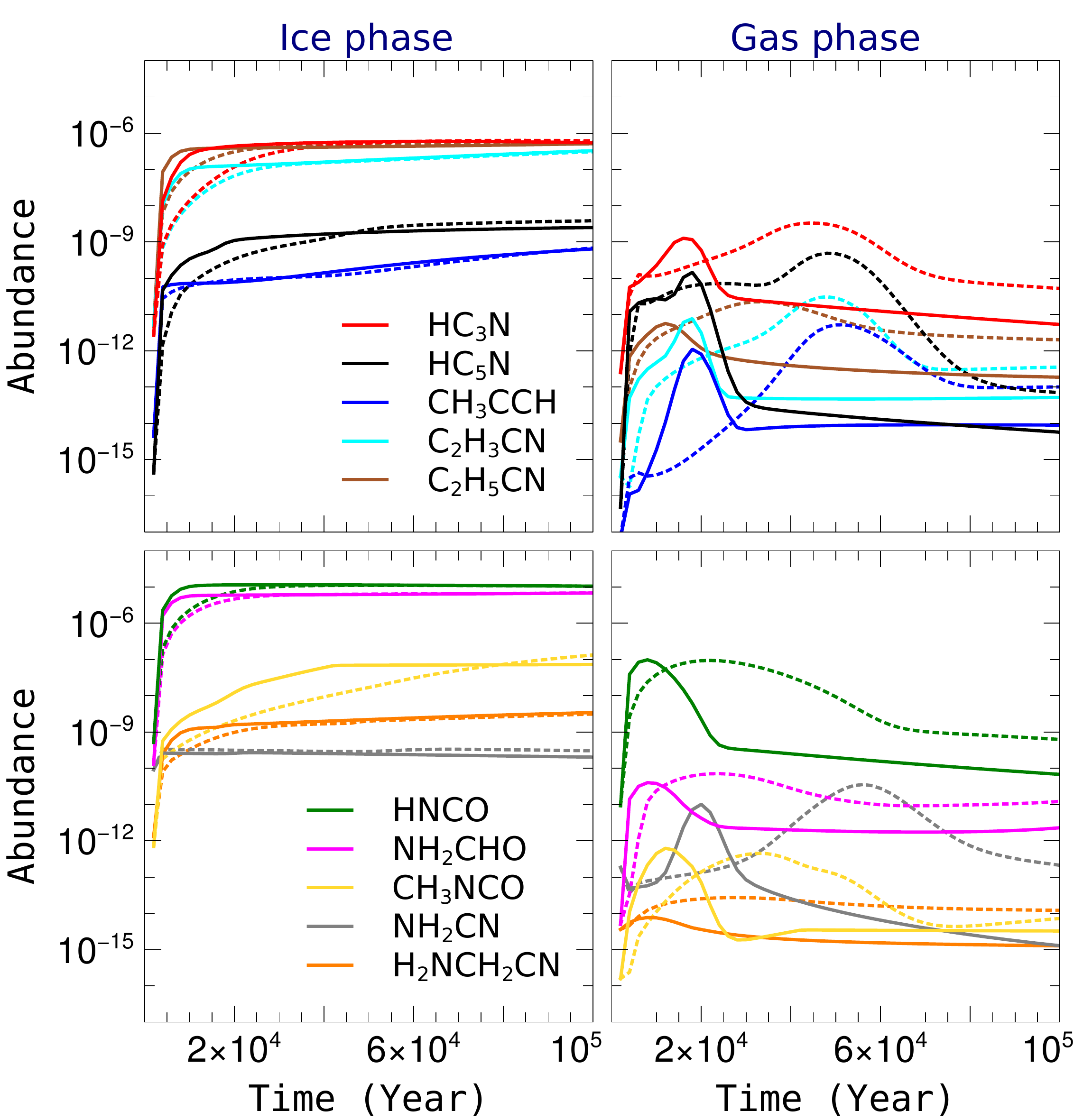}
\caption{Impact of peak H density (solid lines ($10^7$ cm$^{-3}$ ) versus dotted lines ($10^6$ cm$^{-3}$ ) on simulated abundances of selected species during the cloud-collapse phase. Plots on the left and the right are for the ice and gas-phase abundances, respectively. Legends in the top left plot are for both plots at the top, and legends in the bottom left plot are for both plots at the bottom. All solid lines are for model N7-Tg200-Td20, and all dotted lines are for model N6-Tg200-Td20 (see Table \ref{tbl:physicalEvo} for details of each model).}
\label{fig:n7n6c}
\end{figure}
\begin{figure*}
\centering
\includegraphics[width=.85\textwidth]{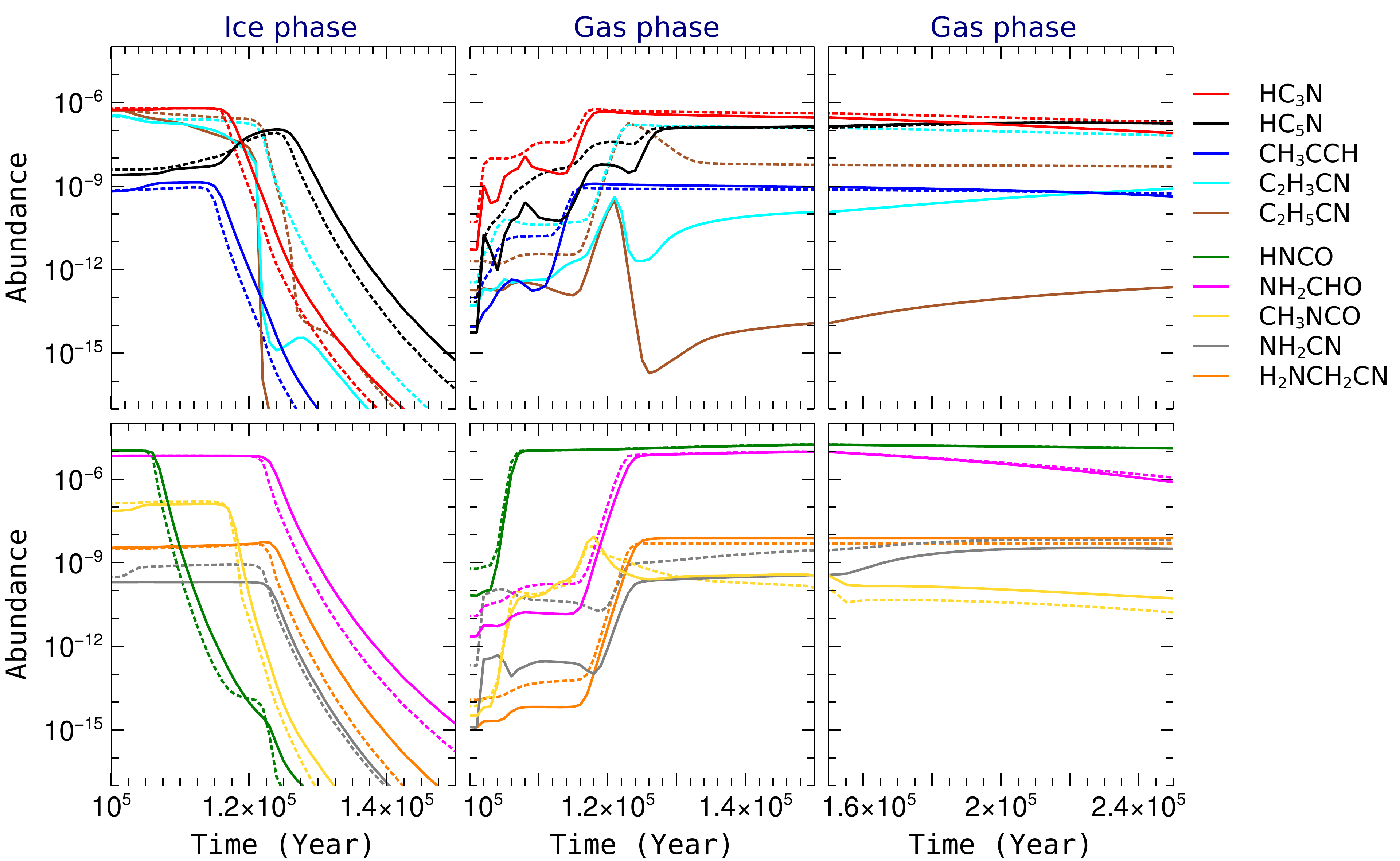}
\caption{Same as Figure \ref{fig:n7n6c}, but for warm-up and post-warm-up stages. Plots on the left show the ice abundance (warm-up phase only). Plots in the middle and right show the gas-phase abundances during the warm-up and the post-warm-up phases, respectively.}
\label{fig:n7n6wup}
\end{figure*}
In Figure \ref{fig:n7n6c}, we show simulated abundances (gas and total ice) as a function of time for nine N-bearing species and CH$_3$CCH (observed in G10) during the collapse stage, 
 where total ice abundance includes the contribution of both surface and mantle.
In all plots, solid lines are for the model N7-Tg200-Td20 (final $n_{\rm H} = 10^7$ cm$^{-3}$) and dotted lines are for the model N6-Tg200-Td20 (final $n_{\rm H} = 10^6$ cm$^{-3}$). In Figure \ref{fig:n7n6c}, we notice a general trend that is followed by all species. In denser clouds (plotted with solid lines), as the cloud collapses and the gas density increases, it takes less than $2 \times 10^4$ years for all species to attain their maximum abundance values in %both 
the gas phase. %and the ice. 
This is followed by a rapid depletion of these species before 
 a relatively stable state is reached; this state is maintained until the start of warm-up phase. At this stage, the abundances of different species continue to evolve, but at a very low rate.

During the isothermal collapse, the dust temperature is only 20K, which is too low for thermal desorption of these complex organic species from the dust surface. We further checked to verify that nonthermal desorption does not contribute to this rapid growth in gas phase species.  In fact, a rapid increase in the gas phase abundances is almost entirely driven by the gas-phase formation of these species. During this phase, HC$_5$N is mostly produced in the gas phase through H$_2$C$_5$N$^+$ + e$^- \rightarrow$ H + HC$_5$N  and N + C$_6$H $\rightarrow$ C + HC$_5$N. Similarly, two major reactions contributing to the rapid growth of HC$_3$N in the gas phase are N + C$_4$H $\rightarrow$ C + HC$_3$N  and  C + H$_2$CCN $\rightarrow$ H + HC$_3$N.

As expected, in model  N6-Tg200-Td20, with a lower peak H density (plotted with dotted lines), gas phase formation and destruction rates are relatively small compared to model N7-Tg200-Td20, and it takes longer (up to $6 \times 10^4$ years) to reach a similar state of chemical evolution to that seen in model  N7-Tg200-Td20.
%and at higher abundance values.

During the warm-up stage (see Figure \ref{fig:n7n6wup}), the gas-phase abundances rise rapidly due to desorption from the dust surface as surface temperature rises. HNCO is the first to desorb (just above 60 K) from the surface. All other species start rapid desorption when the surface temperature rises above 100 K. %During this stage all species, 
%with exception of NS, %each abundances above $10^{-10}$. 
The best fit time for model with lower peak H density also lies during the early stages of warm-up phase when abundances of most species lies in the range of $10^{-10}$ to $10^{-9}$. 

We note that the model with lower peak H density ($10^6$ cm$^{-3}$) results in higher abundances of the considered species at the end of the cloud-collapse stage. However, this difference rapidly disappears during the warm-up stage (except for C$_2$H$_3$CN and C$_2$H$_5$CN). 
%It is clear that smaller value of peak H density can not produce lower observed values of selected species during the post-warm-up stage. 
We also note that the gas-phase abundance is affected by considering a lower value of peak H density, while the ice abundances remain very much unaffected.
\subsubsection{Impact of initial dust temperature}
\label{sec:resultdt}
\begin{figure*}[t]
\centering
\includegraphics[width=.48\textwidth]{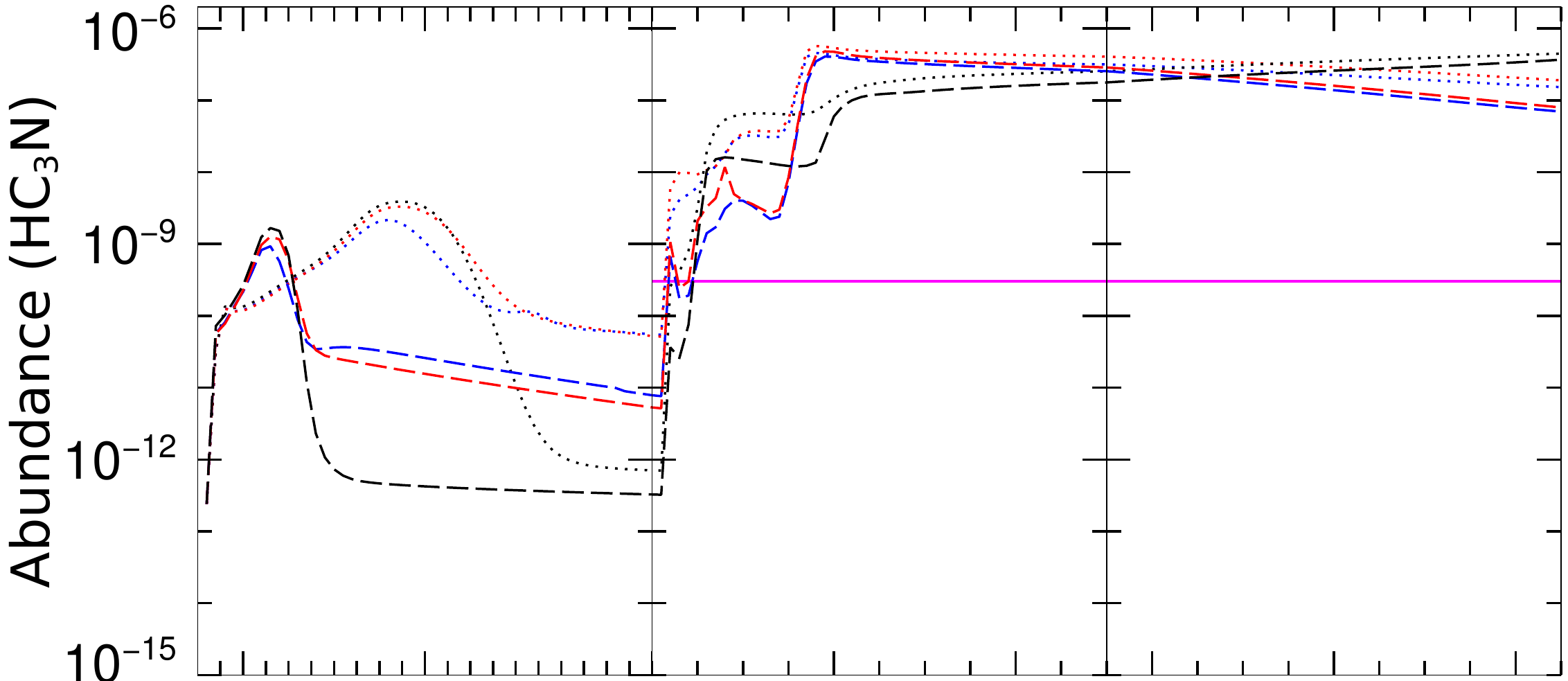}
\includegraphics[width=.48\textwidth]{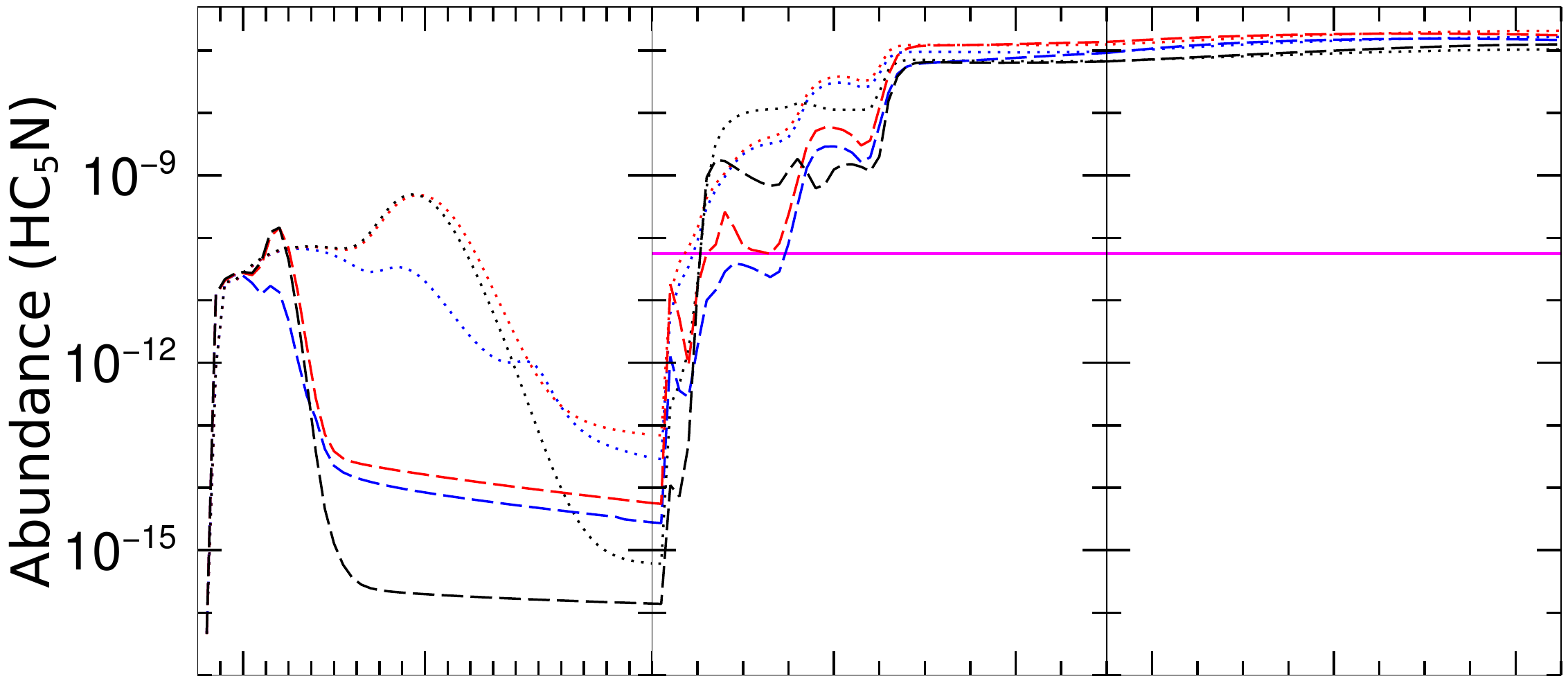}
\includegraphics[width=.48\textwidth]{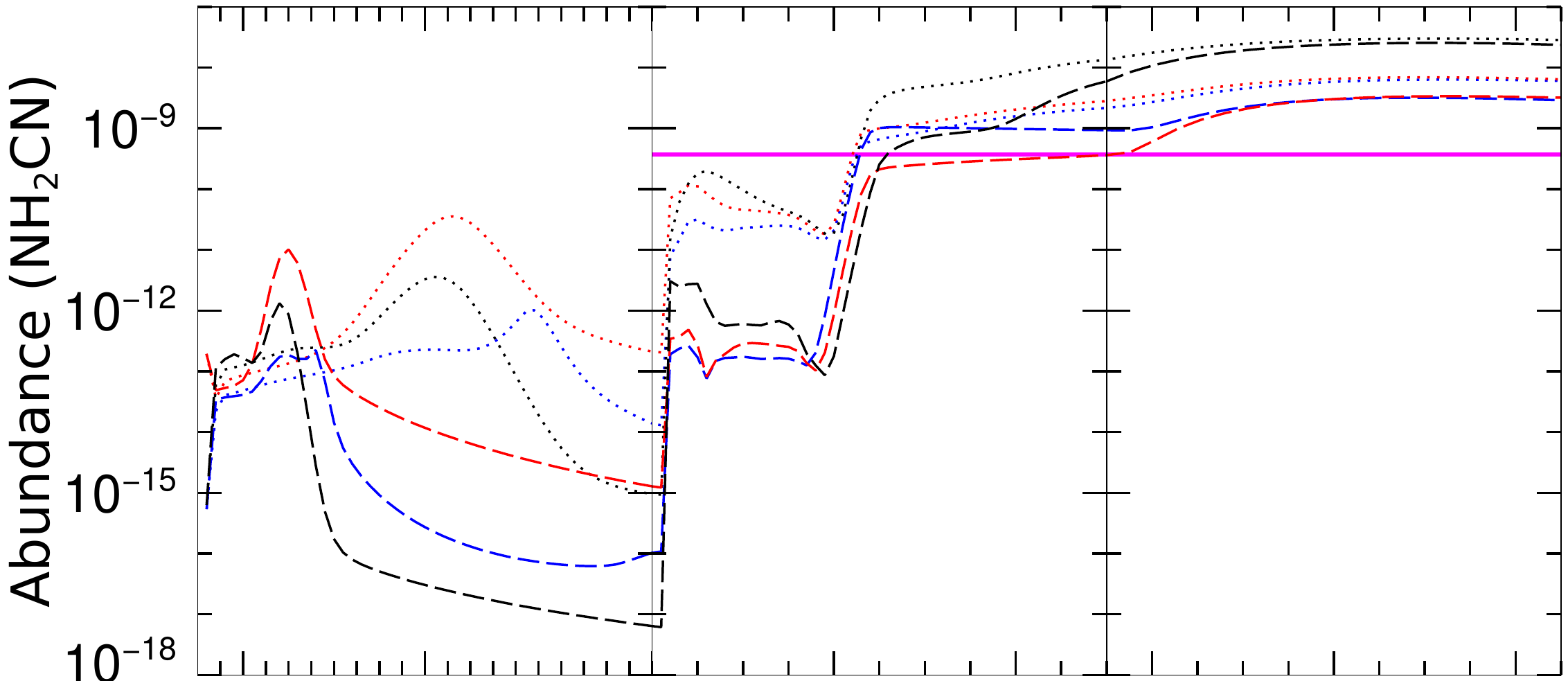}
\includegraphics[width=.48\textwidth]{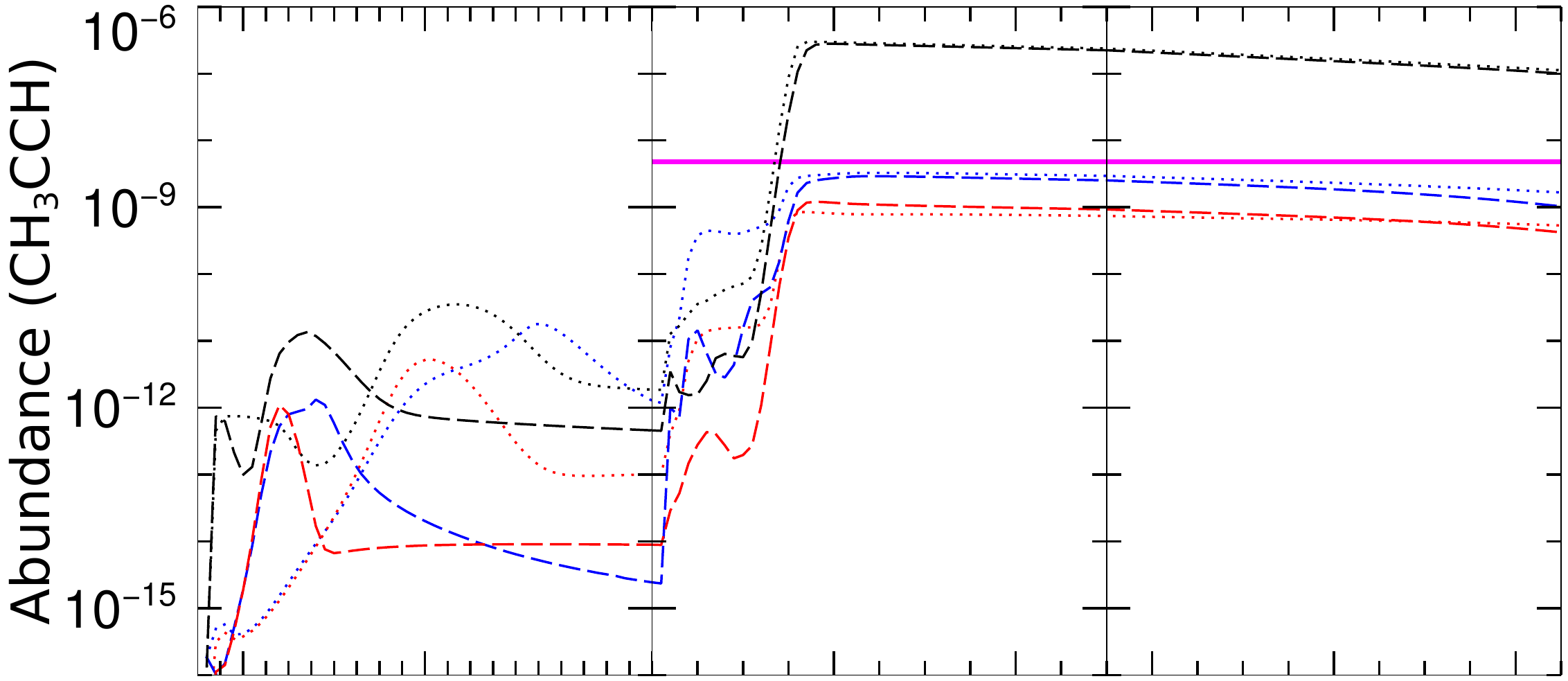}
\includegraphics[width=.48\textwidth]{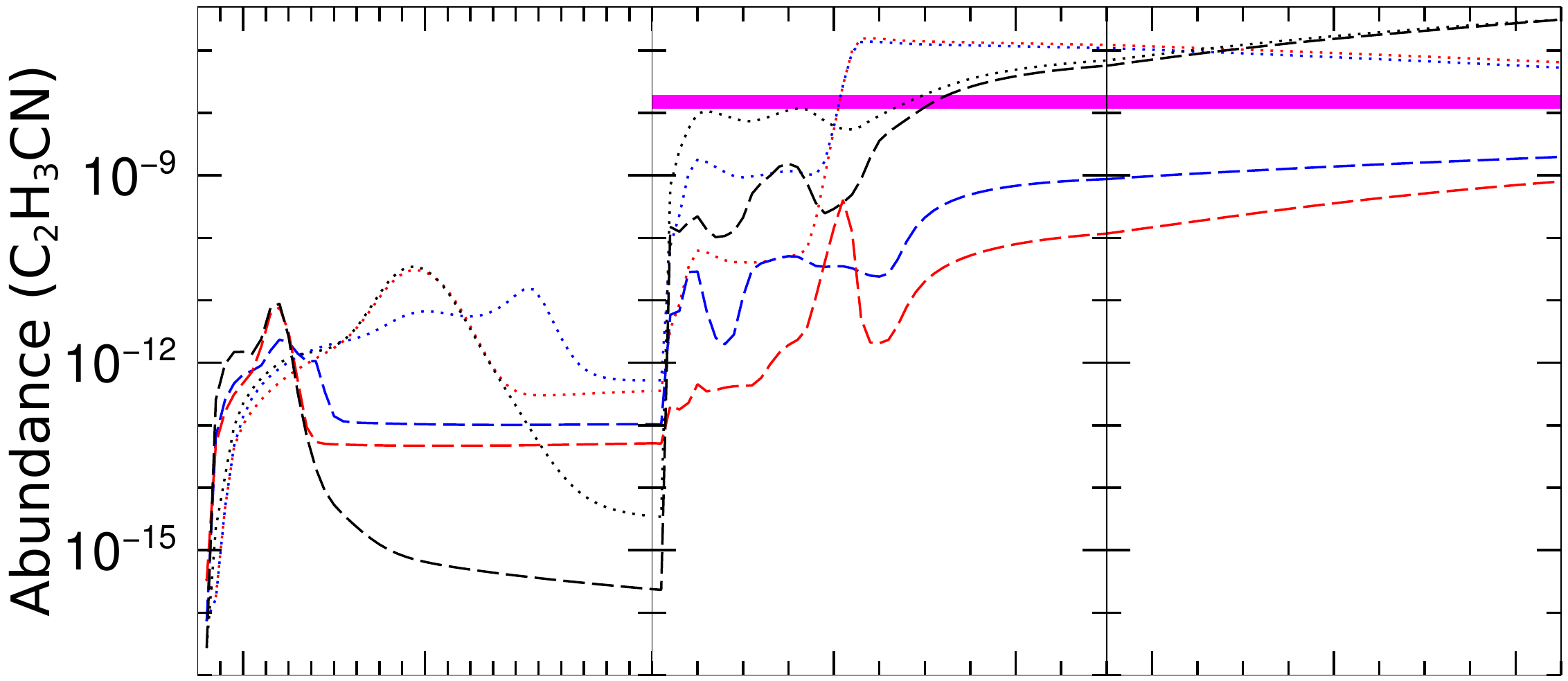}
\includegraphics[width=.48\textwidth]{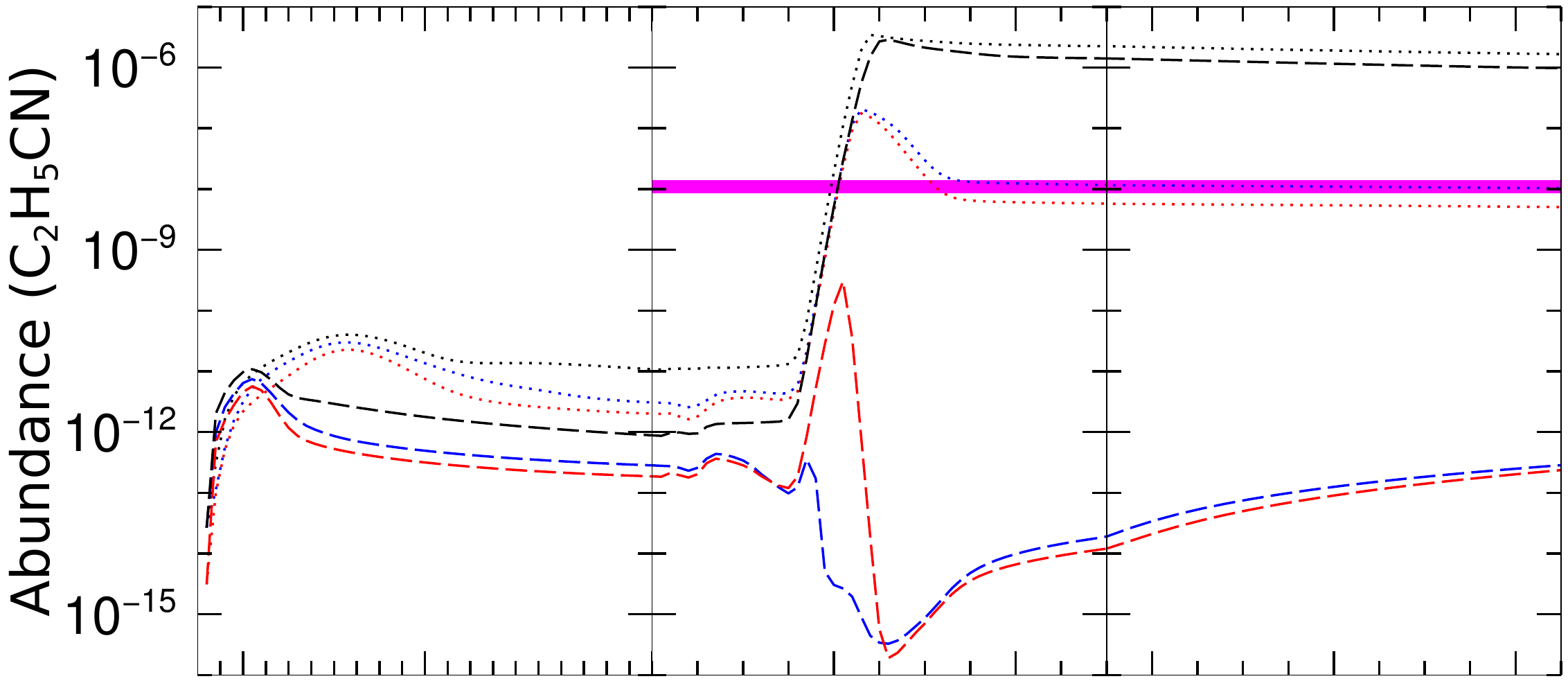}
\includegraphics[width=.48\textwidth]{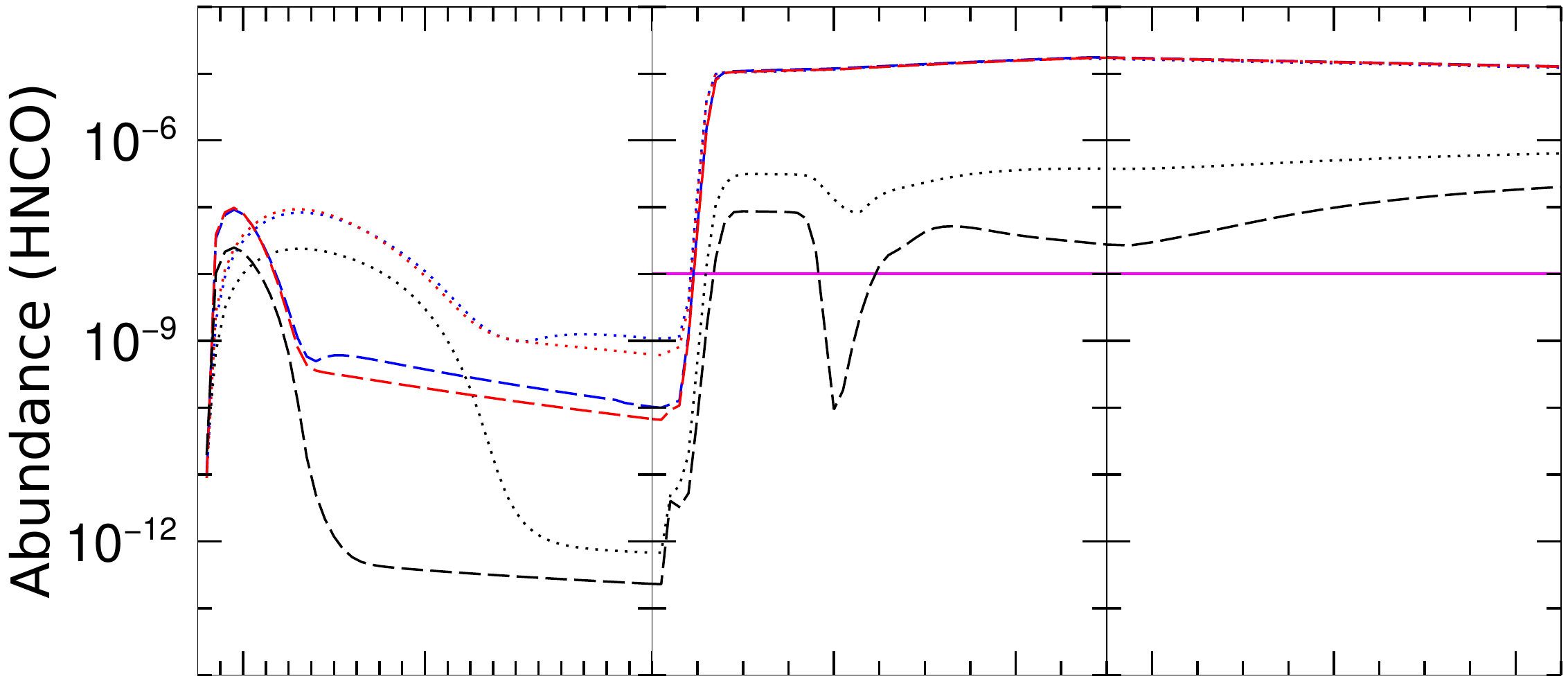}
\includegraphics[width=.48\textwidth]{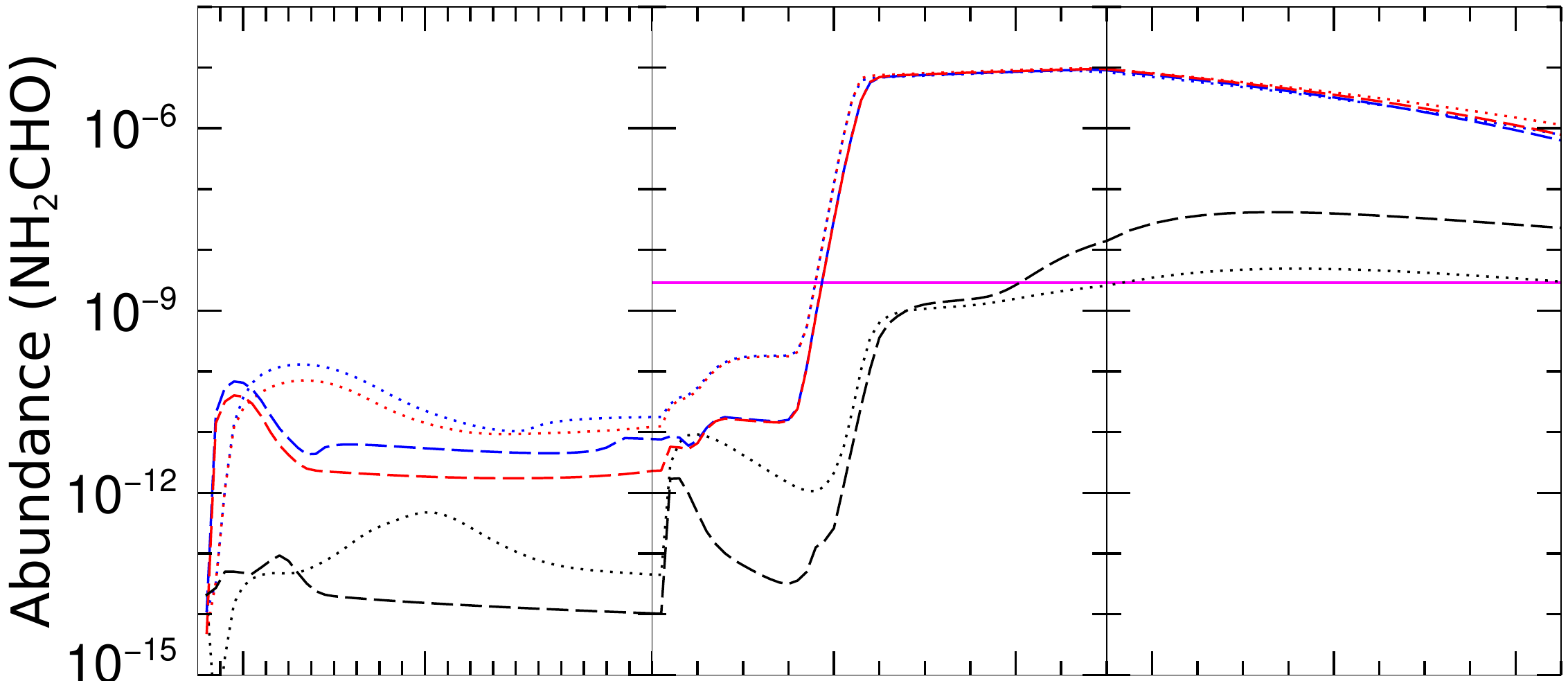}
\includegraphics[width=.48\textwidth]{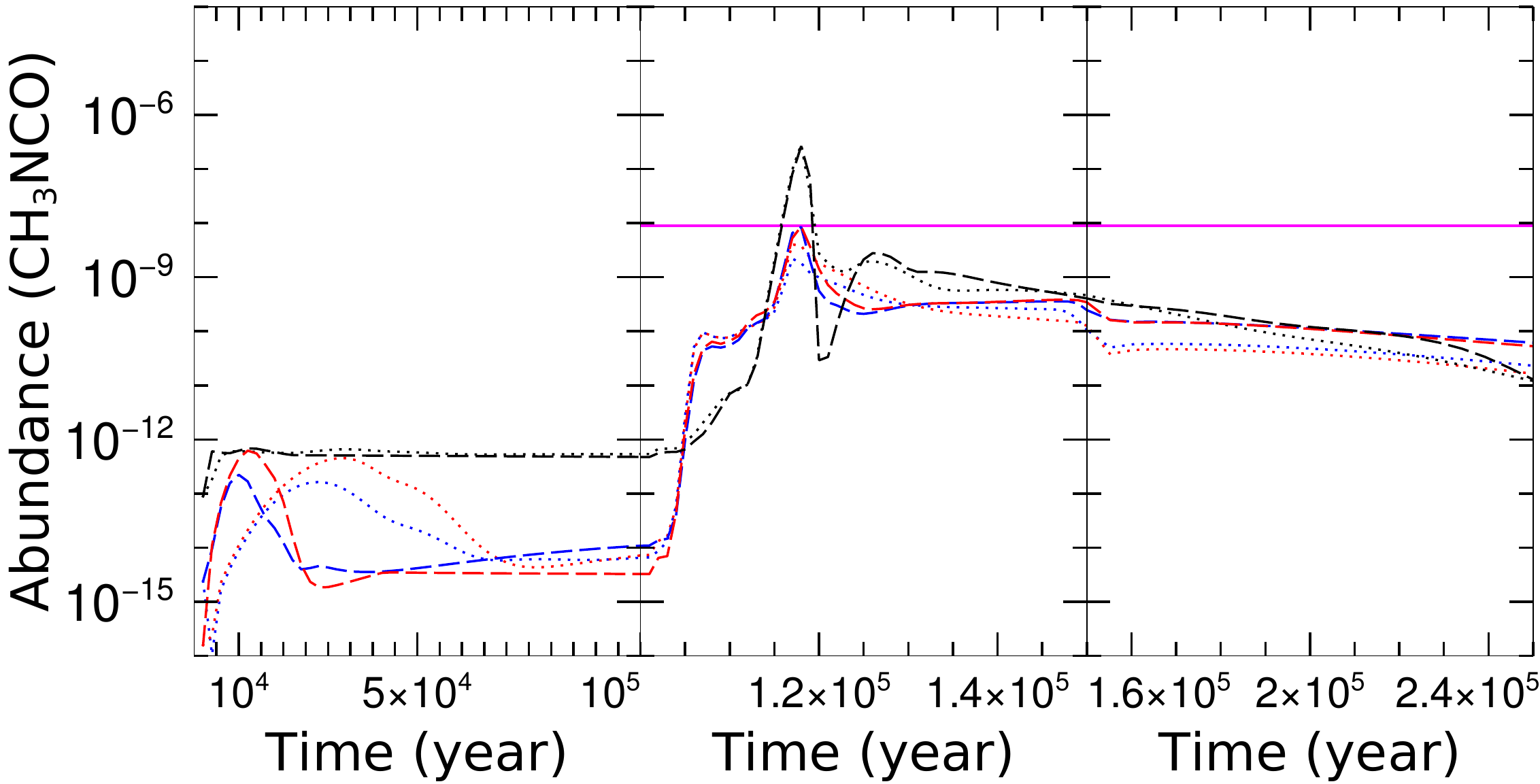}
\includegraphics[width=.48\textwidth]{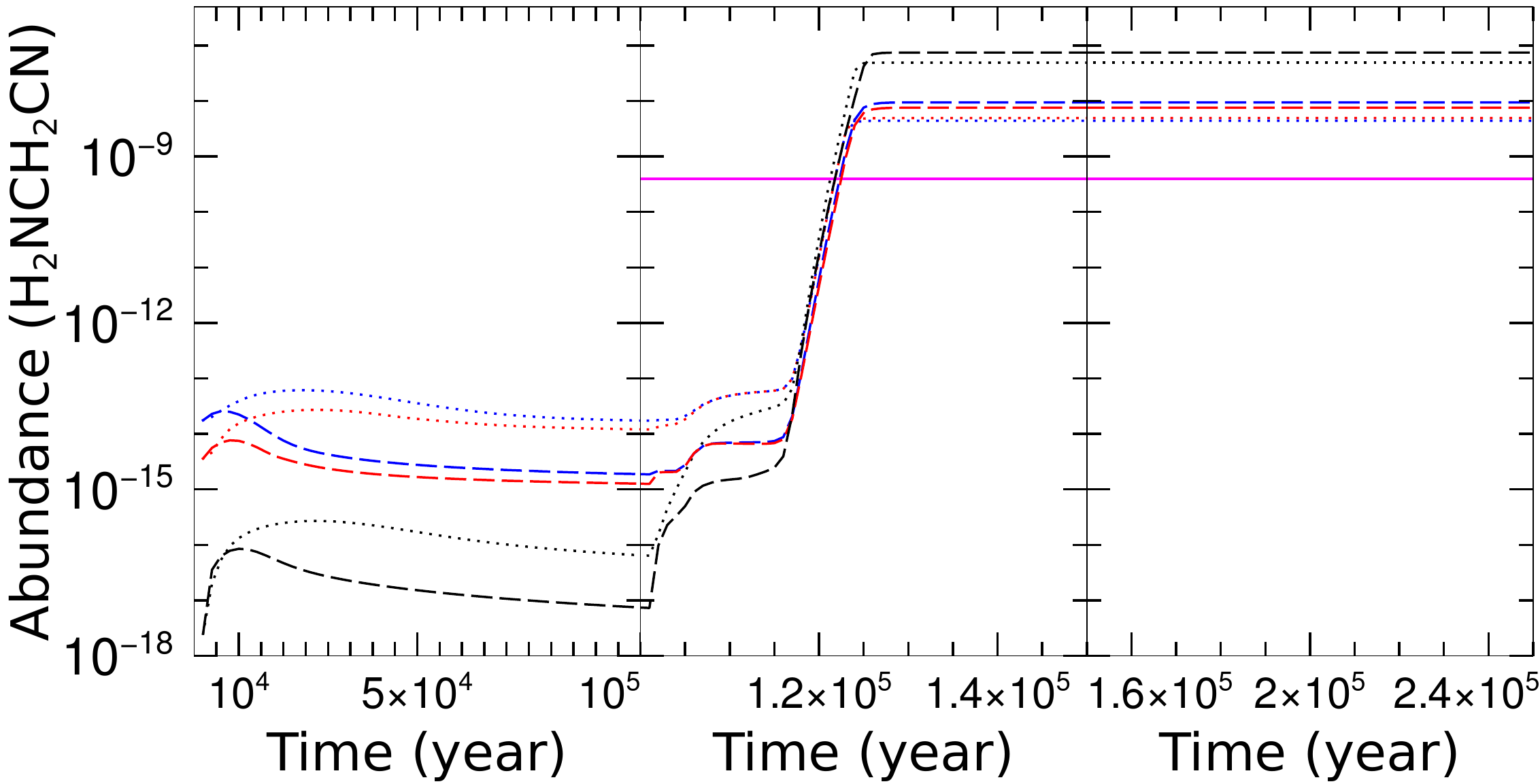}
\caption{Impact of initial dust temperature (black (15 K), red (20 K), and blue (25 k) lines) and final H density (dashed lines ($10^7$ cm$^{-3}$ ), and dotted lines ($10^6$ cm$^{-3}$ )) on simulated abundances of selected species. For all plots, black, red, and blue dashed lines are for models N7-Tg200-Td15, N7-Tg200-Td20, and N7-T2400-Td25, respectively, and black, red, and blue dotted lines are for models N6-Tg200-Td15, N6-Tg200-Td20, and N6-Tg200-Td25, respectively. Pink horizontal bar represents the observed value in G10 (see Table \ref{tbl:physicalEvo} for details of each model).}
\label{fig:r2}
\end{figure*}
Surface chemistry is very sensitive to small changes in surface temperature. Hence, we used our models (N7-Tg200-Td20 and N6-Tg200-Td20) presented in Section \ref{sec:resultnH} and ran them again with initial dust temperatures of 15 and 25K. Thus, in total we ran four additional models: N7-Tg200-Td15 and N7-Tg200-Td25 with a peak H density of $10^7$ cm$^{-3}$, and N6-Tg200-Td15 and N6-Tg200-Td25 with a peak H density of $10^6$ cm$^{-3}$.

%It becomes very important to test how selecting a different initial surface temperature  We run  models, N7-Tg200-Td15, N7-Tg200-Td20, and N7-Tg200-Td25 with peak H density of $10^7$ cm$^{-3}$ during cloud collapse stage. In these three models we keep all parameters same except initial dust temperature which is respectively 15 K, 20 K and 25 K. To verify if the impact of initial dust temperature is similar when peak density of collapsing cloud is $10^6$ cm$^{-3}$ , we run three more models, N6-Tg200-Td15, N6-Tg200-Td20, and N6-Tg200-Td25.

In Figure \ref{fig:r2}, we show the chemical evolution of selected species as a function of time in the gas phase during all three stages of physical structure evolution of a simulated cloud. In this figure, the pink horizontal bar represents the observed abundance. We note that, in general, the species have the highest gas-phase abundance during the cloud-collapse stage and the warm-up stage in the model with the highest dust temperature (25 K), and the lowest abundance when the dust temperature is lowest (15 K), with a few exceptions such as CH$_3$NCO and CH$_3$CCH. In fact, CH$_3$NCO and CH$_3$CCH are produced efficiently to cross the limit of observed abundance in the model with initial dust temperature of only 15 K. In the other two models, abundances of these species are always lower than observed abundances by a small factor. We see that for most of the species, the best-fit time falls during the early to mid warm-up phase of the cloud.
%
%with exception of HNCO and CH$_3$CCH. The time evolution curves for these two species also touches respective observed values during post-warm-up stage (between $1.01 \times 10^6$ to $1.02 \times 10^6$ years {\color{red} correct it after putting the corrected figure}) for the model with initial dust temperature 15 K and peak H density of $10^7$ cm${-3}$. 

We can conclude that initial dust temperature has a major impact on the time evolution of the selected species, with a higher dust temperature resulting in higher abundance values in general. A model with an initial dust temperature of 15 K produces a difference of few orders of magnitude in the abundances of a number of species when compared to a model with an initial dust temperature of 20 K.  This difference is higher during the cloud-collapse stage and lowest during the post-warm-up stage. When we increase the dust temperature to 25 K, the difference remains small compared to the 20 K model. Hence, it is critical to select the initial dust temperature during the cloud-collapse stage, and it may result in a difference of few orders of magnitude if a temperature under 20 K is used.
\subsubsection{Sensitivity to initial gas temperature and the peak gas and dust temperature}
\label{sec:resultTg}
\begin{figure*}[t]
\centering
\includegraphics[width=.48\textwidth]{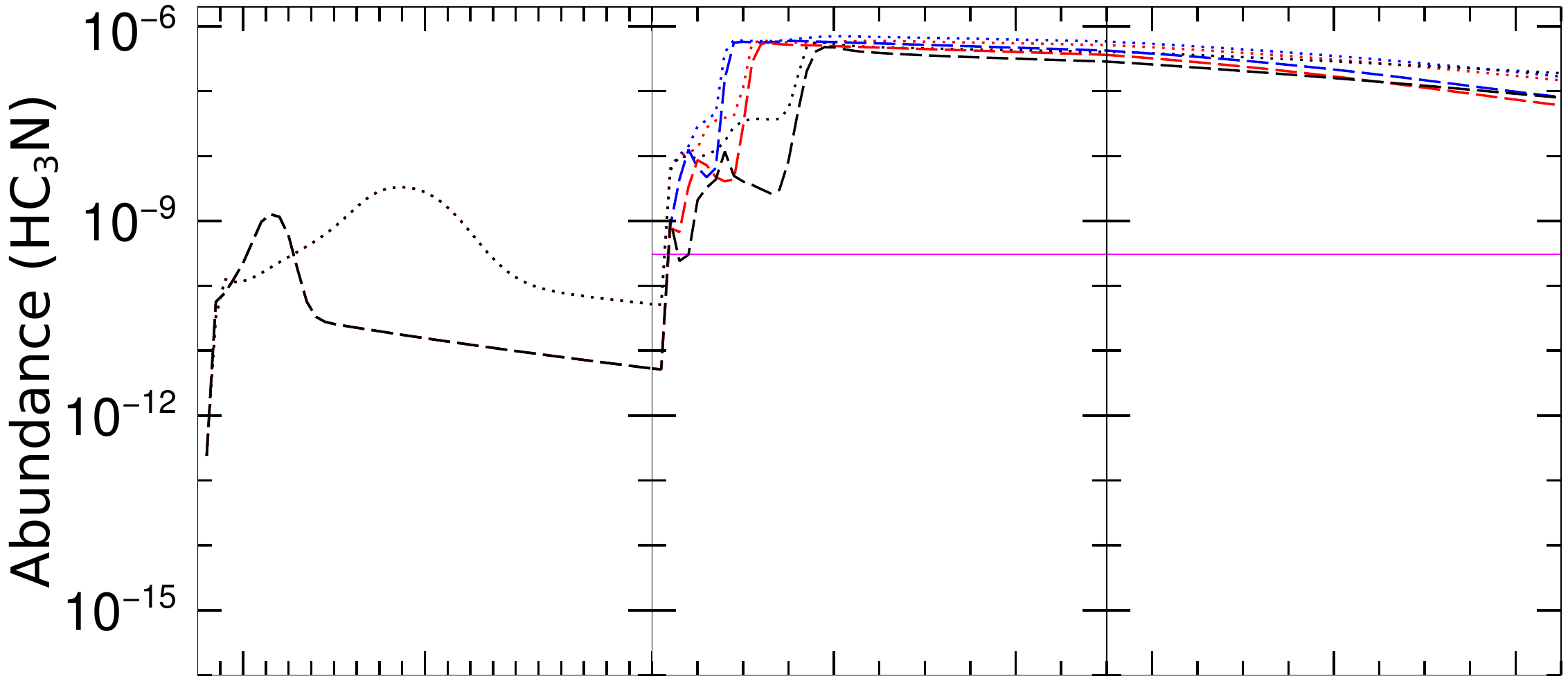}
\includegraphics[width=.48\textwidth]{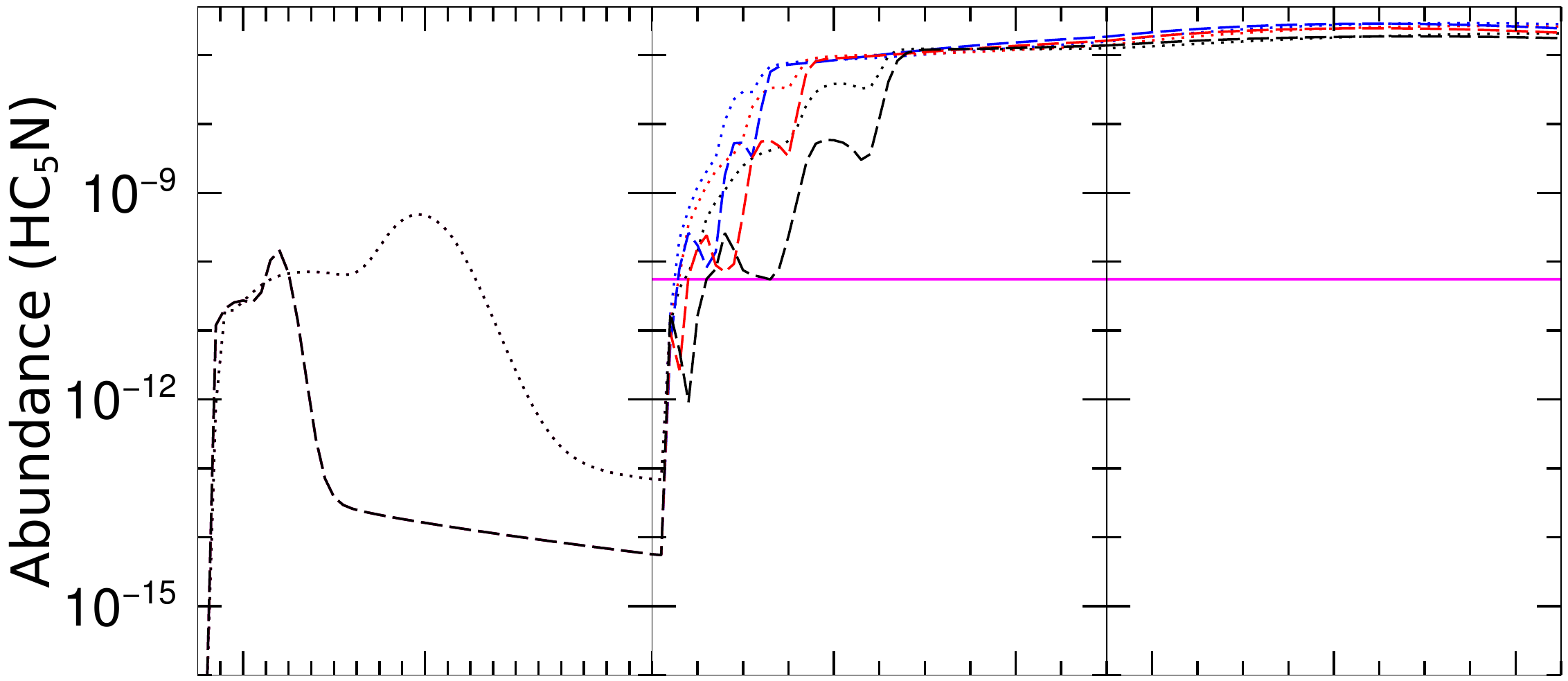}
\includegraphics[width=.48\textwidth]{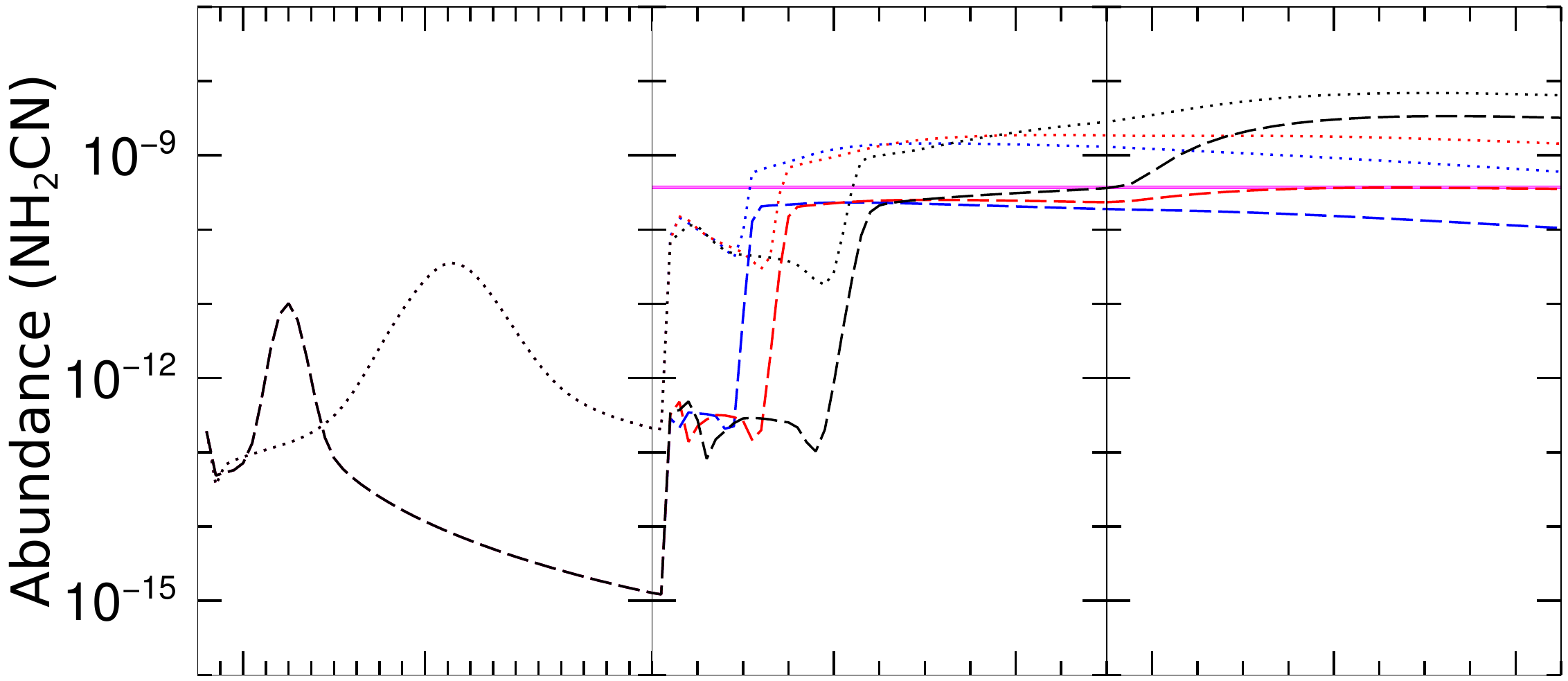}
\includegraphics[width=.48\textwidth]{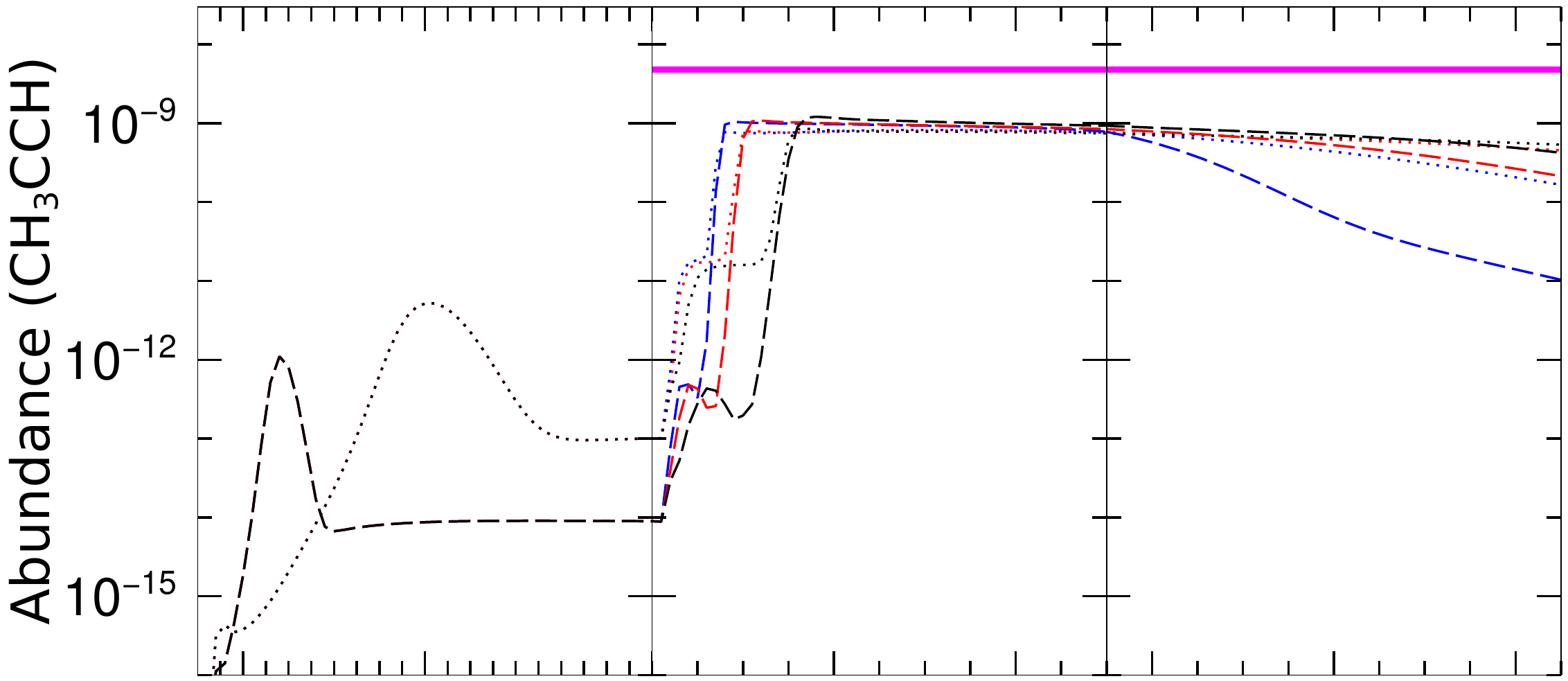}
\includegraphics[width=.48\textwidth]{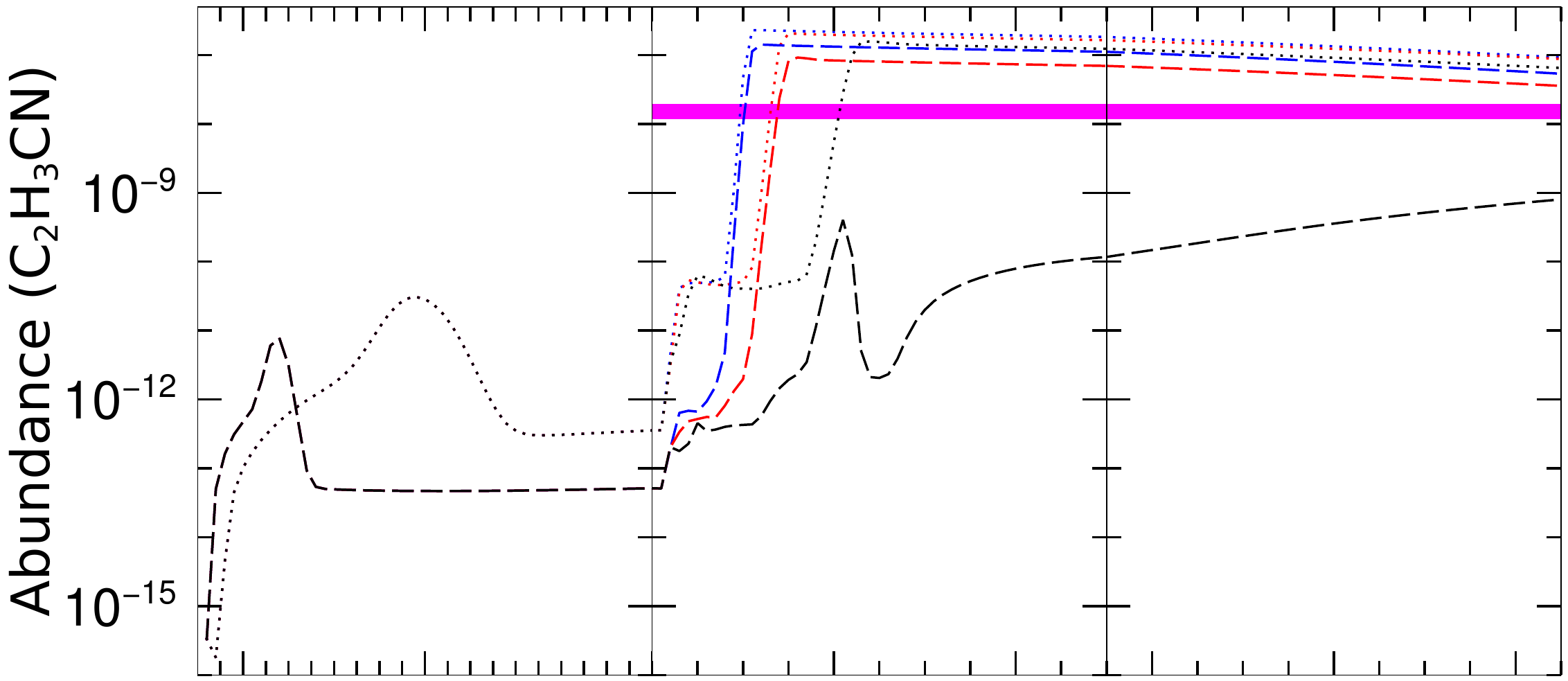}
\includegraphics[width=.48\textwidth]{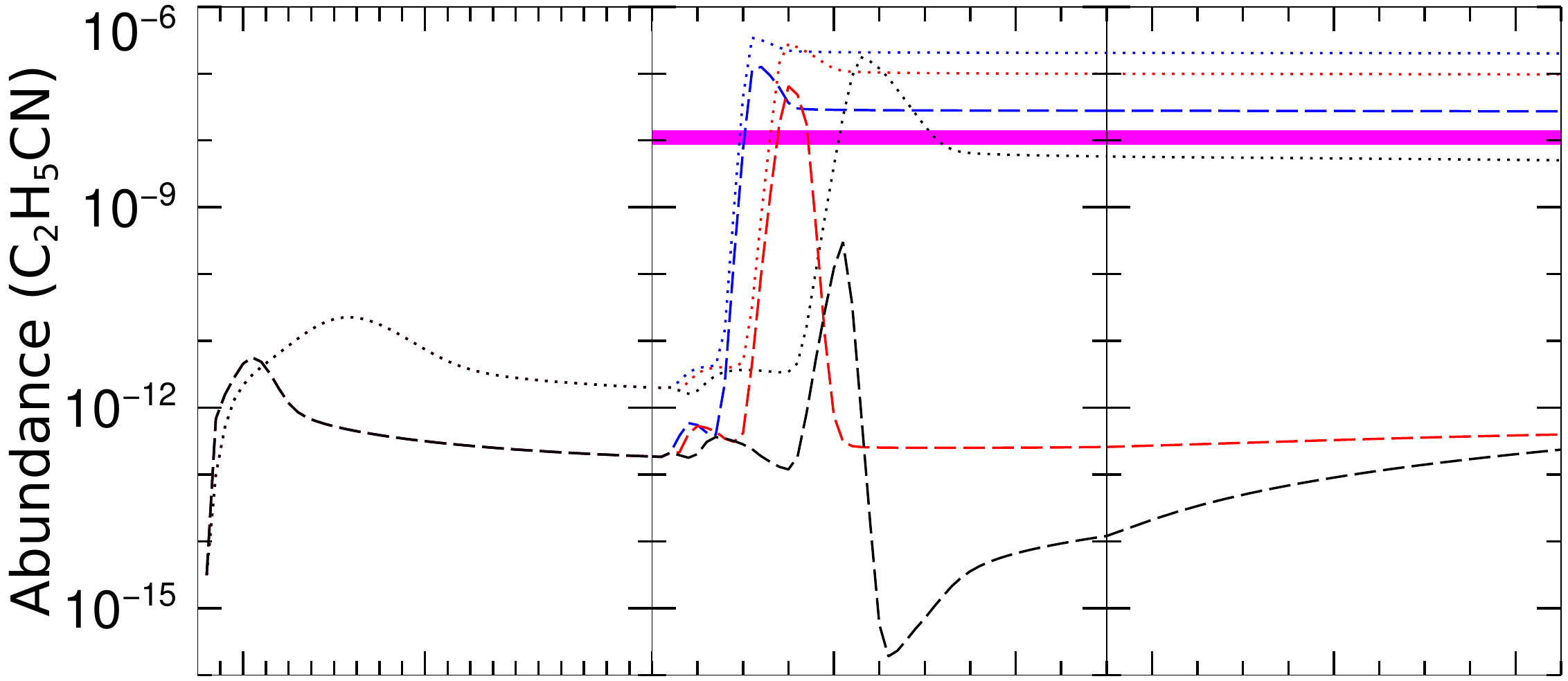}
\includegraphics[width=.48\textwidth]{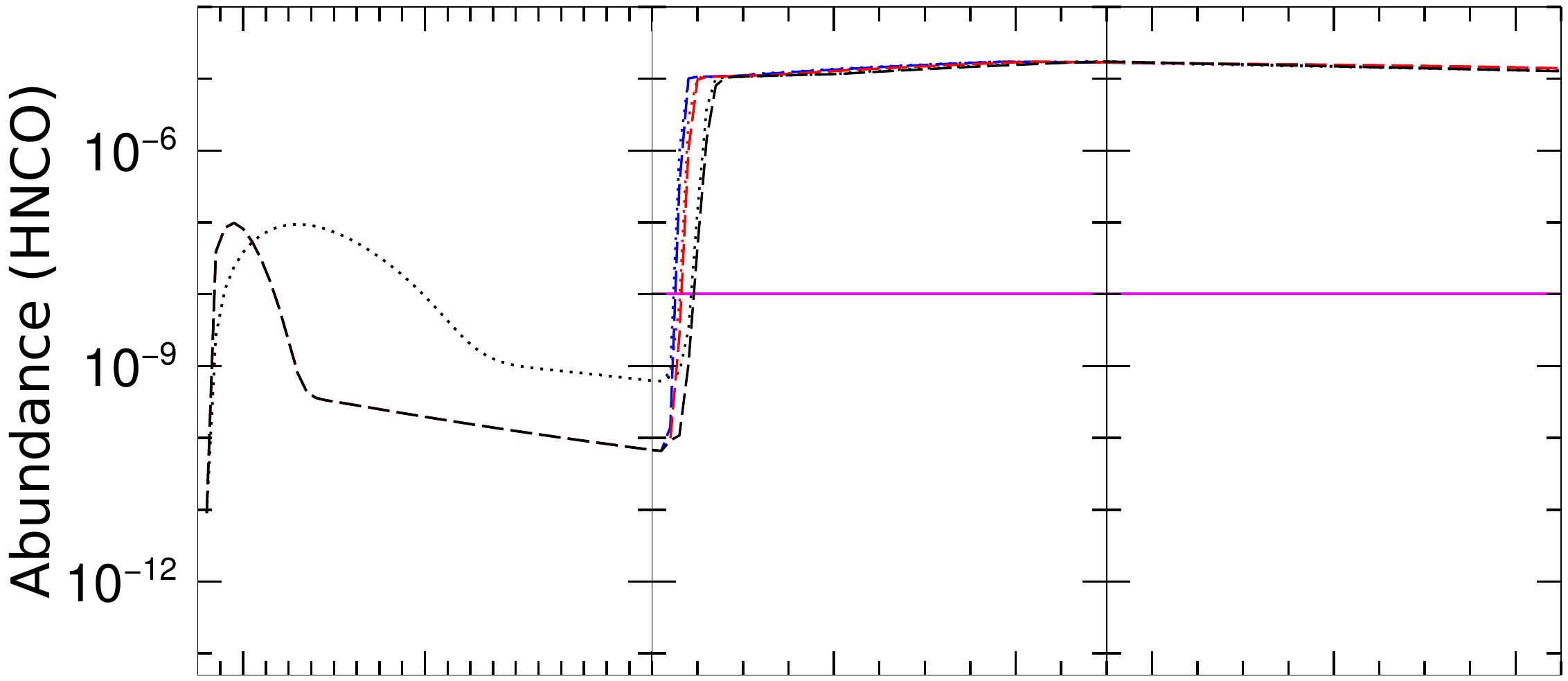}
\includegraphics[width=.48\textwidth]{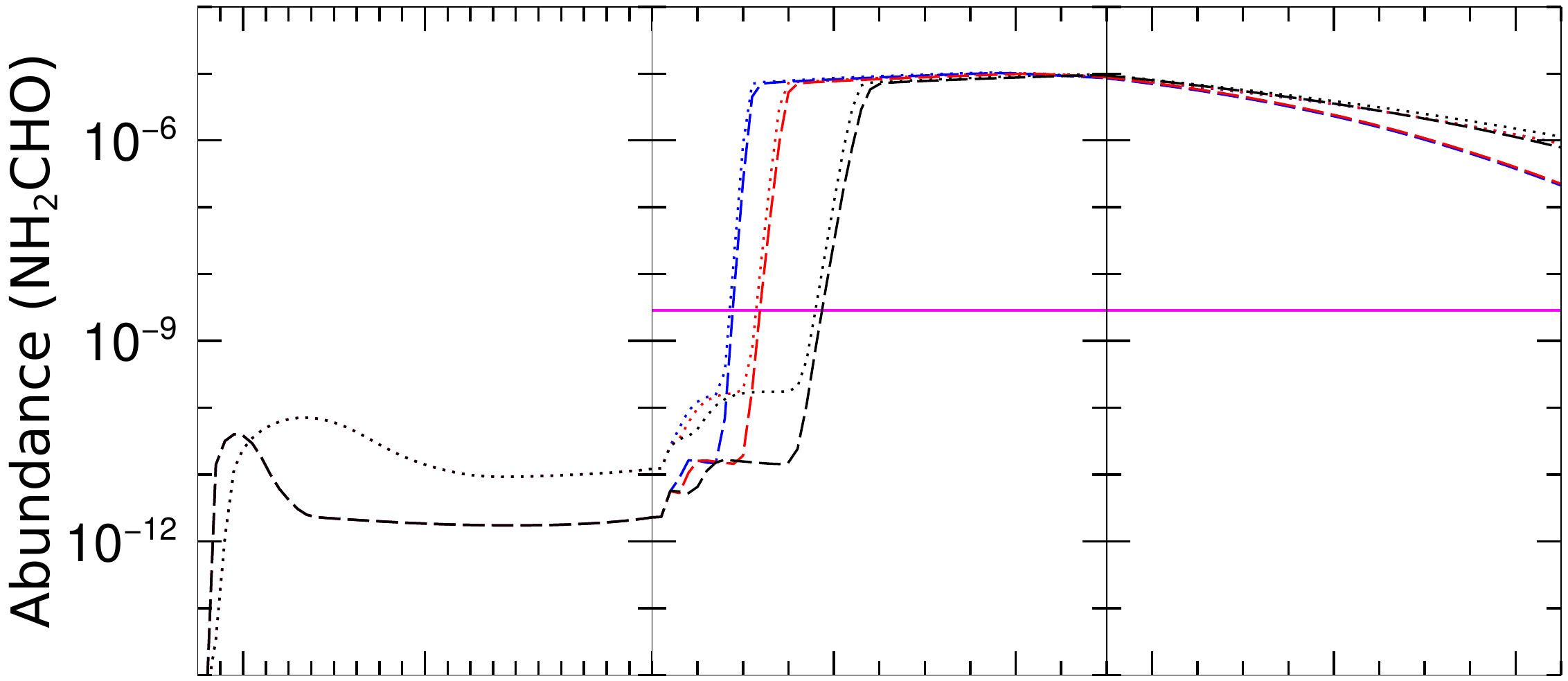}
\includegraphics[width=.48\textwidth]{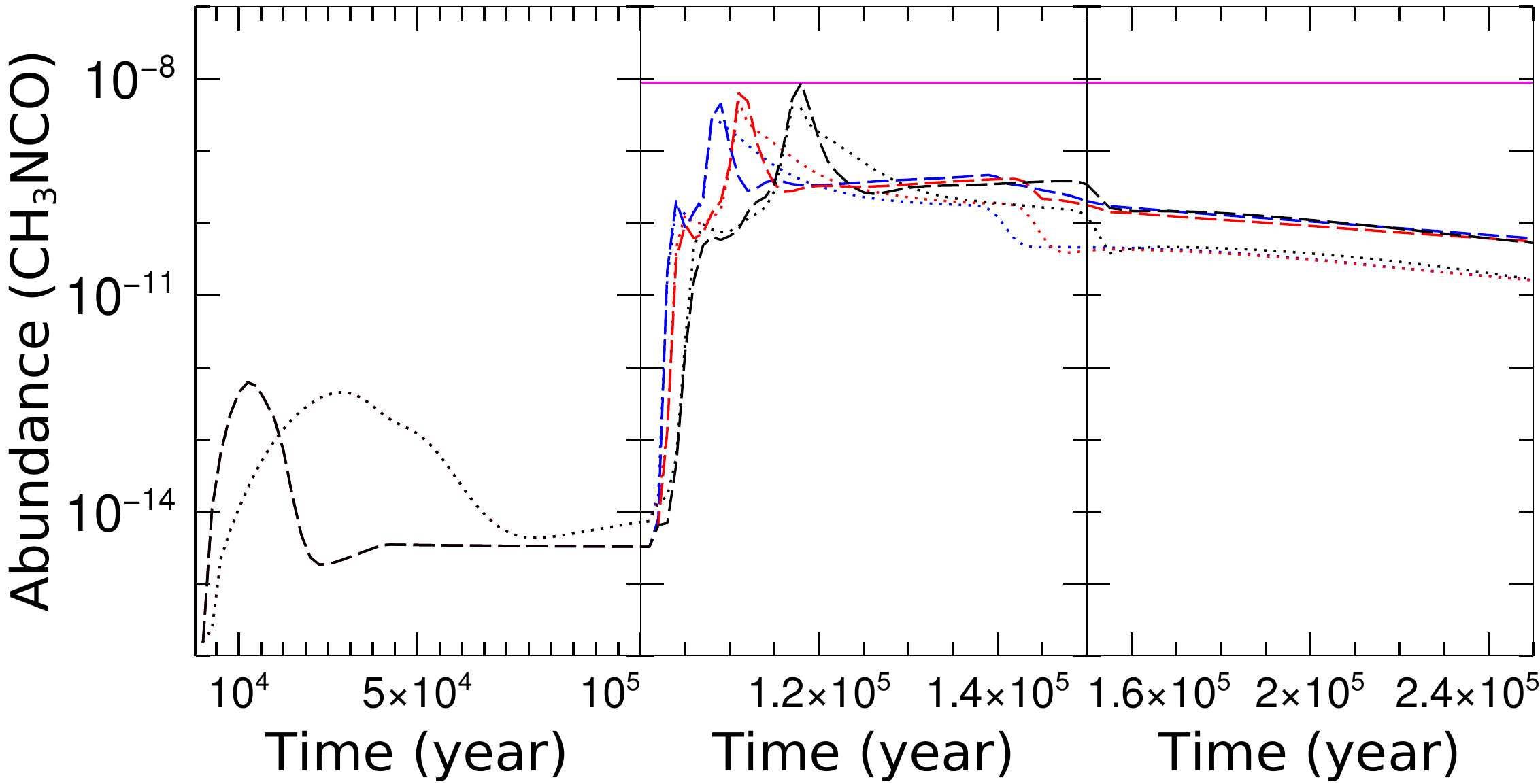}
\includegraphics[width=.48\textwidth]{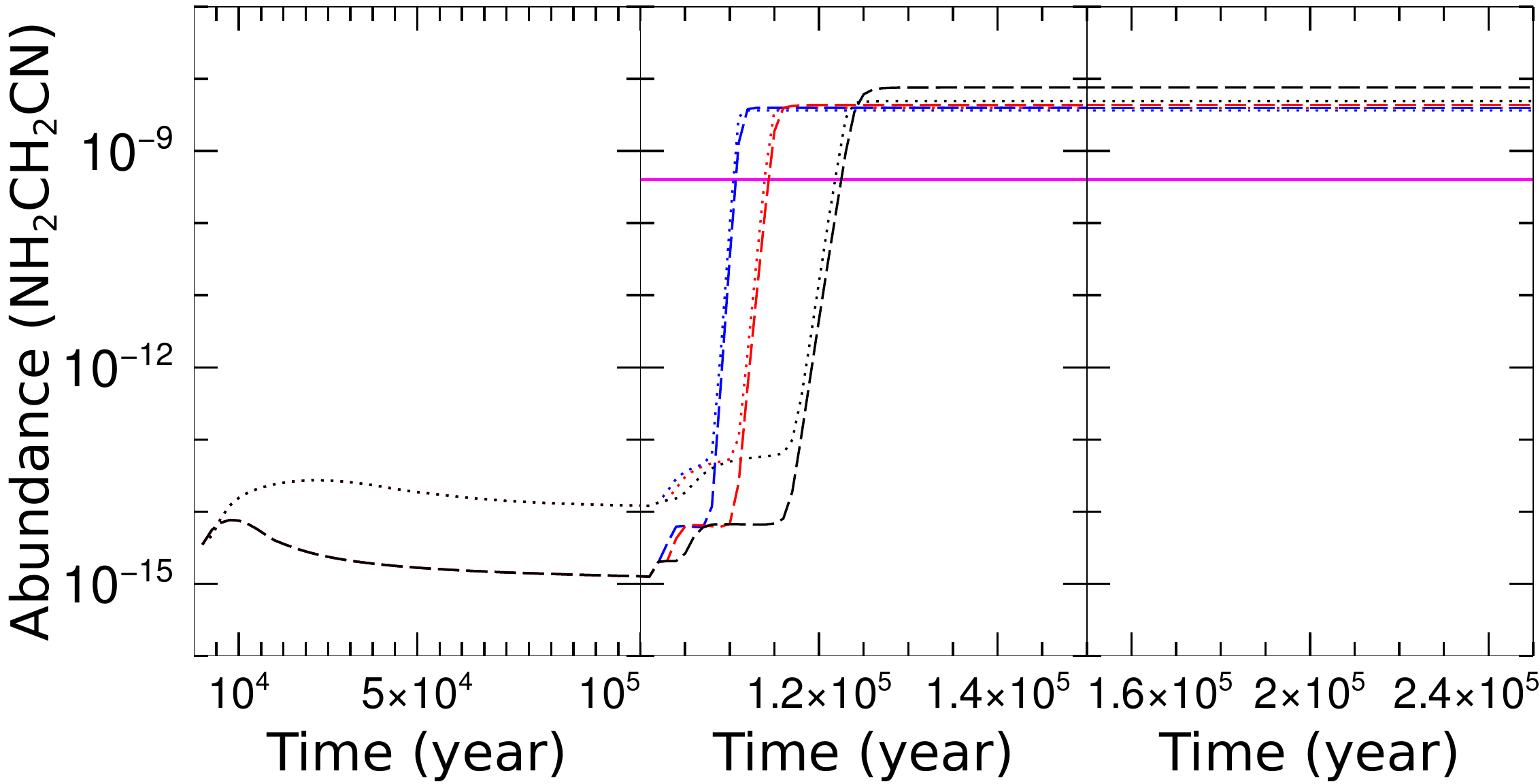}
\caption{Impact of peak gas and dust temperature (black, red, and blue lines) and peak H density (dashed lines ($10^7$ cm$^{-3}$ ) versus dotted lines ($10^6$ cm$^{-3}$ )) on simulated abundances of selected species. For all plots, black, red, and blue dashed lines are for models N7-Tg200-Td20, N7-Tg300-Td20, and N7-Tg400-Td20, respectively, and black, red, and blue dotted lines are for models N6-Tg200-Td20, N6-Tg300-Td20, and N6-Tg400-Td20. Pink horizontal bar represents the observed value in G10 (see Table \ref{tbl:physicalEvo} for details of each model).}
\label{fig:r1}
\end{figure*}

In all models discussed so far, we used an initial gas temperature of 40 K and the peak gas and dust temperature of 200 K. We found that chemistry is not very sensitive to the initial gas temperature if the temperature is below 100 K. So, we do not show results for different initial gas temperatures here. 

However, we found that variations in peak gas temperature do have noticeable impact on chemistry, especially during the warm-up phase. We know that in a collapsed cloud, gas and dust temperatures are coupled due to extreme density, such that $T_d = T_g$. Basically, the peak gas and dust temperatures are the same, and both change with the models presented in this section. Again, we used our reference models (N7-Tg200-Td20 and N6-Tg200-Td20) presented in Section \ref{sec:resultnH} and ran four additional models to show the impact of variations in the peak gas and dust temperatures on chemical abundances of N-bearing species. We ran new models with the peak gas and dust temperatures of 300 K and 400 K; these models are N7-Tg300-Td20 and N7-Tg400-Td20, respectively, with a peak H density of $10^7$ cm$^{-3}$, and N6-Tg300-Td20 and N6-Tg400-Td20, respectively, with a peak H density of $10^6$ cm$^{-3}$.

In Figure \ref{fig:r1}, we show simulated abundances as a function of time for selected species in the gas phase during all the three stages of physical structure evolution of simulated cloud. We see that the variation in peak gas and dust temperature only shifts the steady state abundance time for different species during warm-up stage. This shift in time is on the order of $10^4$ years. 
We also note that, during the warm-up phase, the steady-state condition is reached earlier, and this happens at a higher gas and dust temperature if we increase the peak value of the gas and dust temperatures in different models. Once the gas and dust temperature go above 200 K, there is little difference observed in abundances between different models. Our best-fit time falls again during the warm-up stage and is only shifted to the left by factor of $10^4$ years depending on the peak gas temperature. 
\subsubsection{Impact of warm-up duration}
\label{sec:result-tc}
\begin{figure*}[t]
\centering
\includegraphics[width=.48\textwidth]{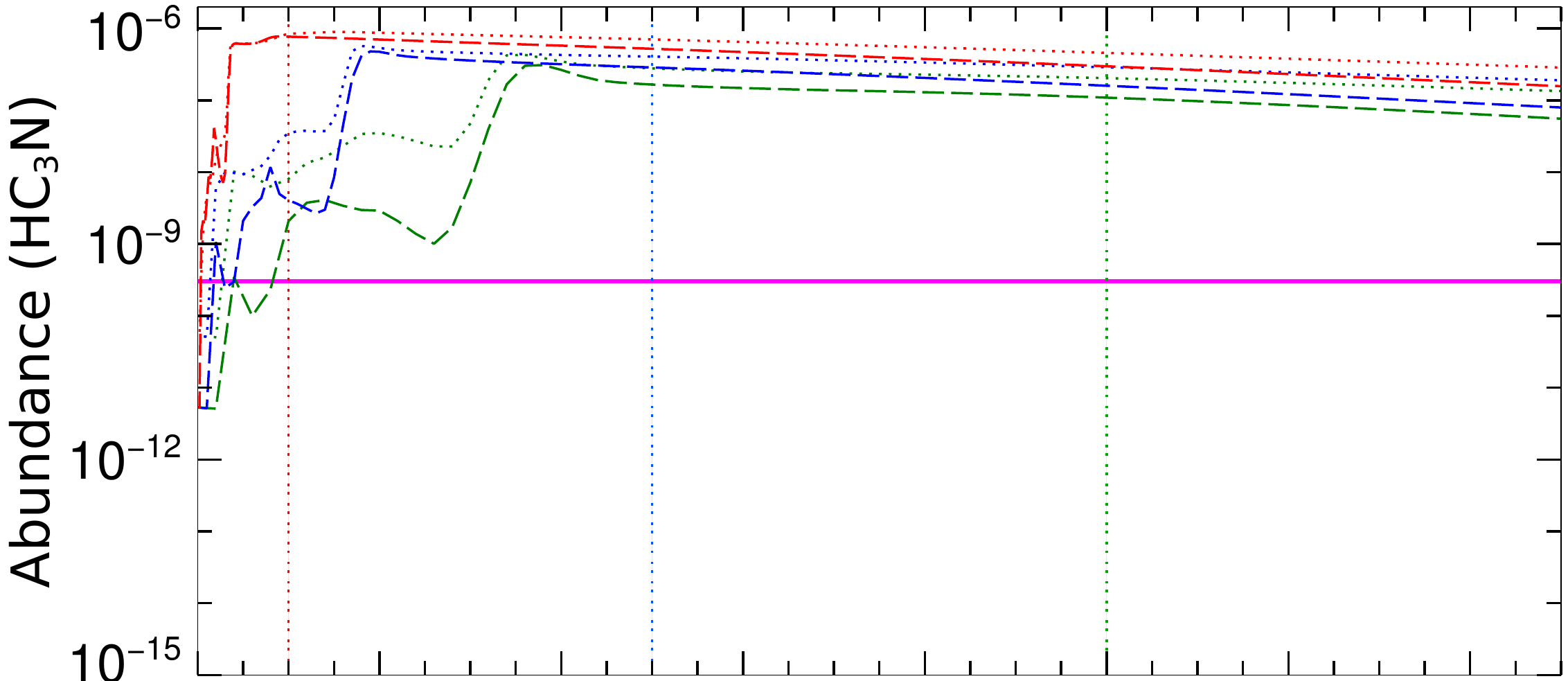}
\includegraphics[width=.48\textwidth]{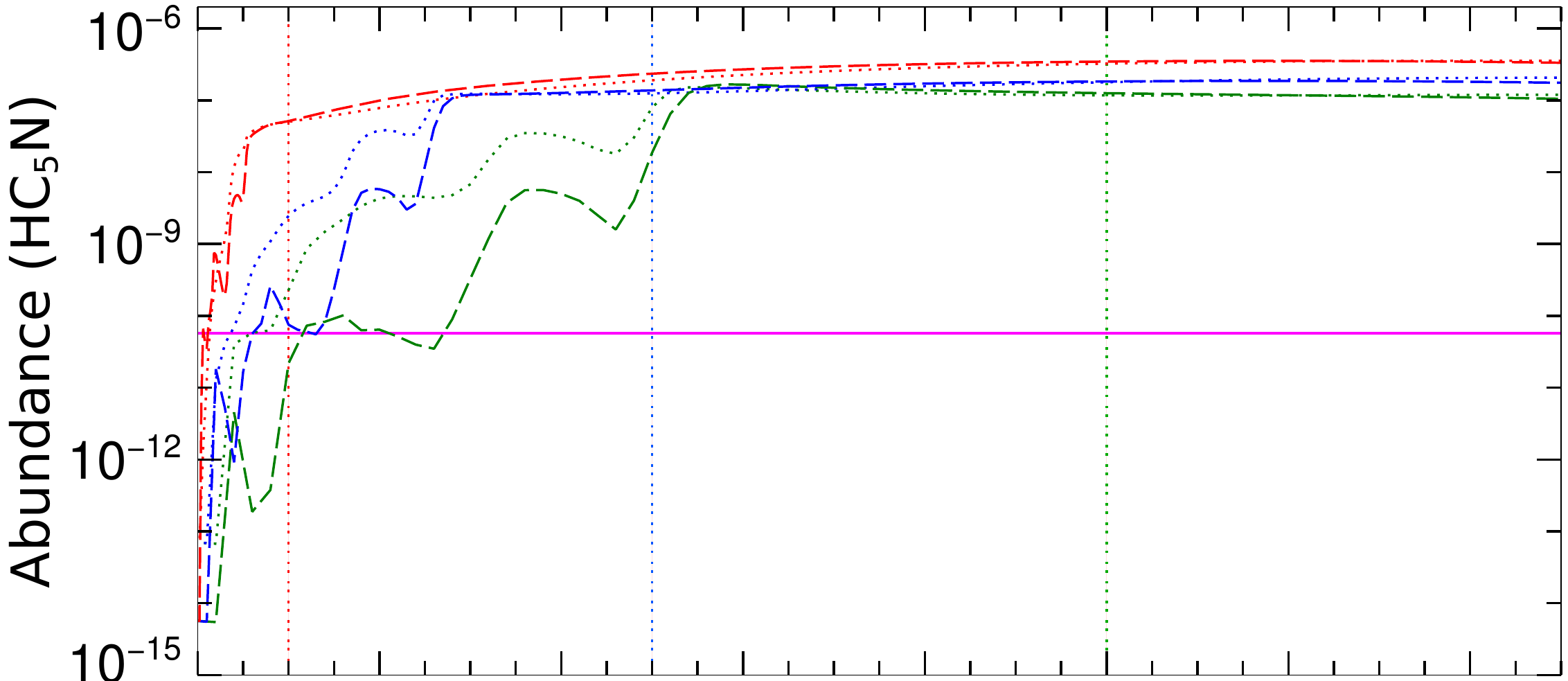}
\includegraphics[width=.48\textwidth]{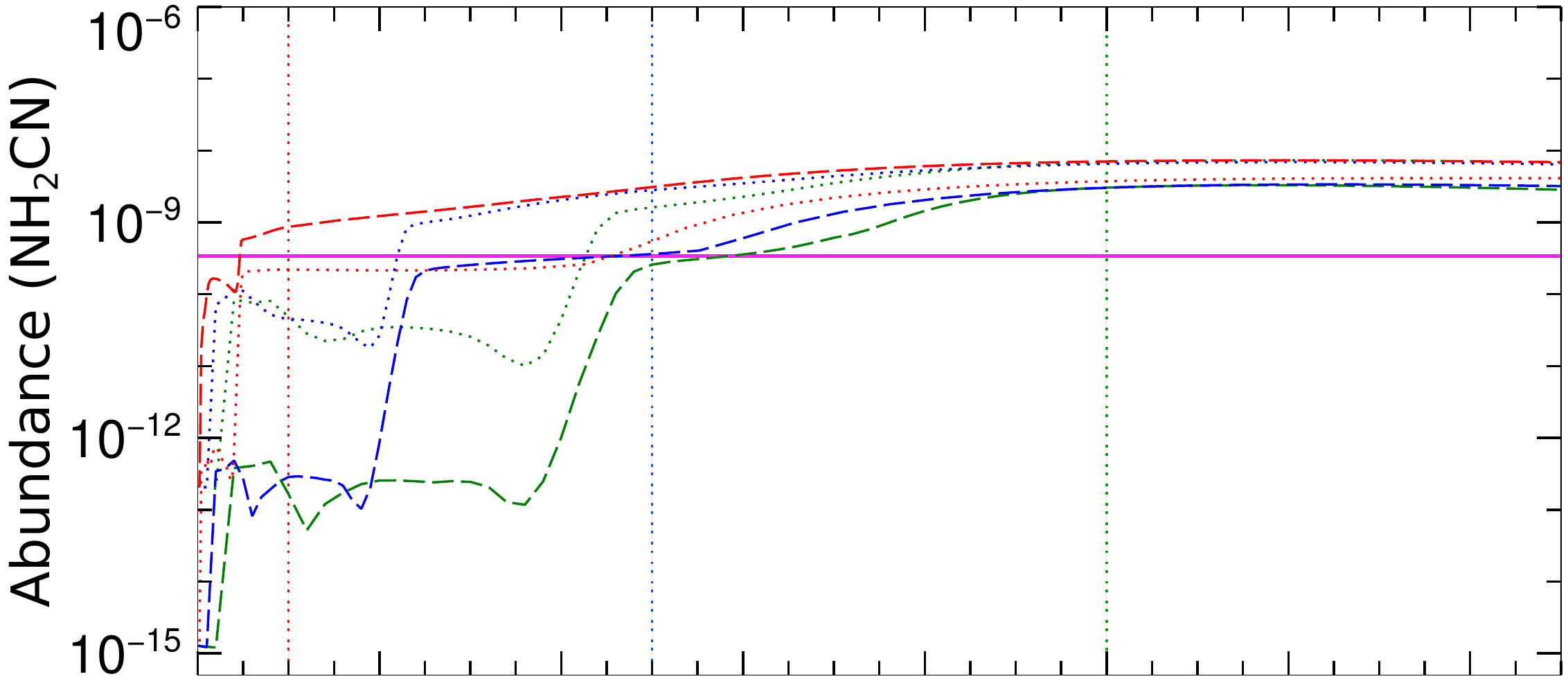}
\includegraphics[width=.48\textwidth]{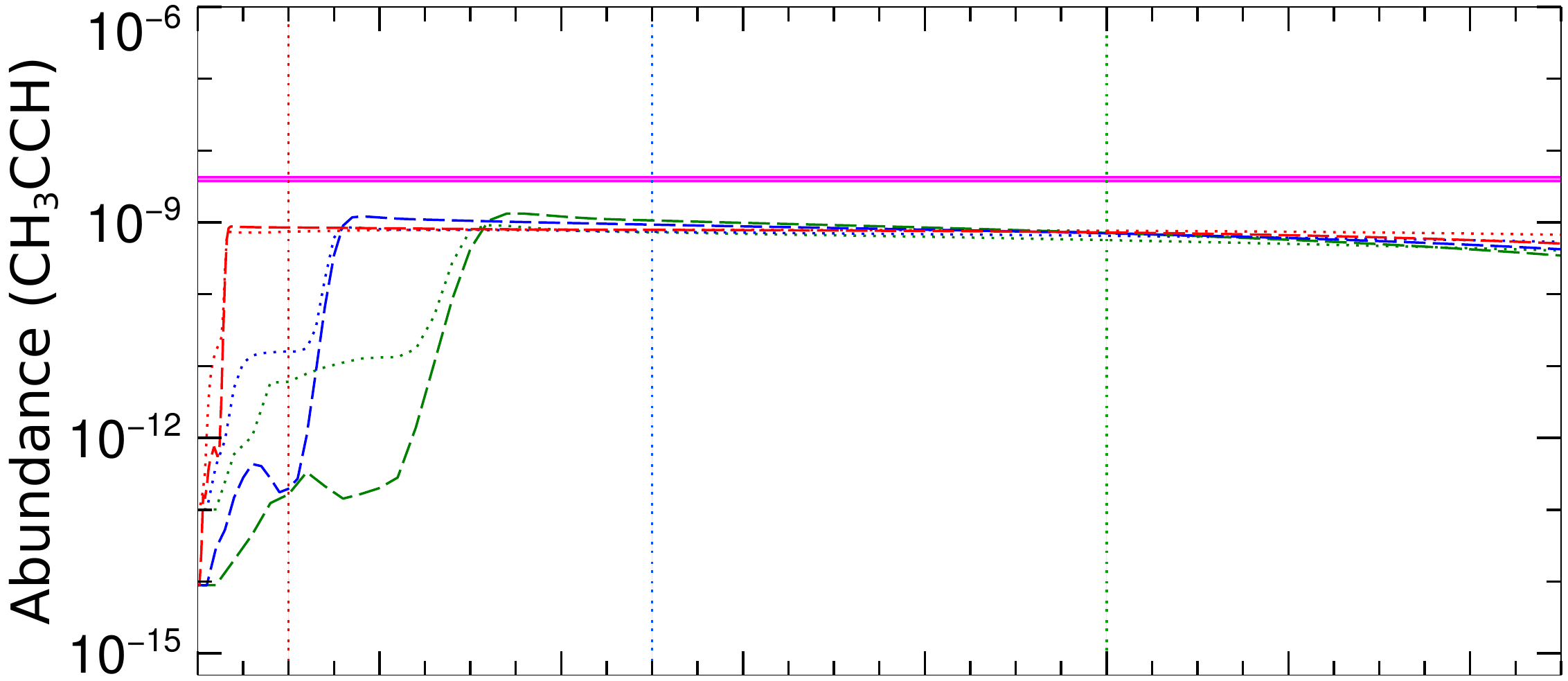}
\includegraphics[width=.48\textwidth]{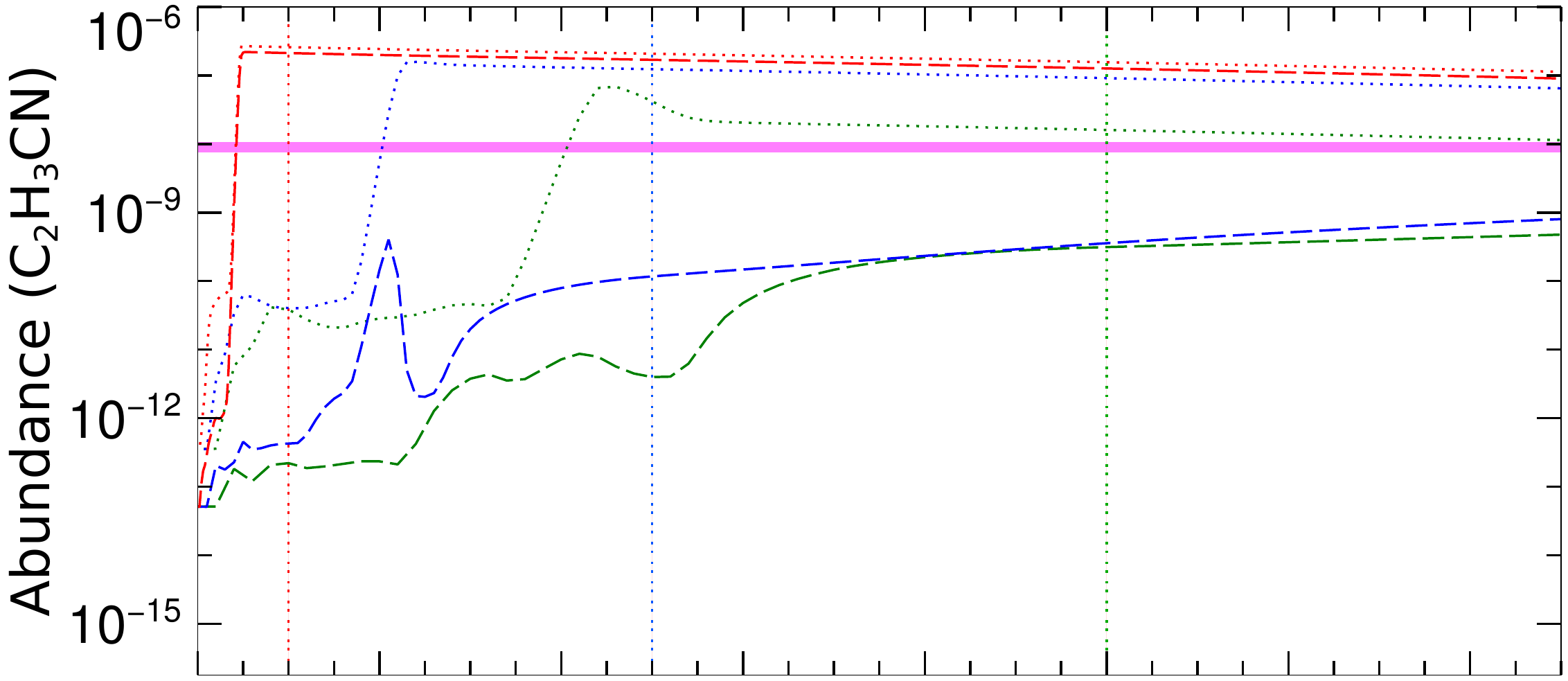}
\includegraphics[width=.48\textwidth]{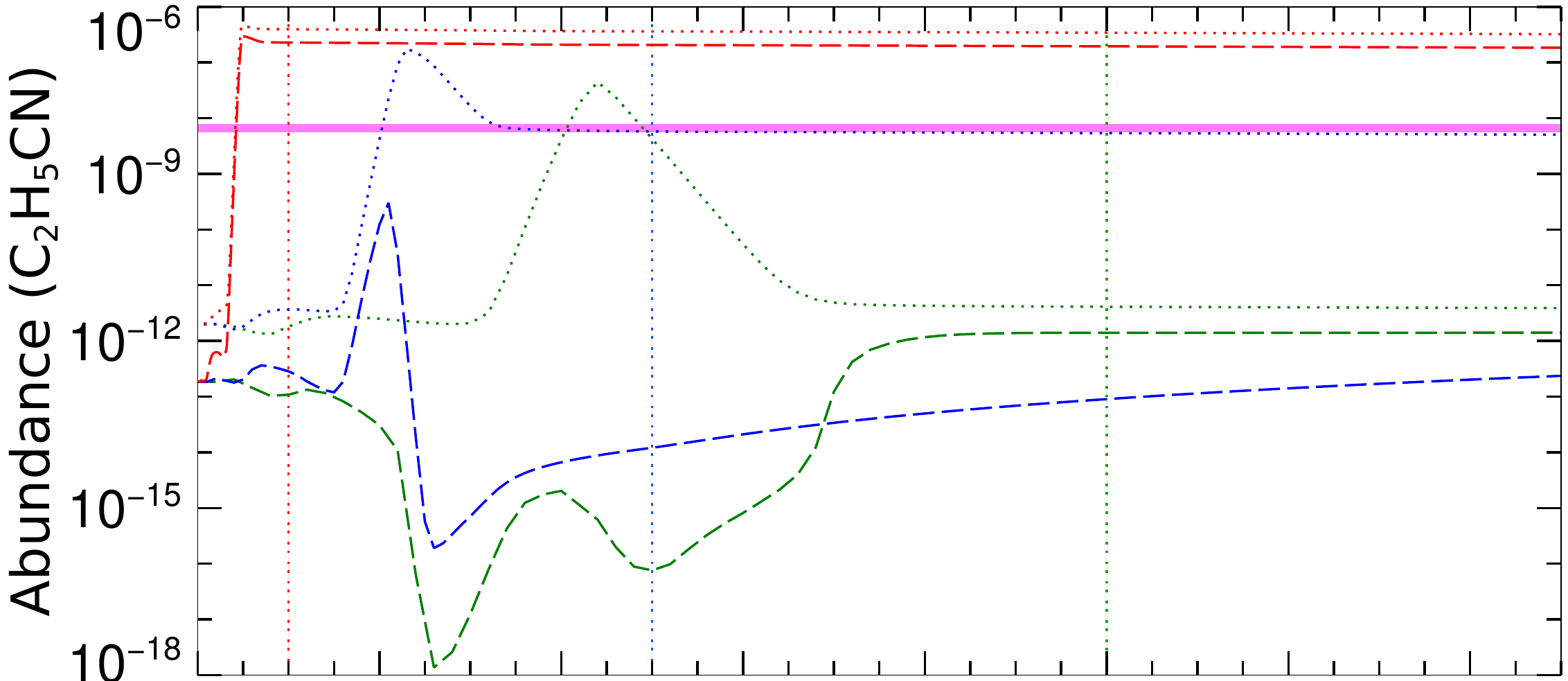}
\includegraphics[width=.48\textwidth]{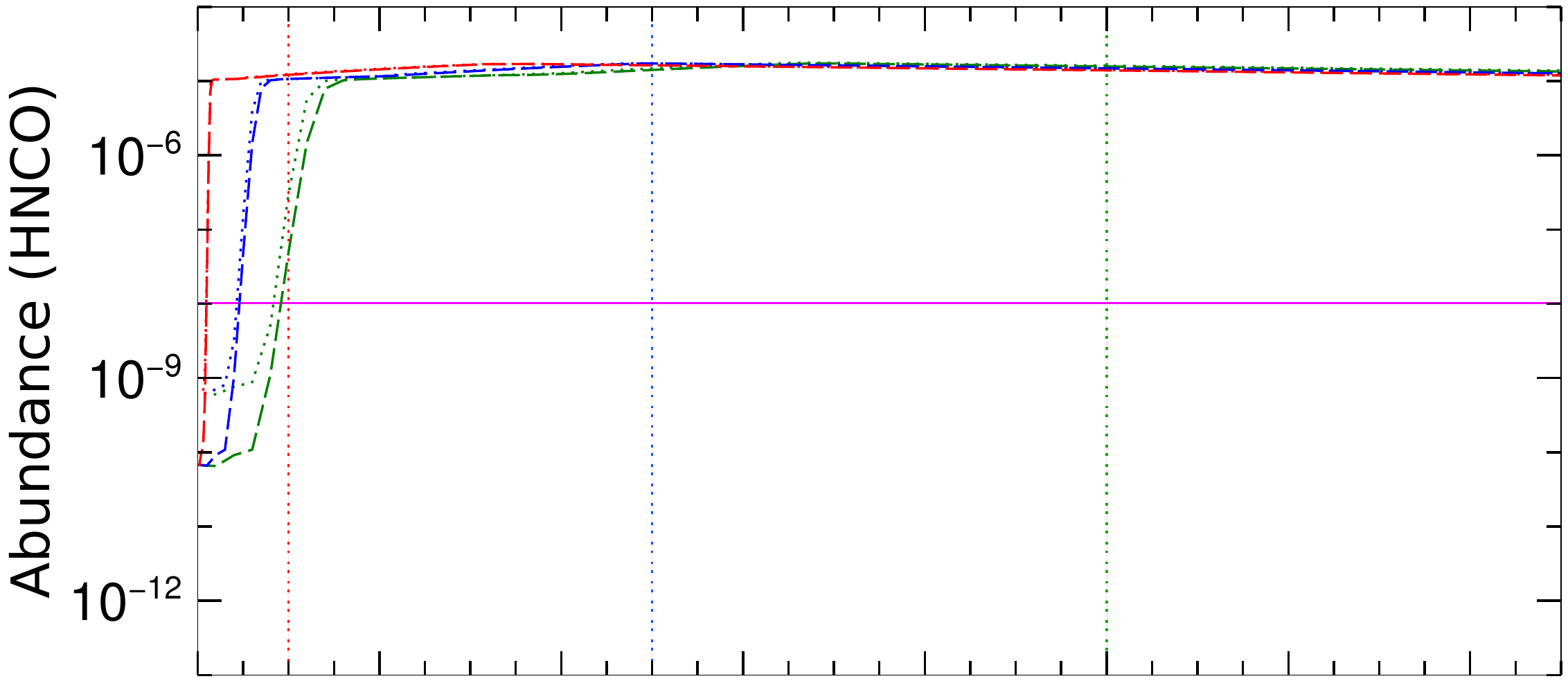}
\includegraphics[width=.48\textwidth]{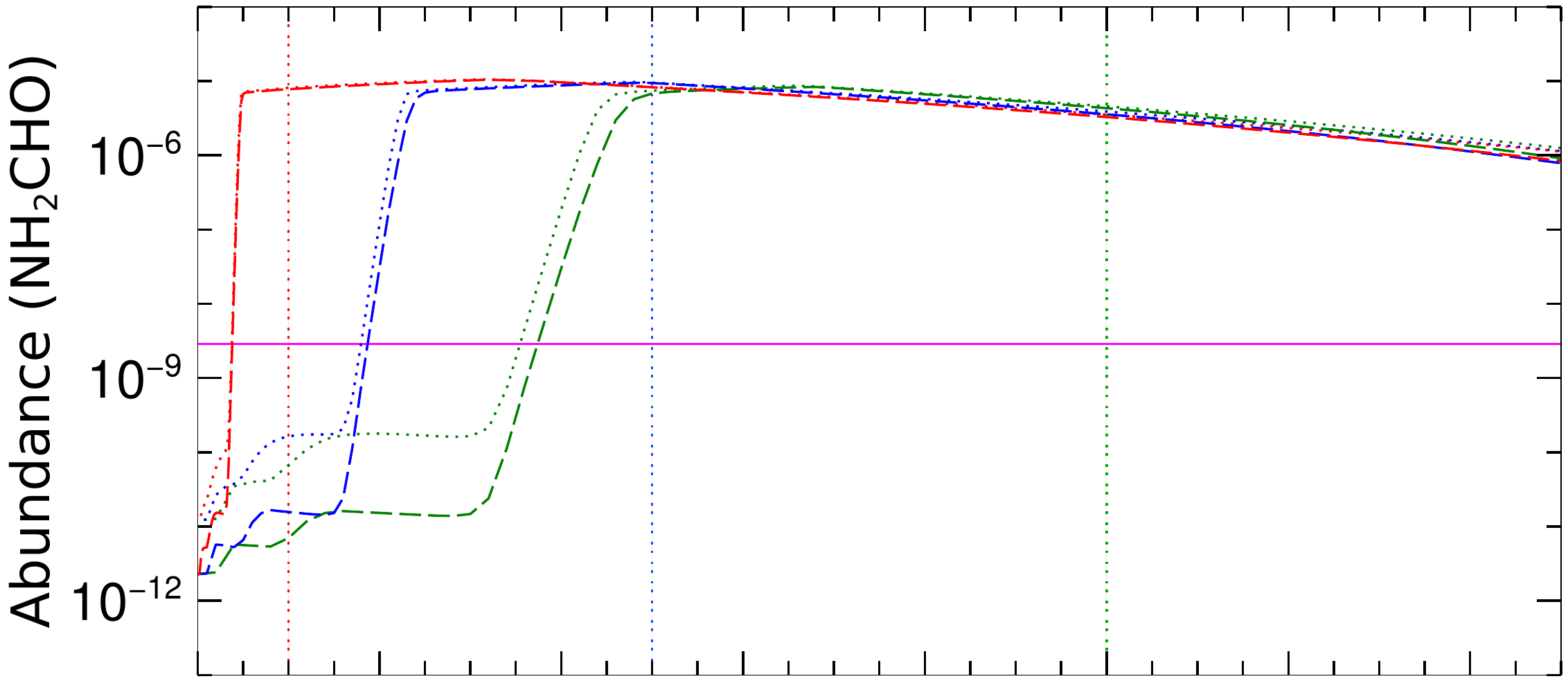}
\includegraphics[width=.48\textwidth]{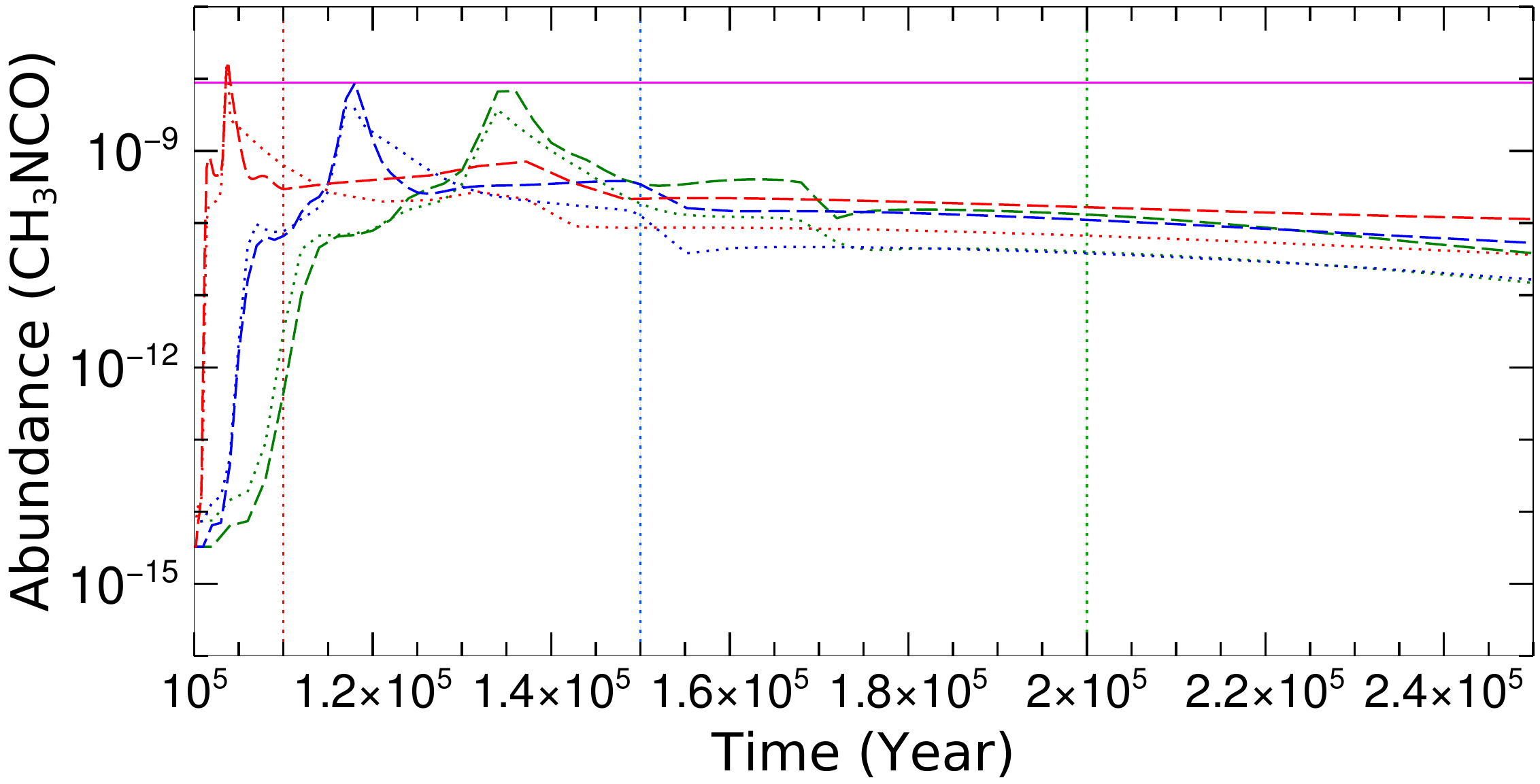}
\includegraphics[width=.48\textwidth]{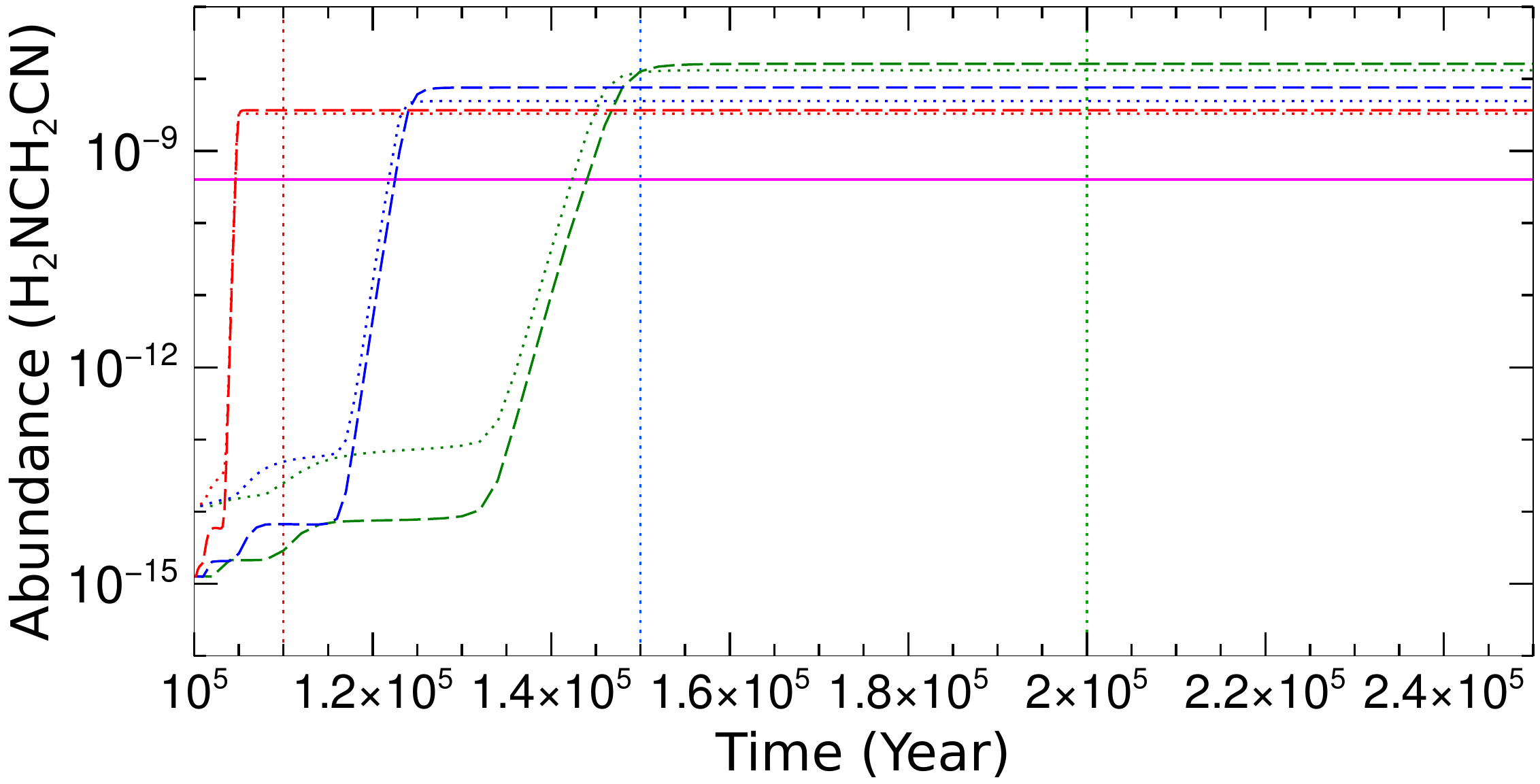}
\caption{Impact of warm-up time (fast, mid, and slow cases are plotted in red, blue, and green, respectively) and final H density (dashed line ($10^7$ cm$^{-3}$) versus dotted line ($10^6$ cm$^{-3}$)) on simulated abundances of selected species. Red, blue, and green dashed lines are for models N7-Tg200-Td20-fast-wup, N7-Tg200-Td20, and N7-Tg200-Td20-slow-wup, respectively, and red, blue, and green dotted lines are for models N6-Tg200-Td20-fast-wup, N6-Tg200-Td20, and N6-Tg200-Td20-slow-wup, respectively. All plots are for both warm-up and post-warm-up stages, with both stages separated by red, blue, and green vertical dotted lines for fast, mid, and slow cases, respectively. Pink horizontal bar represents the observed value in G10 (see Table \ref{tbl:physicalEvo} for details of each model).}
\label{fig:r3}
\end{figure*}
We implemented three different values for the total duration of the warm-up stage: $t_{w}$ = $10^4$, $5\times 10^4$, and $10^5$  years, respectively, for fast, mid, and slow cases. In Figure \ref{fig:r3}, we show the impact of different warm-up durations on the time-dependent chemical evolution of selected species. In Figure \ref{fig:r3}, red, blue, and green vertical dotted lines mark the end of the warm-up period and the start of the post-warm-up stage for fast, mid, and slow models, respectively.  During the warm-up stage, the initial dust temperature is 20K and the gas temperature is 40K. During warm-up, the dust temperature is coupled with the gas temperature and climbs to its peak value (200K). Due to a gradual increase in dust temperature, thermal desorption becomes very efficient even for heavier species with a binding energy above 4000K. Due to efficient thermal desorption, almost all the ice species are evaporated within $10^4$ years once the dust temperature goes above 100K and continues to rise. Soon after that, abundances of gas species reach a steady-state value. In Figure \ref{fig:r3}, we can see that abundances of all species rise fastest in a fast warm-up model, as the rise in temperature is steeper, followed by mid and slow warm-up models. However, there is no significant difference in the steady state abundance values in different models after $2\times 10^5$ years, except for C$_2$H$_3$CN and C$_2$H$_5$CN, as our simulation shows that these two species are sensitive to the warm-up duration. In all three models, simulated abundances of all selected species fit well with observed values during the warm-up stage. CH$_3$CCH is slightly under-produced in all our models. CH$_3$CCH is efficiently produced only when the initial dust temperature is 15K (see Figure \ref{fig:r2}).

\subsection{Results obtained with 1D model \label{model-II}}
\begin{figure*}[t]
\includegraphics[width=.96\textwidth]{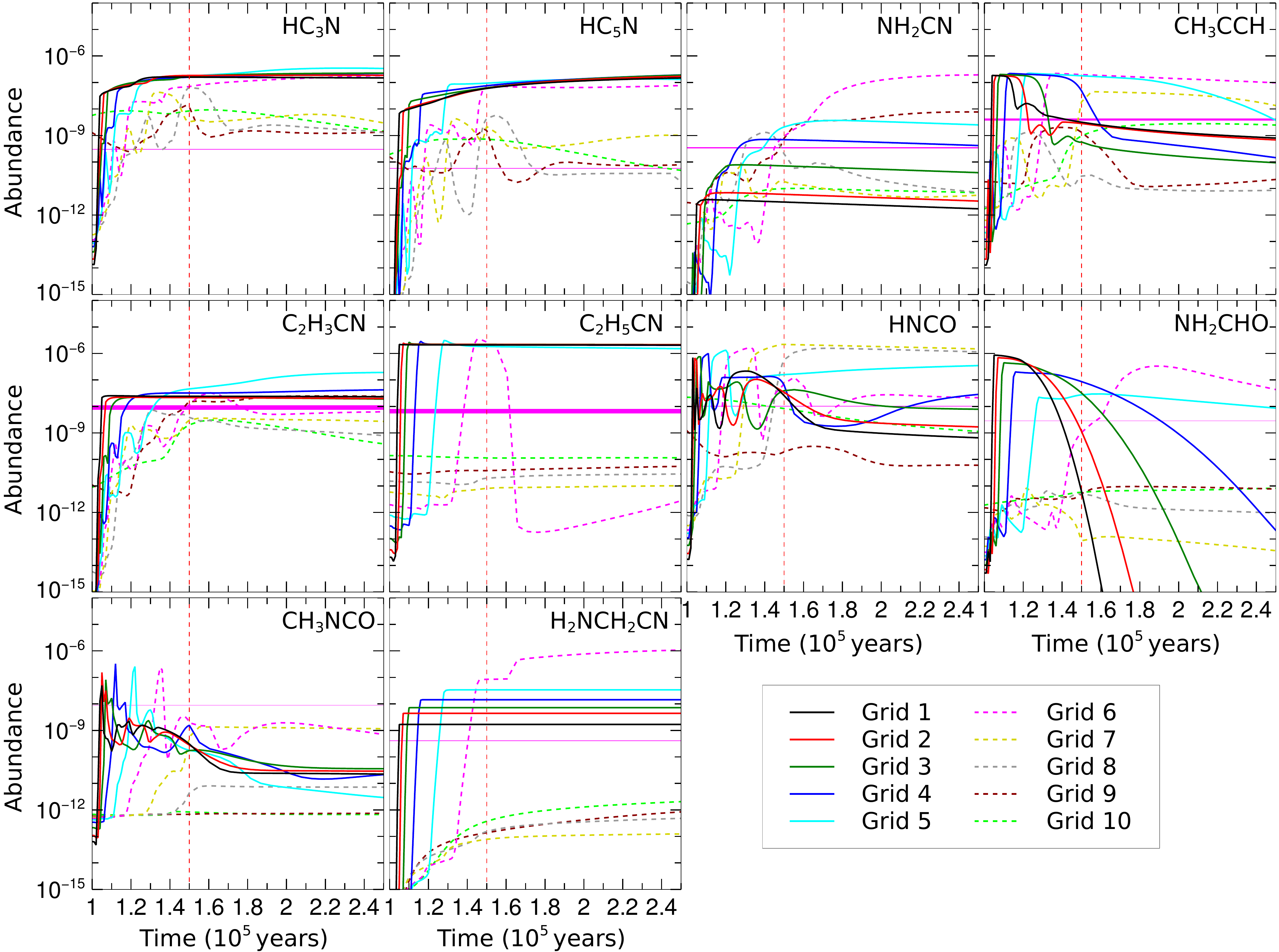}
\caption{Time evolution of abundances of selected species during the warm-up and post-warm-up phases for 1D physical model with initial dust and gas temperature of 15 K. Red dashed vertical line at $1.5 \times 10^5$ years shows the end of the warm-up phase. Pink horizontal bar represents observed abundance range.}
\label{fig:abun-evo-II-15K}
\end{figure*}
\begin{figure*}[t]
\includegraphics[width=.96\textwidth]{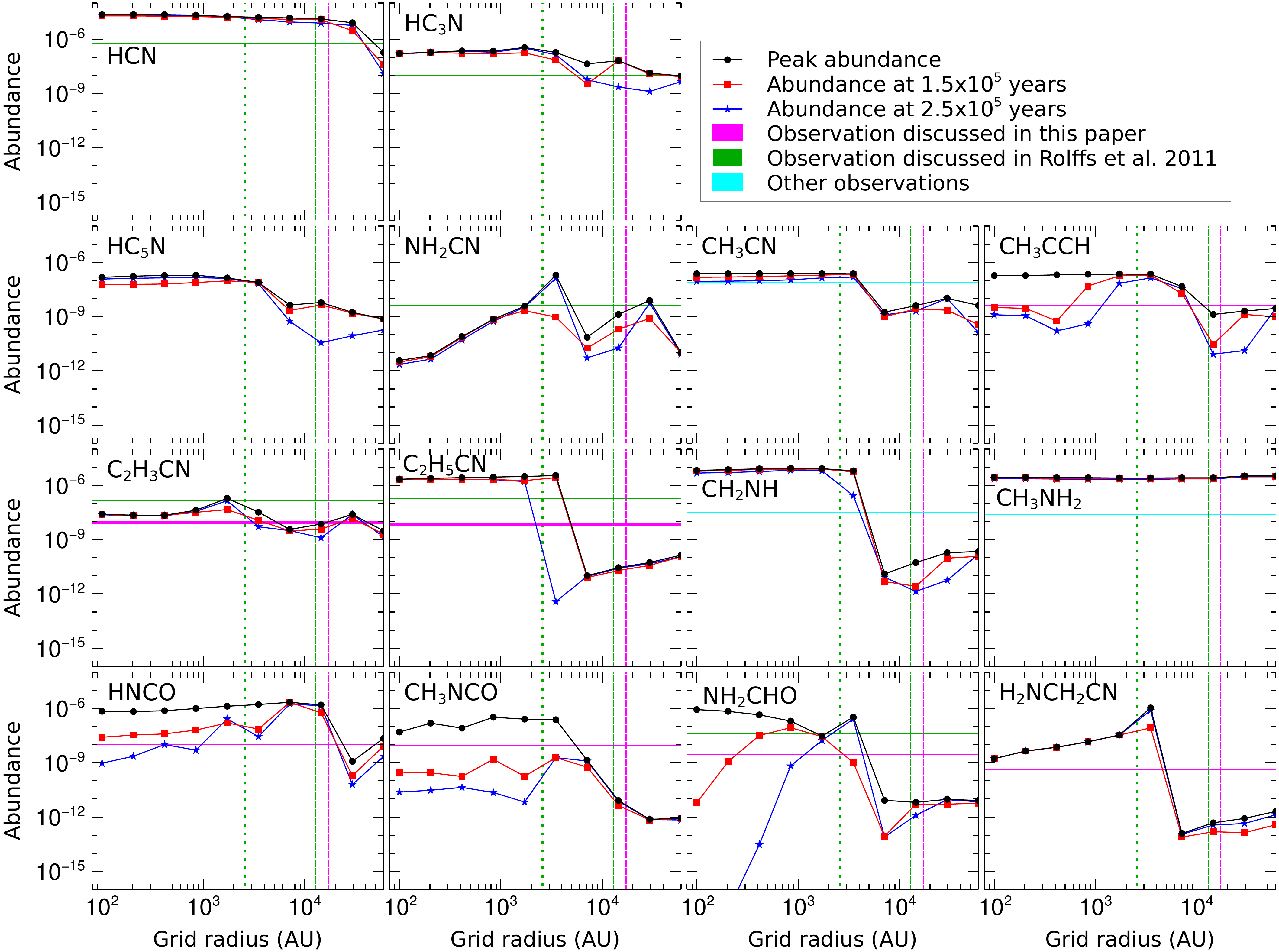}
\caption{Spatial distribution of abundances of selected species at three selected times in simulation. Red, blue, and black curves show the spatial distribution of abundances at  $\sim 1.5 \times 10^5$ years,  $\sim 2.5 \times 10^5$ years, and the peak abundance, respectively. Peak abundance is the max abundance during the warm-up and post-warm-up phases. Here, we considered the initial dust temperature of 15 K. Green dotted (at 2580 AU), green dashed (at 12900 AU), and pink dashed (at 17200 AU) lines represent the highest spatial resolution (0.3$^{''}$) and the moderate resolution (1.5$^{''}$) of \cite{rolf11},  and our resolution ($2^{''}$), respectively. Pink horizontal bar represents observed abundance range.}
\label{fig:peak-model-II-15K}
\end{figure*}
Here, we show the results for 1D simulations, which are essential to studying the possible spatial distribution of N-bearing species in the source G10. We ran two simulations (based on initial dust temperatures of 15 and 20 K) of the 1D model. In each simulation, we have ten grids representing ten spatial points in the source. Details on these spatial points are given in Section \ref{sec:physicalG101D}.  

The results showing the time evolution of the abundances of selected species for all ten grids are shown in  Figure \ref{fig:abun-evo-II-15K}. In this simulation, the initial gas and dust temperature was set at 15 K. The same plot for an initial dust and gas temperature of 20 K is given in Appendix \ref{App-ID}, Figure \ref{fig:abun-evo-II-20K}. The observed abundances are marked with horizontal bars. Grid 1 is the innermost grid, whereas Grid 10 is the outermost. Our model shows that the abundances of all species vary by a few orders of magnitude between the inner and outer grids. Some species such as HC$_3$N and HC$_5$N are most abundant at inner grids, while others such as NH$_2$CN and CH$_3$CCH are most abundant in the outer grids.

For a better comparison with values presented in \cite{rolf11}, we show the spatial distribution of peak abundances of selected species in Figure \ref{fig:peak-model-II-15K}. Again, this plot is for models with initial dust temperatures of 15 K. A similar plot with an initial dust temperature of 20 K is given in Appendix \ref{App-ID}, Figure \ref{fig:peak-model-II-20K}. The dotted and dashed green vertical lines represent the observed spatial resolutions of \cite{rolf11}, whereas the dashed pink vertical line represents the spatial resolution of our observation.  Peak abundances (black curves) are extracted by noting the abundances achieved during warm-up and post-warm-up periods. To avoid the time uncertainty of the peak values, we also show the abundances obtained just after the warm-up phase ($1.5 \times 10^5$ years) and final time ($2.5 \times 10^5$ years). 

As shown in Figure \ref{fig:peak-model-II-15K}, the spatial resolution of the observation presented in \cite{rolf11} is much higher than that of the observations in this paper. We also note that the simulated peak abundances of all species is at its maximum toward the inner grids, though not in the same place. This is important to understand why, for most of the species, \cite{rolf11} recorded a higher maximum abundance compared to us for the source G10. 
Our ID simulations shows a high abundance of HCN and HC$_3$N deeper inside the cloud (i.e., around the high-temperature region), which could not be resolved with observation presented in this paper. However, HC$_3$N is more extended compared to HC$_5$N. This is reflected in our modeling results (Figure \ref{fig:peak-model-II-15K} and Figure \ref{fig:peak-model-II-20K}). They show that HC$_5$N decays in the outer part of the envelope more rapidly compared to HC$_3$N. 
The abundance of vinyl cyanide and ethyl cyanide seems to be heavily dependent on the initial ice temperature. In the model with the initial gas and dust temperature of 20 K, shown in Figure \ref{fig:abun-evo-II-20K}, abundances of vinyl cyanide and ethyl cyanide drop rapidly compared to the model with the initial gas and dust temperature of 15 K, shown in Figure \ref{fig:abun-evo-II-15K}. More interestingly, inside 3000 AU, it depicts a higher abundance of ethyl cyanide than vinyl cyanide. \cite{rolf11} also obtained a higher abundance of ethyl cyanide ($9 \times 10^{17}$ cm$^{-2}$) compared to vinyl cyanide ($7 \times 10^{17}$ cm$^{-2}$). In contrast, we obtained a higher abundance of vinyl cyanide than ethyl cyanide with our resolution. This is also reflected in our model. So, our model at the initial ice phase temperature of 15 K successfully explains a high abundance of ethyl cyanide over vinyl cyanide deep inside, and of vinyl cyanide over ethyl cyanide at the outer part of the envelope. An abundance of CH$_3$NH$_2$ seems roughly invariant across the cloud. The abundances of CH$_2$NH and CH$_3$NH$_2$ are overproduced in our 15 K model, whereas they are underproduced in the 20 K model. Results presented here can explain the observed abundances of the peptide-like bond containing species (HNCO, CH$_3$NCO, and NH$_2$CHO). With the initial gas and dust temperature of 15 K, CH$_3$CCH shows a good match with the observational results (see Figure \ref{fig:peak-model-II-15K}).

\begin{figure}
\includegraphics[width=0.48\textwidth]{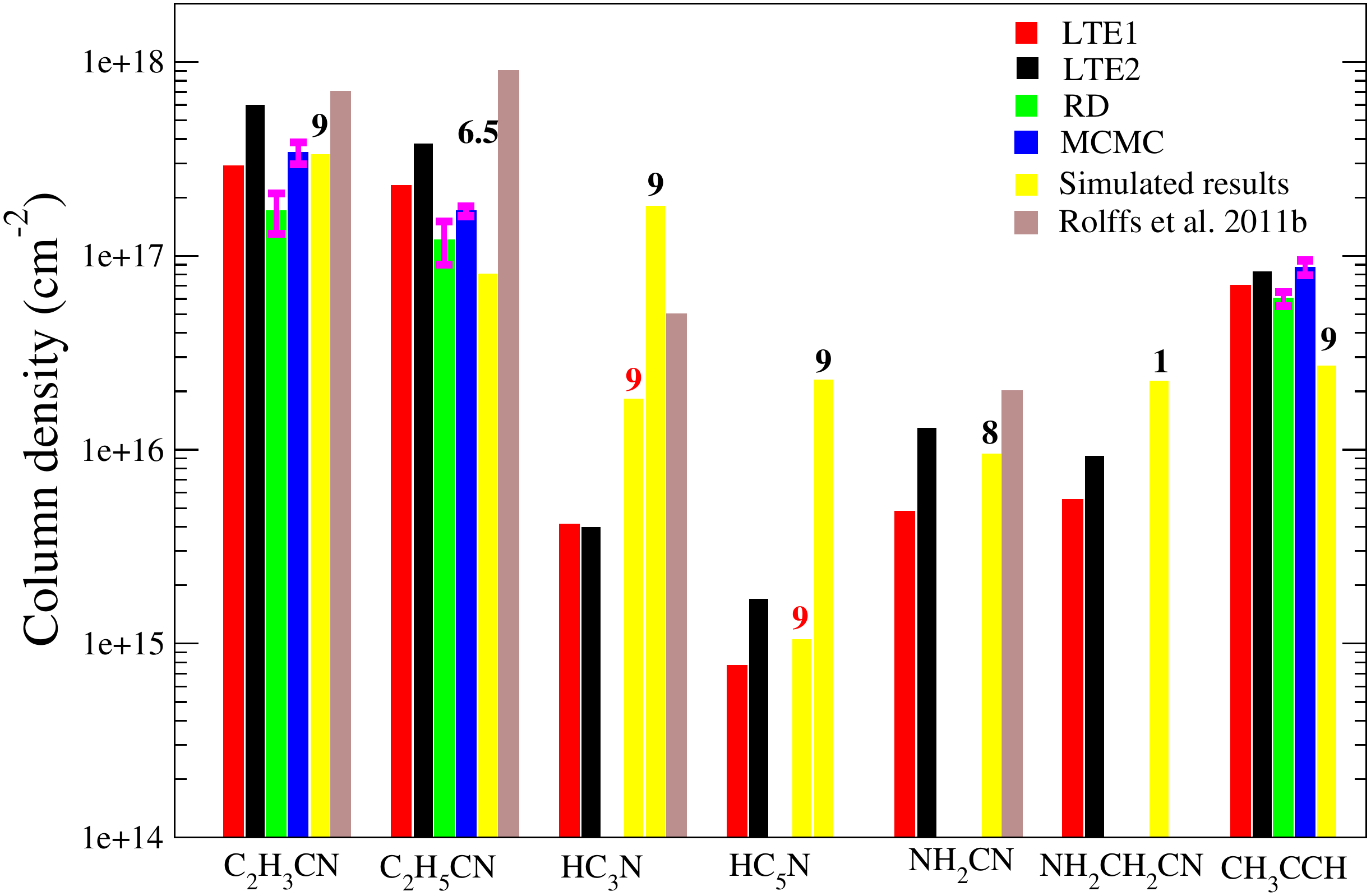}
\caption{Observed and simulated column density of the species are shown. LTE1 represents column density obtained with a constant source size of $2^{''}$, whereas LTE2 represents the cases with the average emitting regions of each species. Simulated column densities (obtained in best-fit 1D model) are shown in yellow. The grid number (noted in black) of the modeled column density represents the peak column density of that grid, which matches well with the observed results.
For HC$_3$N and HC$_5$N, the modeled column density at  $\sim 2.5 \times 10^5$ years is additionally shown for the same grid number noted in red. Error bars obtained with the rotational diagram and MCMC methods are shown in magenta.}
\label{fig:abun_com}
\end{figure}

Simulated (obtained in best-fit 1D model) and observed column densities of the identified species are shown in  Figure \ref{fig:abun_com}.
Various methods are implemented to derive the observed column density of the species. RD (shown in green) represents the column density obtained with the rotational diagram method, and MCMC (shown in blue) illustrates the MCMC Method. LTE1 (shown in red) represents the cases when a fixed source size of $\sim 2^{"}$ is used, whereas LTE2 represents the circumstances for which the average emitting region of species are used (shown in black).
Figure \ref{fig:abun_com} shows that all the methods of deriving the observed column density of a species agree well among themselves. The simulated column densities of species are shown in yellow. The grid number (noted in black) represents the peak column density of that grid, for which we have a good match with the observed results. Additionally, for the cases of HC$_3$N and HC$_5$N, we include the results obtained for the same grid at $\sim 2.5 \times 10^5$ years. 

Figure \ref{fig:abun_com} summarizes our findings. For example, we obtain a good match between the observed and simulated results for NH$_2$CH$_2$CN in the innermost grid and for NH$_2$CN in the fourth grid. It signifies the presence of these species in the inner part of the hot core.
Figure \ref{fig:peak-model-II-15K} shows that in the case of C$_2$H$_5$CN, we obtain a dramatic change in column density between grid numbers 6 and 7. Thus, the column density obtained at the middle of 6 and 7 (i.e., 6.5) is considered from the column density profile of C$_2$H$_5$CN.
Overall, we note that the 
C$_2$H$_3$CN emission and CH$_3$CCH emission are more extended (these match well with the ninth grid) than that of C$_2$H$_5$CN ($\sim 6.5^{th}$ grid).  For HC$_3$N and HC$_5$N, we obtain a good match with the observation during the latter stages of evolution ($\sim 2.5 \times 10^5$) in the nineth grid. However, a higher column density (simulated and observed) of HC$_3$N compared to HC$_5$N implies a relatively extended emission of HC$_3$N.

\section{Conclusions}
This paper reports the identification of some nitrogen-bearing species in the high-mass protostellar envelope of G10 using the archival ALMA data. We also used the Nautilus gas-grain code to investigate the evolution of Nitrogen chemistry in a hot molecular source such as G10. For our chemical modeling, we utilized a physical structure similar to that observed in G10. Our chemical network contains both the gas- and dust-phase chemistry. The main highlights of this paper are presented below.\\

$\bullet$ We identify numerous transitions of nitrogen-bearing species in G10. Among them, vinyl cyanide and its one $^{13}$C isotopologue ($^{13}$CH$_2$CHCN) and ethyl cyanide and its one $^{13}$C isotopologue ($^{13}$CH$_3$CH$_2$CN) are identified. Two other $^{13}$C isotopologues of vinyl cyanide (CH$_2$$^{13}$CHCN and CH$_2$CH$^{13}$CN), one deuterated form of vinyl cyanide (CH$_2$CDCN), another $^{13}$C isotopologue of ethyl cyanide (CH$_3$CH$_2$$^{13}$CN), an isomer of ethyl cyanide (C$_2$H$_5$NC), cyanoacetylene and its two $^{13}$C isotopologues (HC$^{13}$CCN and HCC$^{13}$CN), cyanodiacetyle and its one $^{13}$C isotopologue (HCC$^{13}$CCCN), cyanamide, and aminoacetonitrile are tentatively identified. Additionally, one $^{13}$C isotopologue of ($\rm{{CH_3}^{13}CCH}$) is tentatively identified with our investigation.
\\

$\bullet$ Furthermore, we also identify multiple transitions of methyl acetylene in this source, which is a valuable tracer of physical condition. From our analysis, a kinetic temperature of 133-193 K is estimated. \\

$\bullet$ Based on our detection of vinyl cyanide and one of its $^{13}$C isotopologues, a $\frac{^{12}C}{^{13}C}$ ratio of $45-64$ is obtained. Similarly, with the ethyl cyanide and its one $^{13}$C isotopologue, a $\frac{^{12}C}{^{13}C}$ ratio of $12-27$ is obtained.
%Moreover, by considering all the detection (confirmed and tentative both), a $\frac{^{12}C}{^{13}C}$ ratio of $0.6-79.3$ is estimated. 
Additionally, a ratio of $57.5$ is obtained between ethyl cyanide and ethyl isocyanide. \\

%$\bullet$ Both our 0D and 1D simulation results can reproduce the observed abundances of Nitrogen bearing species discussed in this paper.\\

$\bullet$ Our 0D model with an initial dust temperature of 15 K and a cloud density of $10^7$ cm$^{-3}$ best describes the observed abundances in the source. Our best-fit model is consistent with the source properties of hot and dense sources similar to G10. \\

$\bullet$ 0D simulations show that the chemistry of certain N-bearing species is very sensitive to initial local conditions such as density or dust temperature or both, and it is not very sensitive to initial gas temperature.\\

$\bullet$ In our 1D model, simulated higher abundances of species such as HCN, HC$_3$N, and HC$_5$N toward the inner shells of the source confirm the observational findings.\\

$\bullet$ We obtained a lower abundance of vinyl cyanide, ethyl cyanide, cyanoacetylene, and cyanamide compared to \cite{rolf11}. We note that the emission of most of these species is compact. Since \cite{rolf11} have a higher spatial resolution (highest resolution $\sim 0.3^{''}$ corresponds to 2580 AU and average resolution $\sim 1.5^{''}$ corresponds to 12900 AU) than ours ($\sim 2^{''}$ corresponds to 17200 AU), they obtained a higher abundance of these species.\\ 

$\bullet$ Results obtained with our 1D model, shown in Figures \ref{fig:peak-model-II-15K} and \ref{fig:peak-model-II-20K}, show that the abundance of most of the species increases as we move toward the denser and warmer inner-regions of the source. \\
 In this work, we have identified various nitrogen-bearing species and investigated the  N-bearing chemistry in G10. In addition, we have discussed how the chemistry of N-bearing molecules can vary with initial local conditions and investigated these species' spatial distribution. Since most of the transitions were marginally resolved, higher-angular-resolution observations would be needed to know their spatial distribution more accurately.\\

\begin{acknowledgements} 
This paper makes use of the following ALMA data: ADS/JAO.ALMA$\#$2016.1.00929.S. ALMA is a partnership of ESO (representing its member states), NSF (USA) and NINS (Japan), together with NRC (Canada), MOST and ASIAA (Taiwan), and KASI (Republic of Korea), in cooperation with the Republic of Chile. The Joint ALMA Observatory is operated by ESO, AUI/NRAO and NAOJ. S.K.M.  acknowledges CSIR fellowship (File no. 09/904(0014)/2018-EMR-I). W.I. acknowledges the fellowship of China Postdoctoral Science Foundation (Grant number: 2020M683386). P.G. acknowledges support from a Chalmers Cosmic Origins postdoctoral fellowship. BB gratefully acknowledges the DST, India's Government, for providing financial assistance through the DST-INSPIRE Fellowship [IF170046] scheme.   

%RTG acknowledges support from the National Science Foundation (grant No.
%AST 19-06489).
\end{acknowledgements}
\appendix
%\restartappendixnumbering
\section{\bf Observed line parameters}

%
%
%
%
%
%%%%%%%%%%%%%%%%%%%%%%%%%%%%%%%%%%%%%%%%%%%%%%
\begin{table*}
\tiny{
\begin{center}
\caption{Summary of line parameters of observed molecules toward G10.47+0.03. \label{tab:dataobs}}
%\centering
%\addtolength{\leftskip} {-2cm}
%\addtolength{\rightskip}{-2cm}
\begin{tabular}{|c|c|c|c|c|c|c|c|c|c|}
\hline
%\hline
Species&${\rm J^{'}_{K_a^{'}K_c^{'}}}$-${\rm J^{''}_{K_a^{''}K_c^{''}}}$ &Frequency&E$_u$&$g_{up}$&$\Delta$V&I$_{max}$&V$_{LSR}$&S$\mu^{2}$&$\int{T_{mb}}dV$\\
&&(GHz)&(K)&&(km s$^{-1}$)&(K)&(km s$^{-1}$)&(Debye$^{2}$)&(K km s$^{-1}$)\\
\hline
%\hline
\multicolumn{10}{|c|}{}\\
\multicolumn{10}{|c|}{\it Vinyl cyanide and its isotopologues}\\
%\multicolumn{10}{|c|}{}\\
\hline
&&&&&&&&&\\
C$_2$H$_3$CN&$6_{2,4}\rightarrow6_{1,5}$,v=0&130.763576&18.2&39&9.5$\pm$0.5&3.9$\pm$0.2&68.1$\pm$0.2&8.0&39.2$\pm$3.9\\
&$29_{2,27}\rightarrow29_{1,28}$,v=0&131.168737&209.5&177&7.6$\pm$0.2&7.5$\pm$0.2&67.2$\pm$0.1&54.3&    61.2$\pm$3.3\\
&$14_{0,14}\rightarrow13_{0,13}$,v=0&131.267473$^b$&47.5&87&10.3$\pm$0.4&33.5$\pm$0.9&66.7$\pm$0.1&610.5&       370.7$\pm$23.2\\
&$16_{1,16}\rightarrow15_{1,15}$,v=0&147.561701$^b$&62.5&99&9.0$\pm$0.1&21.3$\pm$0.1&66.8$\pm$0.1&695.6&205.5$\pm$1.8\\
&$18_{0,18}\rightarrow17_{1,17}$,v=0&147.865229$^a$&77.1&111&&&&28.7&\\ 
&$10_{2,9}\rightarrow10_{1,10}$,v=0     &148.028678$^a$&33.6&63&&&&11.5&\\      
&$12_{2,11}\rightarrow12_{1,12}$,v=0&153.421750&44.1&75&8.5$\pm$0.7&    4.5$\pm$0.3&67.9$\pm$0.3&13.4&40.9      $\pm$5.9\\
&$16_{2,14}\rightarrow15_{2,13}$,v=0&153.518944$^b$&70.9&99&11.9$\pm$0.7&26.5$\pm$1.0&67.4$\pm$0.2&687.7&336.1  $\pm$32.2\\
&$16_{1,15}\rightarrow15_{1,14}$,v=0&154.724540$^b$&65.4&99&10.2$\pm$0.4&15.1$\pm$0.5&66.8$\pm$0.2&695.5&       167.2$\pm$13.1\\
&$17_{0,17}\rightarrow16_{0,16}$,v=0&158.657428$^b$&69.1&105&10.2$\pm$0.2&40.3$\pm$0.5& 67.2$\pm$0.1&741.2&     436.5$\pm$13.1\\
&$19_{0,19}\rightarrow18_{1,18}$,v=0&158.773785&85.5&117&7.3$\pm$0.6&8.3$\pm$0.5&       66.8$\pm$0.2&31.1&65.2$\pm$9.7\\
&$26_{1,25}\rightarrow26_{0,26}$,v=0&159.753858$^a$&164.7&159&&&&25.6&\\
&&&&&&&&&\\
$^{13}$CH$_2$CHCN&$16_{5,11}\rightarrow15_{5,10}$&147.907794$^a$&113.7&99&&&&630.4&\\
&$16_{6,11}\rightarrow15_{6,10}$&147.909047$^a$&137.2&99&&&&600.30&\\
&$16_{7,9}\rightarrow15_{7,8}$&147.927607&164.8&99&9.7$\pm$0.5&4.2$\pm$0.2&66.9$\pm$0.2&564.8&42.9$\pm$3.8\\
&$16_{8,9}\rightarrow15_{8,8}$&147.957689$^a$&196.7&99&&&&523.9&\\
&$16_{3,14}\rightarrow15_{3,13}$&147.986686&79.5&99&9.1$\pm$0.5&5.9$\pm$0.2&66.0$\pm$0.2&673.9&57.9$\pm$4.7\\
&$16_{9,7}\rightarrow15_{9,6}$&147.996640$^a$&232.8&99&&&&477.5&\\
&$16_{2,14}\rightarrow15_{2,13}$&149.423715&69.2&99&9.8$\pm$2.5&5.8$\pm$1.1&67.4$\pm$0.9&687.6& 61.7$\pm$26.7\\
&$17_{0,17}\rightarrow16_{0,16}$&154.577621$^a$&67.2&105&&&&741.2&\\
&$17_{1,16}\rightarrow16_{1,15}$&159.954635&71.2&105&7.4$\pm$0.4&8.2$\pm$0.3&66.9$\pm$0.2&739.4&64.7$\pm$5.9\\
&&&&&&&&&\\
CH$_2$$^{13}$CHCN&$14_{0,14}\rightarrow13_{0,13}$,v=0&130.481700$^a$&47.2&87&&&&610.5&\\
&$16_{0,16}\rightarrow15_{0,15}$,v=0&148.643016$^a$&61.1&99&&&&697.6&\\
&$16_{1,15}\rightarrow15_{1,14}$,v=0&153.925606&65.1&99&6.9$\pm$0.1&3.9$\pm$0.1&67.4$\pm$0.2&   695.5&29.1$\pm$1.2\\
&$17_{2,16}\rightarrow16_{2,15}$,v=0&159.872596$^a$&77.59&105&&&&731.7&\\
&&&&&&&&&\\
CH$_2$CH$^{13}$CN&$14_{0,14}\rightarrow13_{0,13}$,v=0&130.700756$^a$&47.3&87&&&&610.6&\\
&$16_{0,16}\rightarrow15_{0,15}$,v=0&148.916702$^a$&61.1&99&&&&697.7&\\
&$16_{1,15}\rightarrow15_{1,14}$,v=0&154.034285&65.1&99&7.6$\pm$0.1&3.9$\pm$0.1&66.6$\pm$0.1&695.5&31.9$\pm$0.8\\
&$17_{2,16}\rightarrow16_{2,15}$,v=0&160.051637$^a$&77.8&105&&&&731.8&\\
&&&&&&&&&\\
CH$_2$CHC$^{15}$N&$16_{2,14}\rightarrow15_{2,13}$&148.8982$^a$&61.9&33&&&&223.1&\\
&$17_{1,16}\rightarrow16_{1,15}$&159.436449$^a$&71.2&35&&&&240.4&\\
&&&&&&&&&\\
CH$_2$CDCN&$14_{6,9}\rightarrow13_{6,8}$&130.637285$^a$&108.3&29&&&&166.4&\\    
&$14_{5,10}\rightarrow13_{5,9}$&130.661515$^a$&89.6&29&&&&177.8&\\      
&$14_{9,5}\rightarrow13_{9,4}$&130.645695$^a$&184.7&29&&&&119.6&\\      
&$14_{4,11}\rightarrow13_{4,10}$&130.720066&74.2&29&11.2$\pm$0.5&2.0$\pm$0.1&67.6$\pm$0.1&      23.6&187.1$\pm$1.6\\
&$14_{4,10}\rightarrow13_{4,9}$&130.725273$^a$&74.3&29&&&&187.1&\\      
&$17_{0,16}\rightarrow16_{0,16}$&154.309753$^a$&67.4&35&&&&246.8&\\     
&$17_{3,14}\rightarrow16_{3,13}$&159.383849&83.9&35&6.7$\pm$0.2&3.2$\pm$0.1&68.2$\pm$0.1&       23.2&239.7$\pm$0.9\\
\hline
\multicolumn{10}{|c|}{}\\
\multicolumn{10}{|c|}{\it Ethyl cyanide and its isotopologues and isomer}\\
%\multicolumn{10}{|c|}{}\\
\hline
&&&&&&&&&\\
C$_2$H$_5$CN&$14_{1,14}\rightarrow13_{0,13}$,v=0&129.768140&44.8&29&9.8$\pm$0.8&15.5$\pm$0.9&65.6$\pm$0.3&15.1&162.8$\pm$22.2\\
&$15_{1,15}\rightarrow14_{1,14}$,v=0&129.795661$^b$&51.1&31&9.4$\pm$0.2&43.1$\pm$0.9&66.9$\pm$0.1&221.2&431.7$\pm$19.4\\
&$21_{3,19}\rightarrow21_{2,20}$,v=0&130.693882&109.4&43&7.9$\pm$0.4&11.7$\pm$0.5&66.5$\pm$0.2&16.3&99.1$\pm$8.9\\
&$15_{0,15}\rightarrow14_{0,14}$,v=0&130.903902$^b$&50.8&31&10.5$\pm$0.3&44.9$\pm$1.1&67.5$\pm$0.1&221.6&503.9$\pm$28.2\\
&$17_{0,17}\rightarrow16_{0,16}$,v=0&147.756711$^b$&64.6&35&11.8$\pm$1.1&31.9$\pm$2.0&65.5$\pm$0.4&251.13&403.5$\pm$63.8\\
&$24_{4,20}\rightarrow24_{3,21}$,v=0&148.293988&147.1&49&7.5$\pm$0.7&7.8$\pm$0.6&66.8$\pm$0.3&21.1&62.1$\pm$10.1\\
&$10_{2,9}\rightarrow9_{1,8}$,v=0&148.362760&28.1&21&9.8$\pm$1.9&6.4$\pm$0.9&68.2$\pm$0.7&5.4&66.5$\pm$22.7\\
&$33_{2,31}\rightarrow33_{1,32}$,v=0&148.374933$^a$&249.5&67&&&&21.9&\\
&$25_{2,24}\rightarrow25_{1,25}$,v=0&148.677605$^a$&143.0&51&&&&10.3&\\
&$17_{3,14}\rightarrow16_{3,13}$,v=0&153.272198$^b$&75.9&35&9.3$\pm$0.2&26.6$\pm$0.4&66.8$\pm$0.1&244.1&266.5$\pm$7.9\\
&$28_{3,26}\rightarrow28_{2,27}$,v=0&153.589524&184.6&57&8.7$\pm$0.8&5.4$\pm$0.4&67.6$\pm$0.3&19.4&50.4$\pm$8.5\\
&$17_{1,16}\rightarrow16_{1,15}$,v=0&154.244297$^b$&68.2&35&10.5$\pm$0.1&28.9$\pm$0.2&67.9$\pm$0.1&250.6&324.4$\pm$5.9\\
&$21_{4,17}\rightarrow21_{3,18}$,v=0&154.475980&117.2&43&7.6$\pm$0.2&5.9$\pm$0.1&66.9$\pm$0.1&17.6&48.5$\pm$1.9\\
&$26_{2,25}\rightarrow26_{1,26}$,v=0&154.557360&154.0&53&7.1$\pm$0.2&3.5$\pm$0.1&66.4$\pm$0.1&10.4&26.8$\pm$1.0\\
&$20_{8,13}\rightarrow21_{7,14}$,v=0&158.742366&161.4&41&6.3$\pm$0.2&4.8$\pm$0.2&66.5$\pm$0.1&3.3&32.3$\pm$2.4\\
&$17_{4,13}\rightarrow17_{3,14}$,v=0&159.330800&83.6&35&7.3$\pm$0.2&13.2$\pm$0.2&66.8$\pm$0.1&13.6&102.9$\pm$4.1\\
&$18_{1,18}\rightarrow17_{0,17}$,v=0&159.78554$^a$&72.2&37&&&&21.1&\\
&$18_{2,17}\rightarrow17_{2,16}$,v=0&159.888831$^b$&77.6&37&9.8$\pm$0.1&53.6$\pm$0.5&66.9$\pm$0.1&263.3&561.5$\pm$11.8\\
&$16_{4,12}\rightarrow16_{3,13}$,v=0&160.050058$^a$&76.3&33&&&&12.7&\\
&&&&&&&&&\\
$^{13}$CH$_3$CH$_2$CN&15$_{2,14}\rightarrow14_{2,13}$,v=0&130.319847$^a$&54.5&31&&&&214.4&\\
&$15_{8,7}\rightarrow14_{8,6}$,v=0&     131.109474$^a$&120.8&31&&&&156.3&\\
&$15_{6,9}\rightarrow14_{6,8}$,v=0&131.119334$^a$&89.9&31&&&&183.4&\\
\hline
\end{tabular}
\begin{tablenotes}
\item $^a$ Gaussian fits are not performed for the blended
transitions. $^b$ Optically thick line.
\end{tablenotes}
\end{center}}
\end{table*}
\clearpage
%
%
%
%
%
%%%%%%%%%%%%%%%%%%%%%%%%%%%%%%%%%%%%%%%%%%%%%%%%%%%%%%%%%%%%%%%%%%%%%
%\begin{table*}
\tiny{
%\caption{(Continuation of Table \ref{tab:dataobs}).}
%Summary of the line parameters of the observed molecules toward G10.47+0.03 
%\centering 
\begin{center}
%\addtolength{\leftskip} {-2cm}
%\addtolength{\rightskip}{-2cm}
\begin{tabular}{|c|c|c|c|c|c|c|c|c|c|}
%\hline
\hline
Species&${\rm J^{'}_{K_a^{'}K_c^{'}}}$-${\rm J^{''}_{K_a^{''}K_c^{''}}}$ &Frequency&E$_u$&$g_{up}$&$\Delta$V&I$_{max}$&V$_{LSR}$&S$\mu^{2}$&$\int{T_{mb}}dV$\\
&&(GHz)&(K)&&(km s$^{-1}$)&(K)&(km s$^{-1}$)&(Debye$^{2}$)&(K km s$^{-1}$)\\
\hline
%\hline
&$15_{5,10}\rightarrow14_{5,9}$,v=0&131.155447&77.8&31&8.2$\pm$0.6&7.8$\pm$0.3&67.9$\pm$0.2&194.2&68.06$\pm$7.3\\
&$15_{11,4}\rightarrow14_{11,3}$,v=0&131.161428&183.4&31&8.1$\pm$0.1&3.2$\pm$0.1&66.7$\pm$0.1&100.9&28.2$\pm$0.7\\
&$15_{4,12}\rightarrow14_{4,11}$,v=0&131.233679&67.9&31&8.3$\pm$0.5&4.8$\pm$0.1&66.1$\pm$0.1&202.9&42.9$\pm$3.8\\
&$15_{4,11}\rightarrow14_{4,10}$,v=0&131.249984&67.9&31&9.9$\pm$0.7&6.1$\pm$0.2&66.4$\pm$0.2&202.8&63.8$\pm$7.4\\
&$17_{6,11}\rightarrow16_{6,10}$,v=0&148.625244$^a$&103.3&35&&&&216.6&\\
&$17_{11,6}\rightarrow16_{11,5}$,v=0&148.647252$^a$&197.2&35&&&&143.8&\\
&17$_{5,13}\rightarrow16_{5,12}$,v=0&148.682220$^a$&91.7&35&&&&226.1&\\
&$17_{4,13}\rightarrow16_{4,12}$,v=0&148.833416$^a$&81.5&35&&&&233.8&\\
&$18_{1,17}\rightarrow17_{1,16}$,v=0&159.10002&74.2&37&7.7$\pm$0.3&8.6$\pm$0.1&66.6$\pm$0.1&260.7&70.2$\pm$3.6\\
&$12_{4,8}\rightarrow12_{3,9}$,v=0&160.026237$^a$&78.2&25&&&&275.6&\\
&&&&&&&&&\\
CH$_3$$^{13}$CH$_2$CN&$15_{0,15}\rightarrow14_{0,14}$,v=0&130.052706$^a$&50.5&31&&&&217.6&\\
&$18_{2,17}\rightarrow17_{2,16}$,v=0&159.009721$^a$&77.1&37&&&&258.6&\\
&&&&&&&&&\\
CH$_3$CH$_2$$^{13}$CN&$15_{0,15}\rightarrow14_{0,14}$,v=0&130.280373$^a$&50.5&31&&&&217.7&\\
%&5.02$\pm$     0.6&2.7$\pm$0.3&66.2$\pm$0.2&217.7&14.5$\pm$3.1\\
&$17_{1,16}\rightarrow16_{1,15}$,v=0&153.487081$^a$&67.8&35&&&&246.1&\\
&$17_{2,15}\rightarrow16_{2,14}$,v=0&154.609607$^a$&70.7&35&&&&244.2&\\
&$18_{1,18}\rightarrow17_{1,17}$,v=0&154.679342&71.8&37&7.7$\pm$0.7&3.1$\pm$0.1&66.9$\pm$0.2&261.1&25.9 $\pm$3.6\\
&$18_{2,17}\rightarrow17_{2,16}$,v=0&159.105544$^a$&77.2&37&&&&258.6&\\
&&&&&&&&&\\
C$_2$H$_5$C$^{15}$N&$15_{10,5}\rightarrow14_{10,4}$&130.567344$^a$&161.2&31&&&&123.5&\\
&$18_{2,16}\rightarrow17_{2,15}$&159.848501$^a$&76.8&37&&&&236.6&\\
&&&&&&&&&\\
C$_2$H$_5$NC&$16_{1,16}\rightarrow15_{1,15}$,v=0&149.120363&62.2&33&11.9$\pm$0.1&4.8$\pm$0.1&   69.6$\pm$0.1&228.7&61.2$\pm$1.1\\
&$17_{0,17}\rightarrow16_{0,16}$,v=0&159.061678&69.6&35&8.4$\pm$0.1&2.9$\pm$0.1&69.6$\pm$0.1&243.3&26.7$\pm$0.7\\
\hline
\multicolumn{10}{|c|}{}\\
\multicolumn{10}{|c|}{\it Cyanoacetylene and its isotopologue}\\
%\multicolumn{10}{|c|}{}\\
\hline
&&&&&&&&&\\
HC$_3$N&17$\rightarrow$16, v=0&154.657284&66.8&35&13.6$\pm$0.2&36.4$\pm$0.4&66.5$\pm$0.2&236.7&528.8$\pm$23.5\\
&&&&&&&&&\\
HC$^{13}$CCN&17$\rightarrow$16, v=0&154.001217&66.5&35&8.3$\pm$0.1&20.9$\pm$0.2&67.0$\pm$0.03&236.8&183.4$\pm$2.5\\
&&&&&&&&&\\
HCC$^{13}$CN&17$\rightarrow$16,v=0&154.016078&66.5&35&9.3$\pm$0.1&22.38$\pm$0.1&66.9$\pm$0.02&236.8&220.9$\pm$3.1\\
\hline
\multicolumn{10}{|c|}{}\\
\multicolumn{10}{|c|}{\it Cyanodiacetylene and its isotopologue}\\
%\multicolumn{10}{|c|}{}\\
\hline
&&&&&&&&&\\
HC$_5$N&49$\rightarrow$48&130.456437&156.5&297&9.2$\pm$0.4&7.0$\pm$0.1&68.9$\pm$0.2&2755.8&68.8$\pm$5.4\\
&&&&&&&&&\\
HCC$^{13}$CCCN&49$\rightarrow$48&130.321350&156.4&99&8.3$\pm$0.2&8.2$\pm$0.1&69.8$\pm$0.1&918.7&72.1$\pm$3.2\\
\hline
\multicolumn{10}{|c|}{}\\
\multicolumn{10}{|c|}{\it Cyanamide}\\
%\multicolumn{10}{|c|}{}\\
\hline
&&&&&&&&&\\
NH$_2$CN&$8_{1,8}\rightarrow7_{1,7}$,v=0&158.891146&48.8&51&11.6$\pm$1.2&9.4$\pm$0.3&68.1       $\pm$0.2&442.2&116.2$\pm$16.2\\
&$8_{2,7}\rightarrow7_{2,6}$,v=1&159.595146$^a$&162.3&51&&&&401.8&\\
&$8_{2,6}\rightarrow7_{2,5}$,v=1&159.607737$^a$&162.3&51&&&&401.8&\\
&$8_{3,6}\rightarrow7_{3,5}$,v=1&159.749861$^a$&232.0&17&&&&125.2&\\
&$8_{0,8}\rightarrow7_{0,7}$,v=1&159.814663$^a$&105.8&51&&&&438.3&\\
&$8_{3,6}\rightarrow7_{3,5}$,v=0&160.104805&164.9&51&10.2$\pm$0.5&13.4$\pm$0.4&68.5$\pm$        0.1&391.6&146.3$\pm$11.5\\
\hline
\multicolumn{10}{|c|}{}\\
\multicolumn{10}{|c|}{\it Aminoacetonitrile}\\
%\multicolumn{10}{|c|}{}\\
\hline
H$_2$NCH$_2$CN&$17_{7,11}\rightarrow16_{7,10}$,v=0&154.369232$^a$  &127.1&35&&&&93.7&\\                 
&$17_{8,9}\rightarrow16_{8,8}$,v=0&154.373384$^a$    &145.5     &35&&&&87.89&\\                 
&$17_{6,12}\rightarrow16_{6,11}$,v=0&154.382222$^a$    &111.1     &35&&&&98.93&\\                       
&$17_{9,8}\rightarrow16_{9,}7$,v=0&154.388904$^a$     &166.4&35 &&&&81.25&\\                    
&$17_{4,14}\rightarrow16_{4,13}$,v=0&154.517480&86.4&35&8.7$\pm$0.4&2.3$\pm$0.1&65.8$\pm$0.1&   106.6&21.3$\pm$1.5\\
&&&&&&&&&\\
\hline
%\multicolumn{10}{c}{Nitrogen Sulfide}\\
%NS-33&4 -1 3.5 3 3$\rightarrow$3 1 2.5 2      2&159.81309&16.4&&15.7$\pm$0.6&12.2$\pm$0.1&68.9$\pm$0.1&6.3&204.9$\pm$9.7\\
%&4 1 3.5 2 1$\rightarrow$3 -1 2.5 1 0&160.23602&16.5&&11.6$\pm$0.4&6.6$\pm$0.
%2&68.8$\pm$0.1&1.9&81.7$\pm$3.2\\
%&&&&&&&&&\\
%NS-34&4 1 3.5 2.5$\rightarrow$3 -1 2.5 2.5&158.78048&16.4&&10.5$\pm$0.6&4.7$\pm$0.2&67.3$\pm$0.2&0.9&52.9$\pm$5.1\\
%\hline
CH$_3$CCH&9$_6\rightarrow8_6$&153.711520&296.1&38&9.2$\pm$0.9&2.5$\pm$0.2&66.6$\pm$0.4&5.6&24.0$\pm$4.3\\
&9$_5\rightarrow8_5$&153.743800$^a$&216.9&19&&&&3.5&\\
&9$_4\rightarrow8_4$&153.770224$^a$&152.1&19&&&&4.0&\\
&9$_3\rightarrow8_3$&153.790769&101.7&38&10.8$\pm$0.9&8.7$\pm$0.6&67.2$\pm$0.3&9.0&100.3$\pm$14.9\\
&9$_2\rightarrow8_2$&153.805457&65.7&19&10.0$\pm$0.4&6.8$\pm$0.2&67.1$\pm$0.2&4.8&72.5$\pm$5.4\\
&9$_1\rightarrow8_1$&153.814273$^a$&44.9&19&&&&4.9&\\
&9$_0\rightarrow8_0$& 153.817211$^a$&36.1&19&&&&5.0&\\
&&&&&&&&&\\
CH$_3$$^{13}$CCH&9$_3\rightarrow8_3$&153.72701&101.7&38&8.6$\pm$0.1&1.9$\pm$0.1&65.9$\pm$0.1&9.0&       17.7$\pm$0.4\\
&9$_2\rightarrow8_2$&153.741680$^a$     &65.7&19&&&&4.8&\\
&9$_1\rightarrow8_1$&153.750470$^a$     &44.1&19&&&&5.0&\\
&9$_0\rightarrow8_0$&153.753370&36.8&19&9.6$\pm$0.4&1.8$\pm$0.1&67.7    $\pm$0.2&       5.1&18.8$\pm$1.3\\
\hline
\end{tabular}
\begin{tablenotes}
\item $^a$  Gaussian fits are not performed for the blended
transitions. $^b$ Optically thick line.
\end{tablenotes}
\end{center}}
%\end{table*}
%
%
%
%
\clearpage

\begin{table*}
%\hskip -3.5cm
\tiny{
\caption{Summary of the best-fit line parameters obtained using MCMC method considering a source size = 2$^{''}$ and a V$_{lsr}$ = 68 km/s.\label{table:mcmc_lte}} 
%For this fitting, we have used V$_{lsr}=97.0$ km/s. }
\begin{center}
\addtolength{\leftskip} {-2cm}
\addtolength{\rightskip}{-2cm}
\begin{tabular}{|c|c|c|c|c|c|c|c|}
\hline
%\hline
Species&Quantum numbers&Frequency&E$_u$&FWHM&Best fitted column&Best fitted&Optical depth\\
&&(GHz)&(K)&(Km s$^{-1}$)&density (cm$^{-2}$) & T$_{ex}$ (K)&($\tau$)\\
\hline
%\hline
\multicolumn{8}{|c|}{}\\
\multicolumn{8}{|c|}{\it Vinyl cyanide and its isotopologues}\\
%\multicolumn{8}{|c|}{}\\
\hline
$\rm{C_2H_3CN}$&$6_{2,4}\rightarrow6_{1,5}$,v=0&130.763576&18.2&&&&0.02\\
&$29_{2,27}\rightarrow29_{1,28}$,v=0&131.168737&209.5&&&&0.05\\
&$12_{2,11}\rightarrow12_{1,12}$,v=0&153.42175&44.1&8.2&$(3.4\pm0.4 )\times10^{17}$&209.8$\pm$18.02&0.03\\
&$19_{0,19}\rightarrow18_{1,18}$,v=0&158.773785&85.5&&&&0.06\\
&&&&&&&\\
%\hline
$^{13}$CH$_2$CHCN&$16_{7,9}\rightarrow15_{7,8}$&147.927607&164.8&&&&0.03\\
&$16_{3,14}\rightarrow15_{3,13}$&147.986686&79.5&9.0&(5.3$\pm0.4)\times10^{15}$&$125.8\pm$13.67&0.07\\
&$16_{2,14}\rightarrow15_{2,13}$&149.423715&69.2&&&&0.07\\
&$17_{1,16}\rightarrow16_{1,15}$&159.954635&71.2&&&&0.08\\
\hline
\multicolumn{8}{|c|}{}\\
\multicolumn{8}{|c|}{\it Ethyl cyanide with its isotopologues and isomer}\\
%\multicolumn{8}{|c|}{}\\
\hline
C$_2$H$_5$CN&$14_{1,14}\rightarrow13_{0,13}$,v=0&129.768140&44.8&&&&0.20\\
&$21_{3,19}\rightarrow21_{2,20}$,v=0&130.693882&109.4&&&&0.12\\
&$24_{4,20}\rightarrow24_{3,21}$,v=0&148.293988&147.1&&&&0.12\\
&$10_{2,9}\rightarrow9_{1,8}$,v=0&148.36276&28.1&7.8&$(1.7\pm0.10)\times10^{17}$&104.1$\pm$9.9&0.10\\
&$28_{3,26}\rightarrow28_{2,27}$,v=0&153.589524&184.6&&&&0.08\\
&$21_{4,17}\rightarrow21_{3,18}$,v=0&154.47598&117.2&&&&0.14\\
&$26_{2,25}\rightarrow26_{1,26}$,v=0&154.55736&154.0&&&&0.07\\
&$20_{8,13}\rightarrow21_{7,14}$,v=0&158.742366&161.4&&&&0.02\\
&$17_{4,13}\rightarrow17_{3,14}$,v=0&159.3308&83.6&&&&0.15\\
&&&&&&&\\
$^{13}$CH$_3$CH$_2$CN&$15_{5,10}\rightarrow14_{5,9}$,v=0&131.155447&77.8&&&&0.05\\
&$15_{11,4}\rightarrow14_{11,3}$,v=0&131.161428&183.4&&&&0.01\\
&$15_{4,12}\rightarrow14_{4,11}$,v=0&131.233679&67.9&8.4&$(7.8\pm1.1)\times10^{15}$&130.8$\pm$23.2&0.06\\
&$15_{4,11}\rightarrow14_{4,10}$,v=0&131.249984&67.9&&&&0.06\\
&$18_{1,17}\rightarrow17_{1,16}$,v=0&159.10002&74.2&&&&0.08\\
\hline
\multicolumn{8}{|c|}{}\\
\multicolumn{8}{|c|}{\it Methylacetylene and its isotopologue}\\
%\multicolumn{8}{|c|}{}\\
\hline
CH$_3$CCH&9$_6\rightarrow8_6$&153.71152&296.1&&&&0.015\\
&9$_3\rightarrow8_3$&153.790769&101.7&10.0&$(8.7\pm0.7)\times10^{16}$&219.1$\pm$14.8&0.06\\
&9$_2\rightarrow8_2$&153.805457&65.7&&&&0.04\\
\hline
\end{tabular}
\end{center}}
\end{table*}

\section{Modeled line parameters}
\begin{figure*}[t]
\centering
%\hskip -1.2cm
\includegraphics[width=10cm,height=18cm,angle =270]{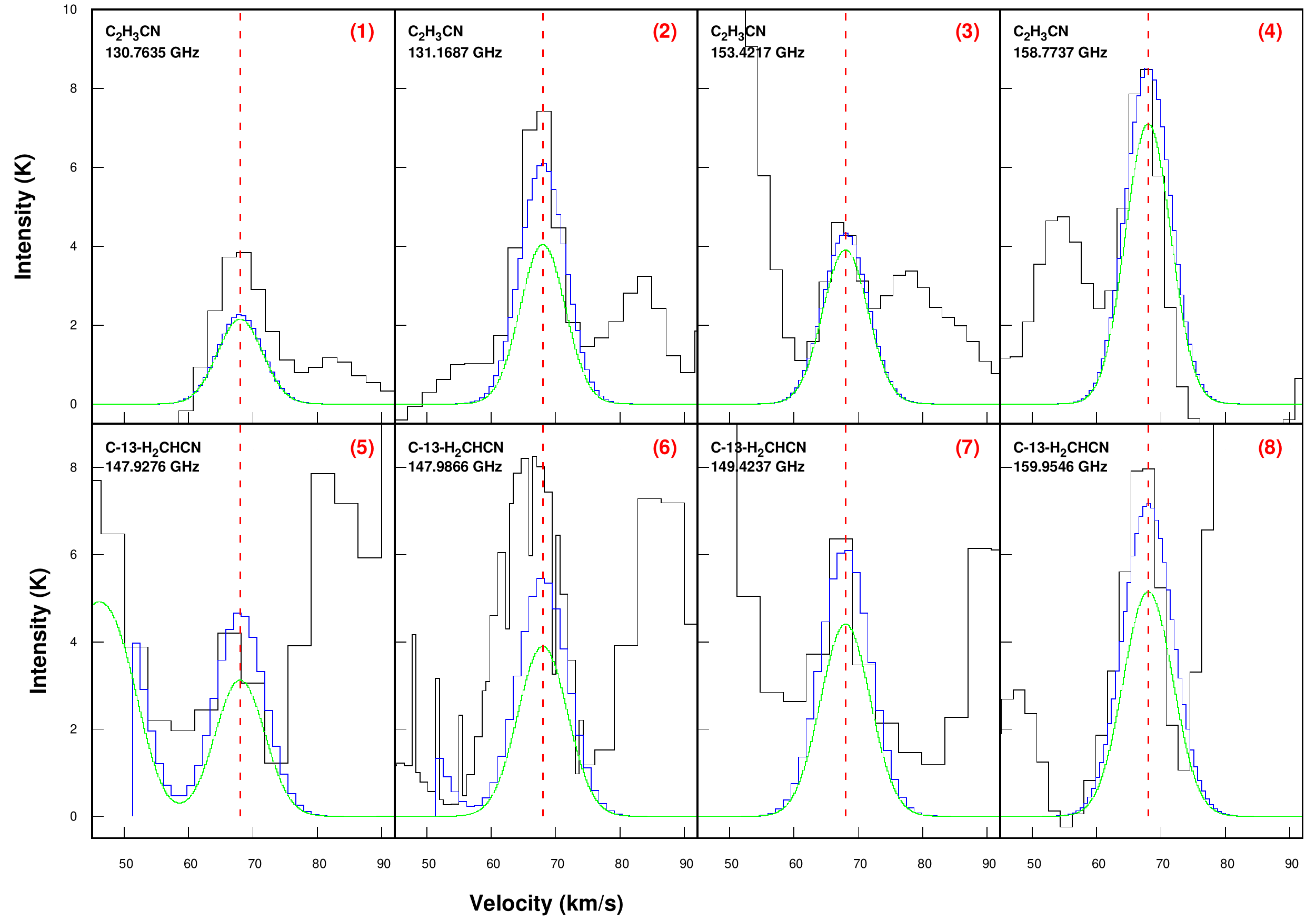}
\caption{MCMC fit of observed, unblended, optically thin transitions of vinyl cyanide and its isotopologues (solid green lines). The blue lines represent the modeled spectra, whereas the observed spectra are shown in black. The vertical red dashed line shows the position of V$_{lsr} = 68$ km/s. The green solid lines represent the LTE spectra of some species by considering the rotational temperature and column density obtained from the rotational diagram analysis (see  Figure \ref{fig:rot-dia}).} 
%An average FWHM of the observed unblended and optically thin transitions.}
\label{fig:c2h3cn-mcmc}
\end{figure*}
%%%%%%%%%%%%%%%%%%%%%%%%%%%%%%%%%%%%%%%%%%%%%%%%%%%%%%%%%%%%%%%%%%%%%%%

\begin{figure*}[t]
%\hskip -1cm
\includegraphics[width=13cm,height=18cm,angle =270]{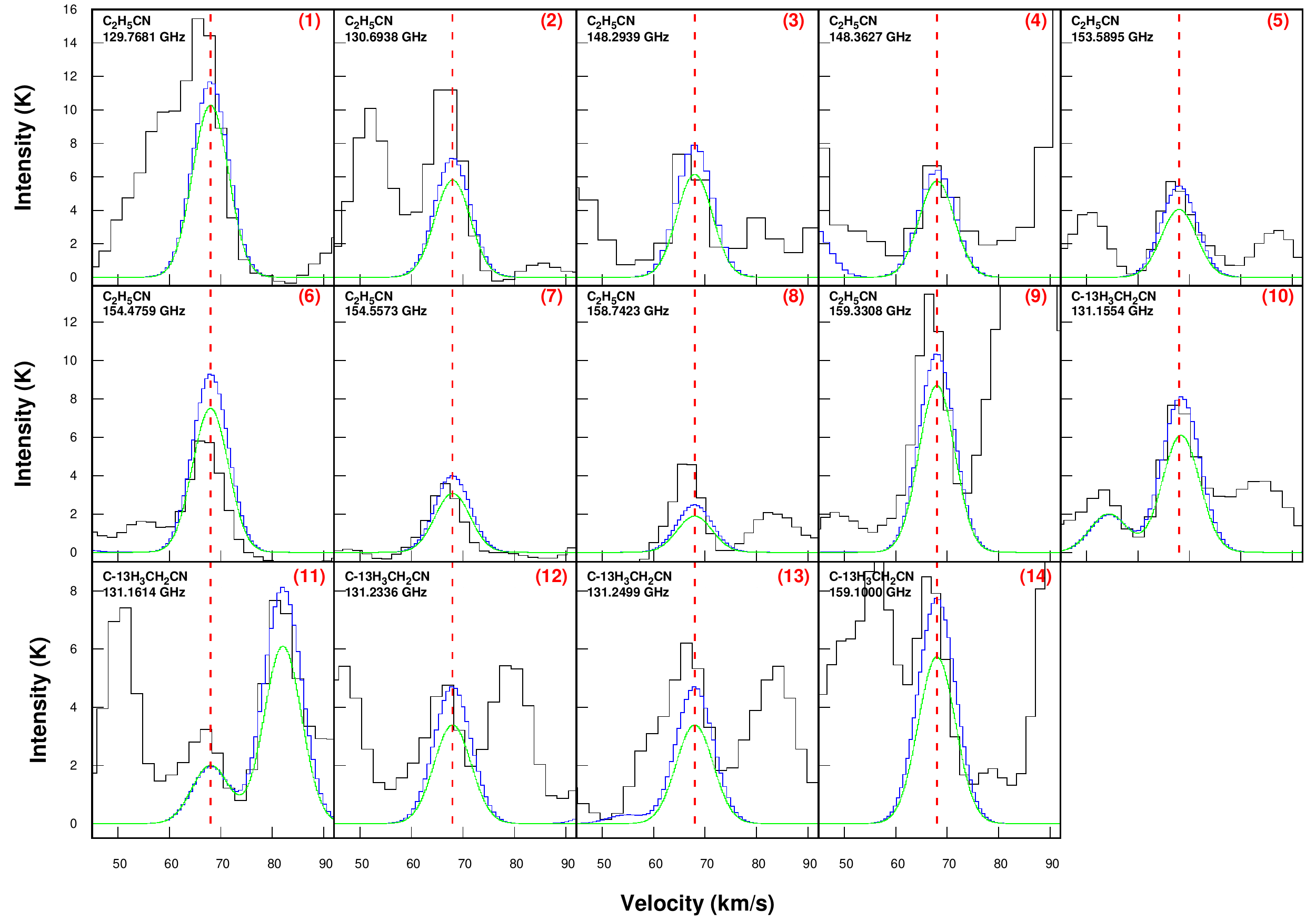}
\caption{MCMC-fit spectra (same as Figure \ref{fig:c2h3cn-mcmc}) of the observed, unblended, optically thin transitions of ethyl cyanide, its one isotopologue.}
\label{fig:c2h5cn-mcmc}
\end{figure*}

\begin{figure}
%\hskip 0.8cm
\includegraphics[width=9cm,height=9cm]{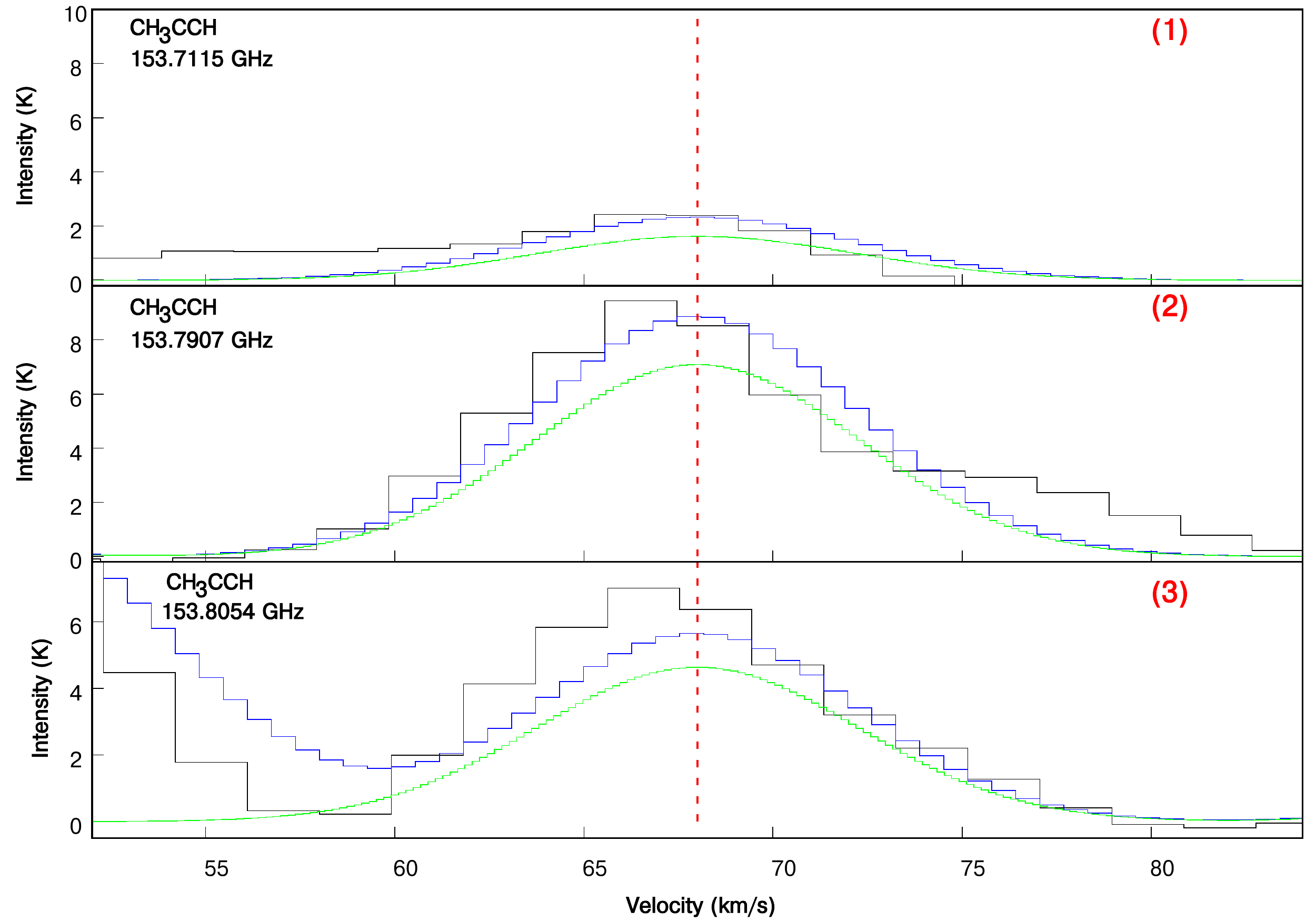}
\caption{MCMC-fit spectra of CH$_3$CCH transitions.}
\label{fig:ch3cch-mcmc}
\end{figure}

\section{Line maps}
\begin{figure*}[t]
\centering
\includegraphics[height=15cm,width=18cm]{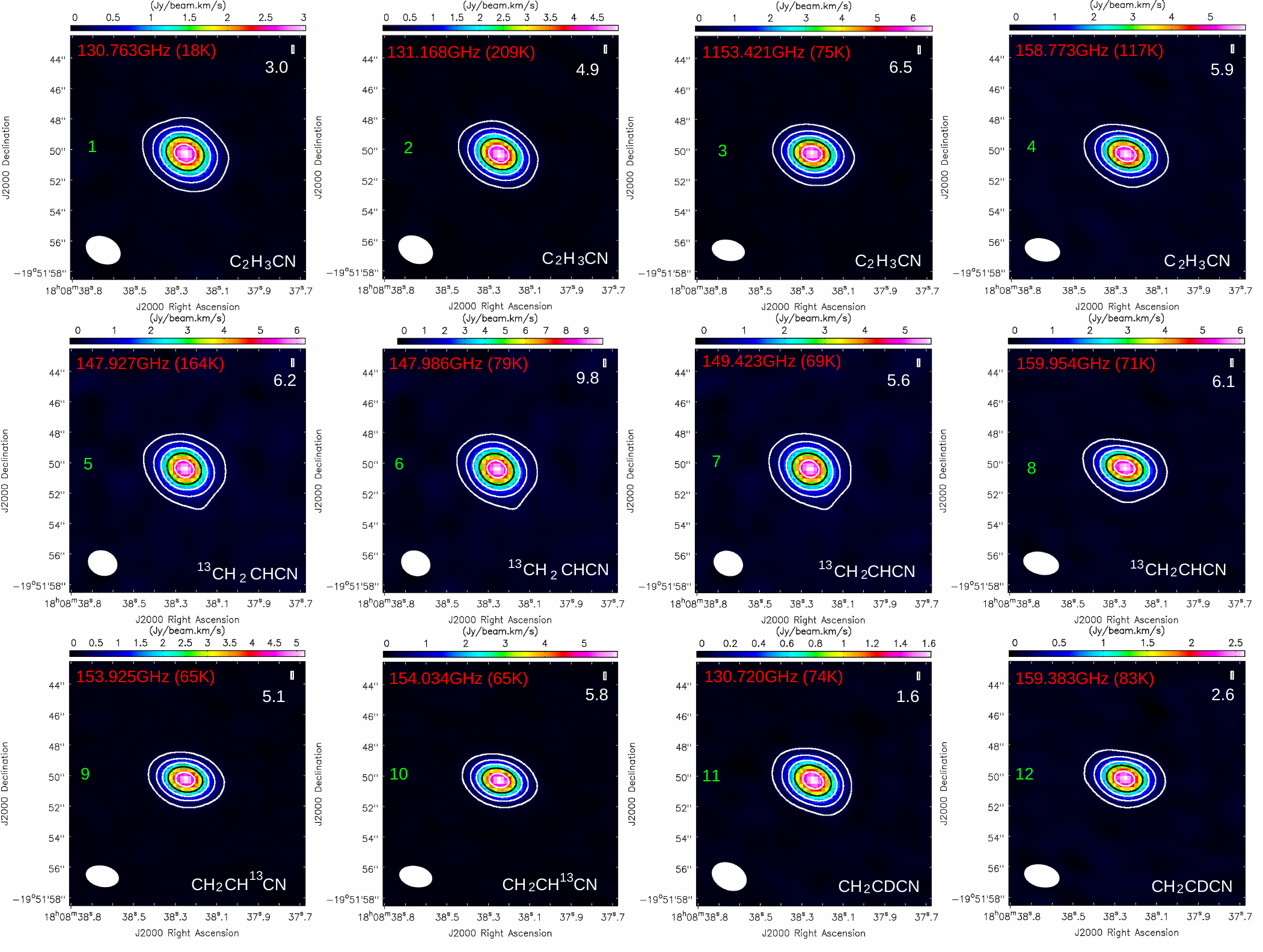}
\caption {Moment maps of unblended transition of C$_2$H$_3$CN and its isotopologues and isomer.
Contours are drawn at $5\%$, $15\%$, $30\%$, $50\%$, and $80\%$ of the maximum intensity (in Jy/beam.km/s noted in each panel) obtained for each transition.The black contour represents  $50\%$ of the maximum intensity.} 
\label{fig:mm_vc} 
\end{figure*}

\begin{figure*}[t]
\centering
\includegraphics[width=18cm]{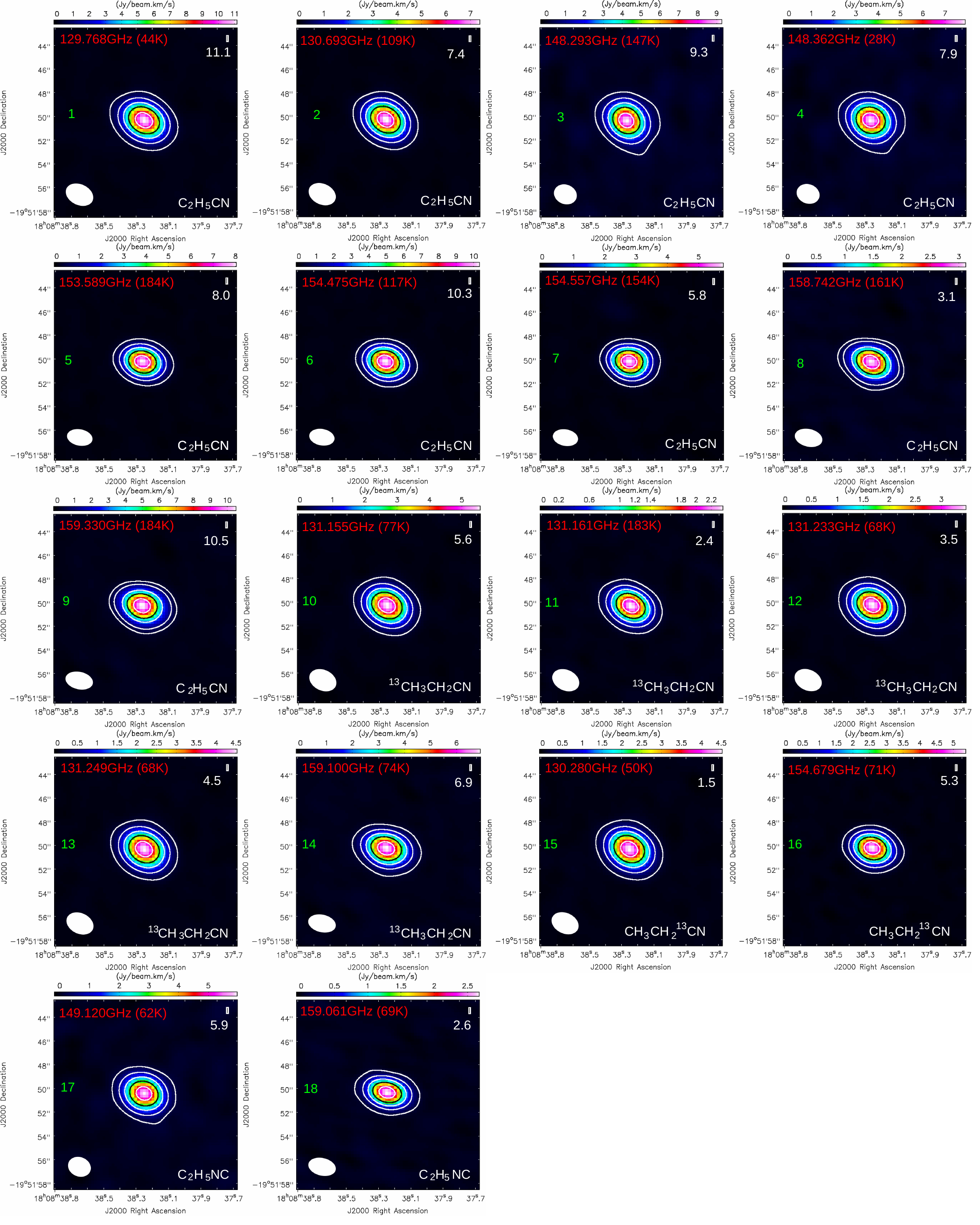}
\caption{Moment maps of transitions of C$_2$H$_5$CN, its isotopologues, and isomer (same as Figure \ref{fig:mm_vc}).}
\label{fig:mm-ec}
\end{figure*}

\begin{figure*}[t]
\centering
\includegraphics[width=18cm,height=10cm]{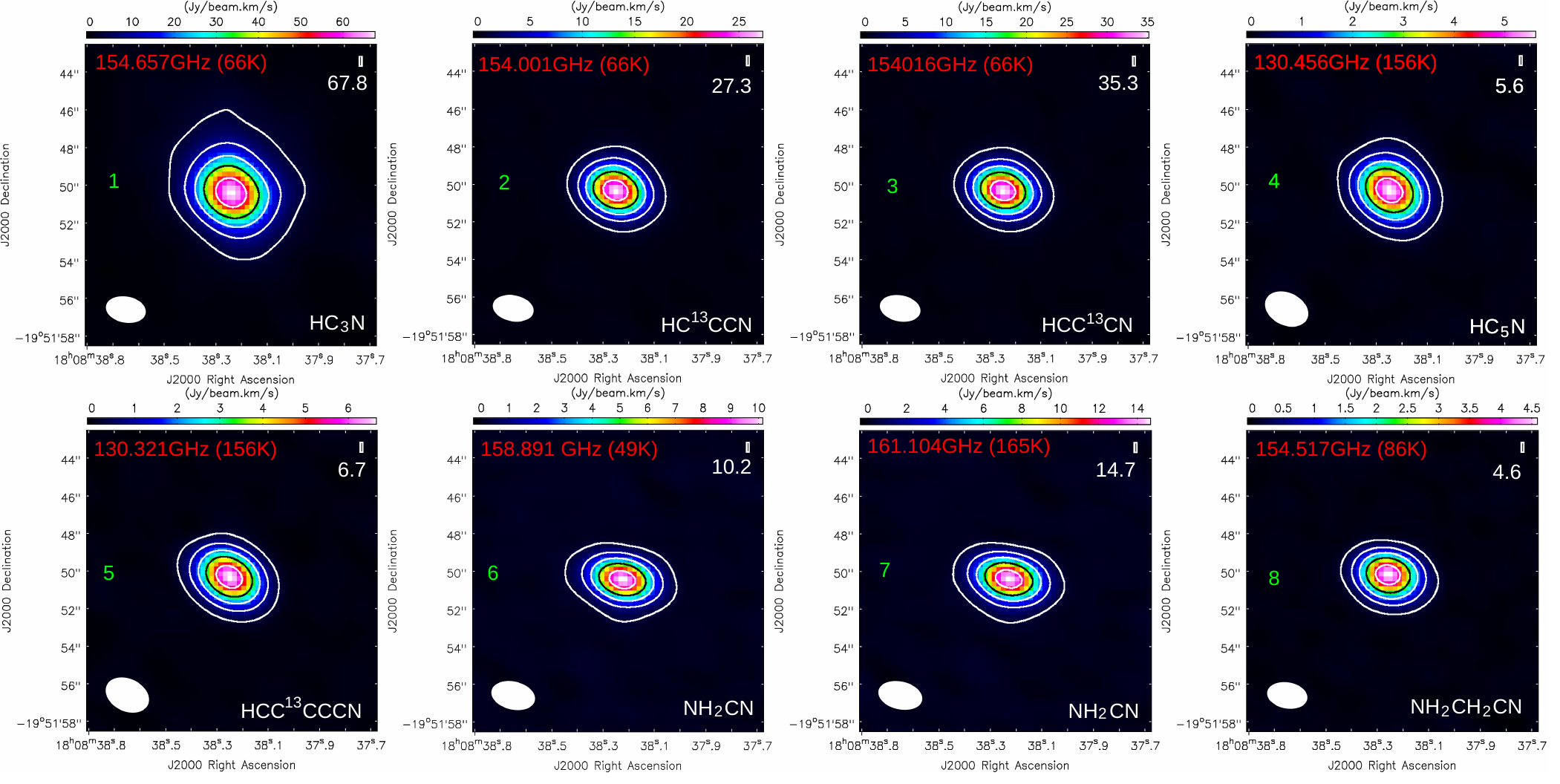}
\caption{Moment maps of unblended transition of HC$_3$N, HC$^{13}$CCN, HC$_5$N, HCC$^{13}$CCCN, NH$_2$CN, and NH$_2$CH$_2$CN (same  as  Figure \ref{fig:mm_vc}). \label{fig:mm-oth}}
\end{figure*}

\begin{figure*}[t]
\centering
\includegraphics[width=13.5cm,height=10cm]{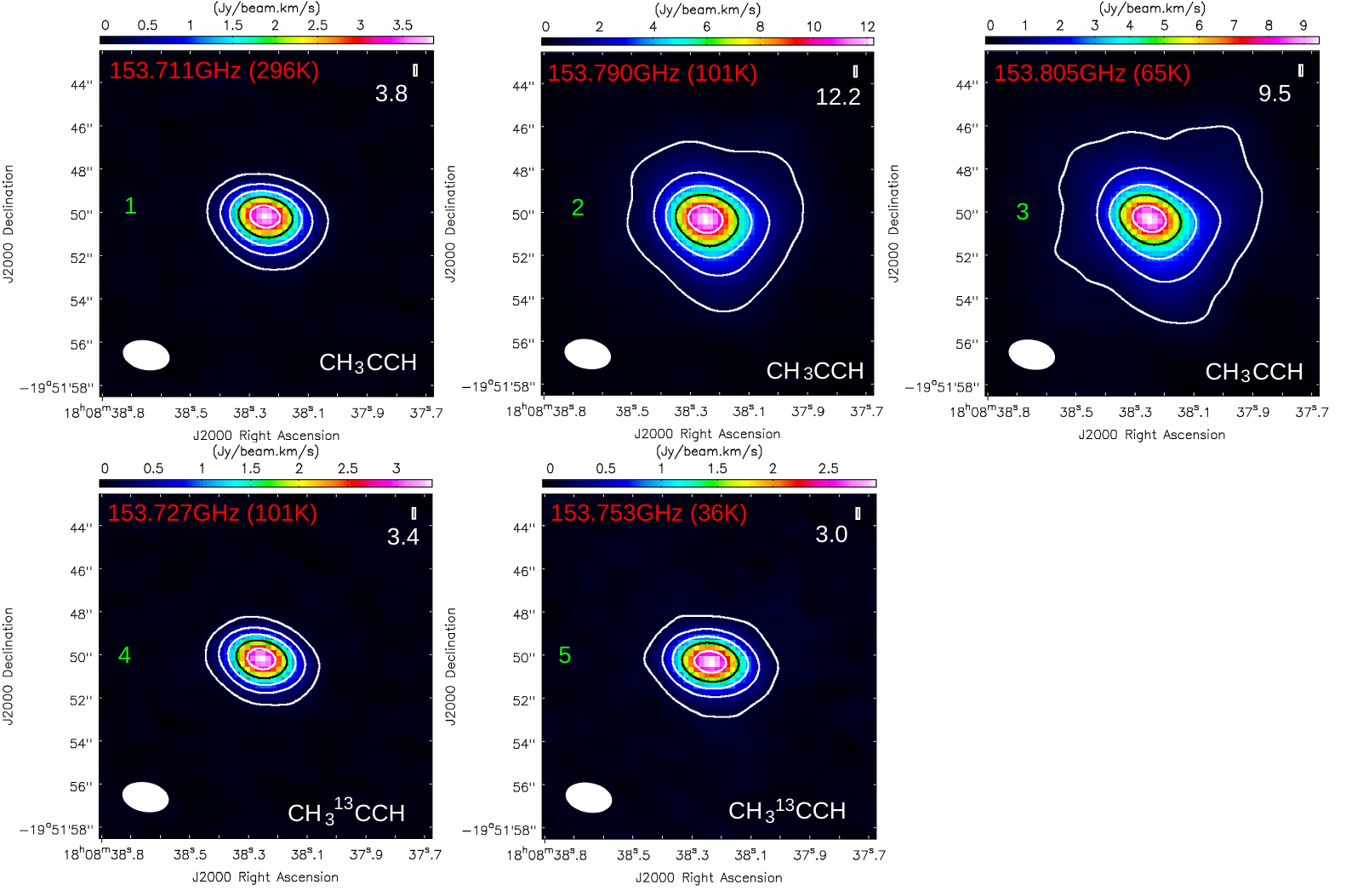}
\caption{Moment maps of unblended transition of CH$_3$CCHand CH$_3$$^{13}$CCH (same  as  Figure \ref{fig:mm_vc}).}
%Contours are drawn at 5$\sigma$, 10$\sigma$, 20$\sigma$, $50\sigma$, and 80$\sigma$ level, where $\sigma$ is the maximum intensity (in Jy/beam km/s noted in each panel) obtained for each transition. 
\label{fig:mm-othh}
\end{figure*}

The integrated intensity distribution of all observed molecules is shown in Figures \ref{fig:mm_vc}, \ref{fig:mm-ec}, \ref{fig:mm-oth}, and  \ref{fig:mm-othh}. The integrated intensity is obtained in the velocity range roughly between 58 km/s and 78 km/s.


%%%%%%%%%%%%%%%%%%%%%%%%%%%%%%%%%%%%%%%%%%%%%%%%%%%%%%%%%%
\section{LTE spectra (LTE2)
}
\begin{figure*}
\centering
\includegraphics[width=18cm]{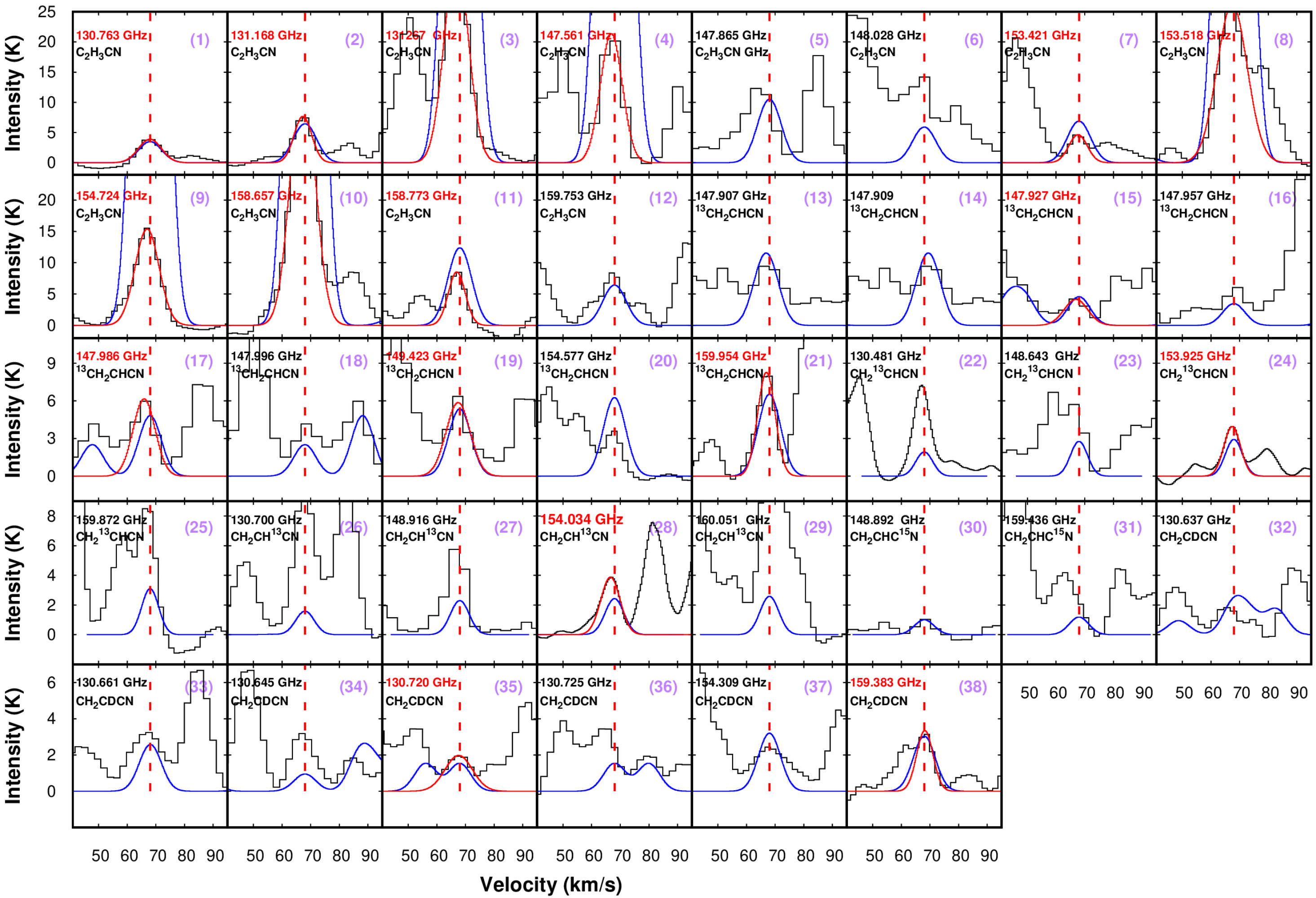}
\caption{Black lines represent observed spectra of vinyl cyanide and its isotopologues ($^{13}$CH$_2$CHCN, {CH$_2$}$^{13}$CCN, CH$_2$CH$^{13}$CN, CH$_2$CNC$^{15}$N, and CH$_2$CDCN). 
Unblended transitions are noted in red, whereas blended transitions are given in black. The solid red lines represent the Gaussian fit of the unblended transitions.
LTE spectra (LTE2) are shown in blue by considering N(H$_2)=1.35 \times 10^{25}$ cm$^{-2}$, an excitation temperature  of$150$ K, average source size (average of emitting regions of unblended transitions), and the average FWHM (obtained from the unblended, optically thin transitions) of the species. The column density of the species is varied until estimated good fits are obtained by eye.  The red dashed lines show the systematic velocity (V$_{LSR}$) of the source at $\sim$ 68 km s$^{-1}$. In addition, the name of the species and their respective transitions (in GHz) are provided in each panel.}
\label{fig:fit_size1}
\end{figure*}

\begin{figure*}
\centering
\includegraphics[width=18cm]{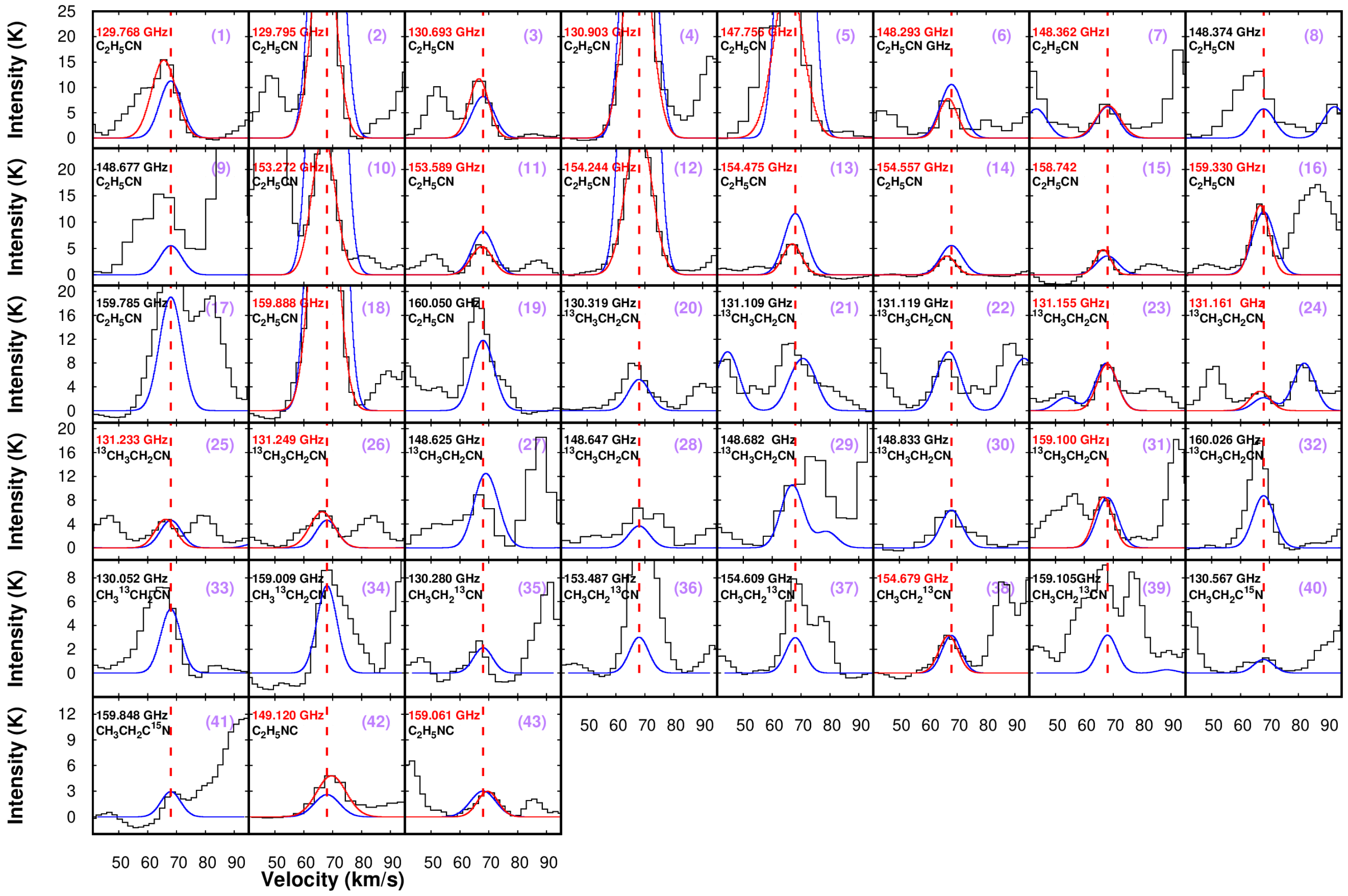}
\caption{Gaussian-fit and -modeled LTE spectra (LTE2; same as Figure \ref{fig:fit_size1}) of the unblended transitions of ethyl cyanide, its isotopologues ($^{13}$CH$_3$CH$_2$CN, CH$_3$$^{13}$CH$_2$CN, CH$_3$CH$_2$$^{13}$CN, C$_2$H$_5$C$^{15}$N), and one of its isomers (C$_2$H$_5$NC).}
\label{fig:fit_size2}
\end{figure*} 

\begin{figure*}
\centering
\includegraphics[width=18cm]{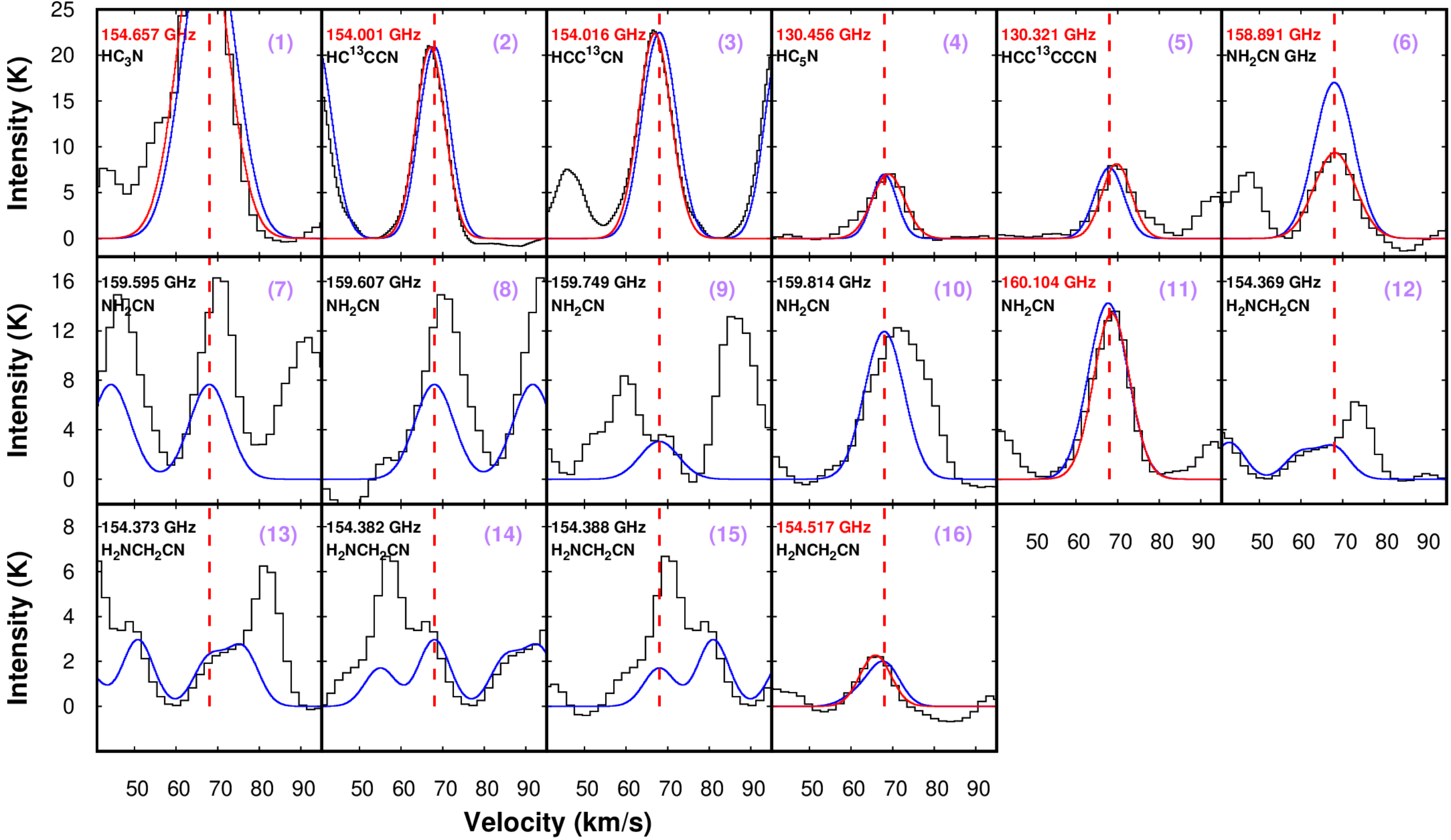}
\caption{Gaussian-fit spectra and LTE spectra (LTE2; same as Figure \ref{fig:fit_size1}) of cyanoacetylene, cyanodiacetylene, cyanamide, and  aminoacetonitrile.}
\label{fig:fit_size3}
\end{figure*} 

\begin{figure}
\includegraphics[width=9cm]{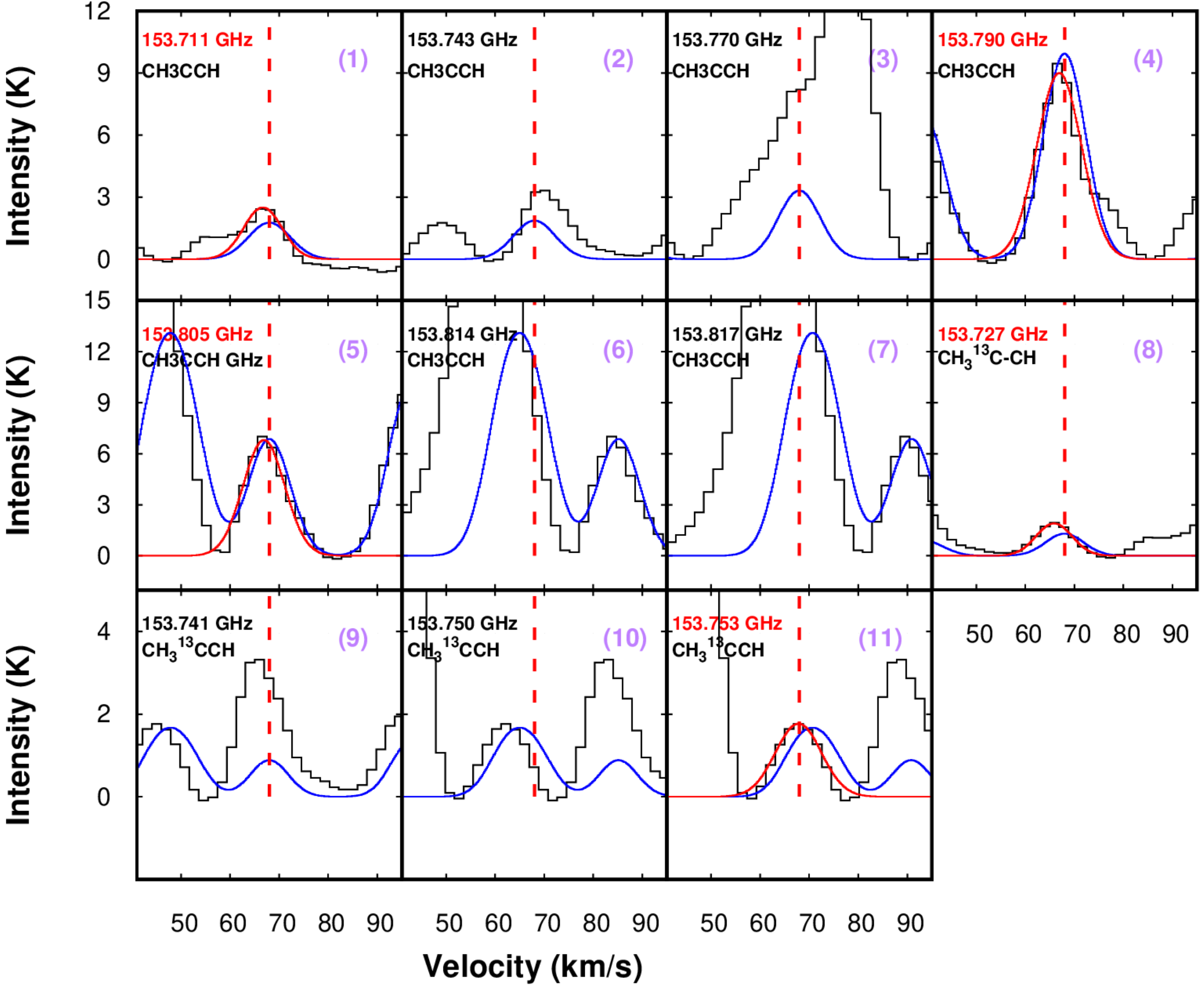}
\caption{Gaussian-fit spectra and LTE spectra (LTE2; same as Figure \ref{fig:fit_size1}) of CH$_3$CCH and its isotopologues.}
\label{fig:fit_size4}
\end{figure}
\clearpage

%%%%%%%%%%%%%%%%%%%%%%%%%%%%%%%%%%%%%%
\section{Extra figures for chemical evolution of selected species in 1D models}
\label{App-ID}

\begin{figure*}[t]
\includegraphics[width=.96\textwidth]{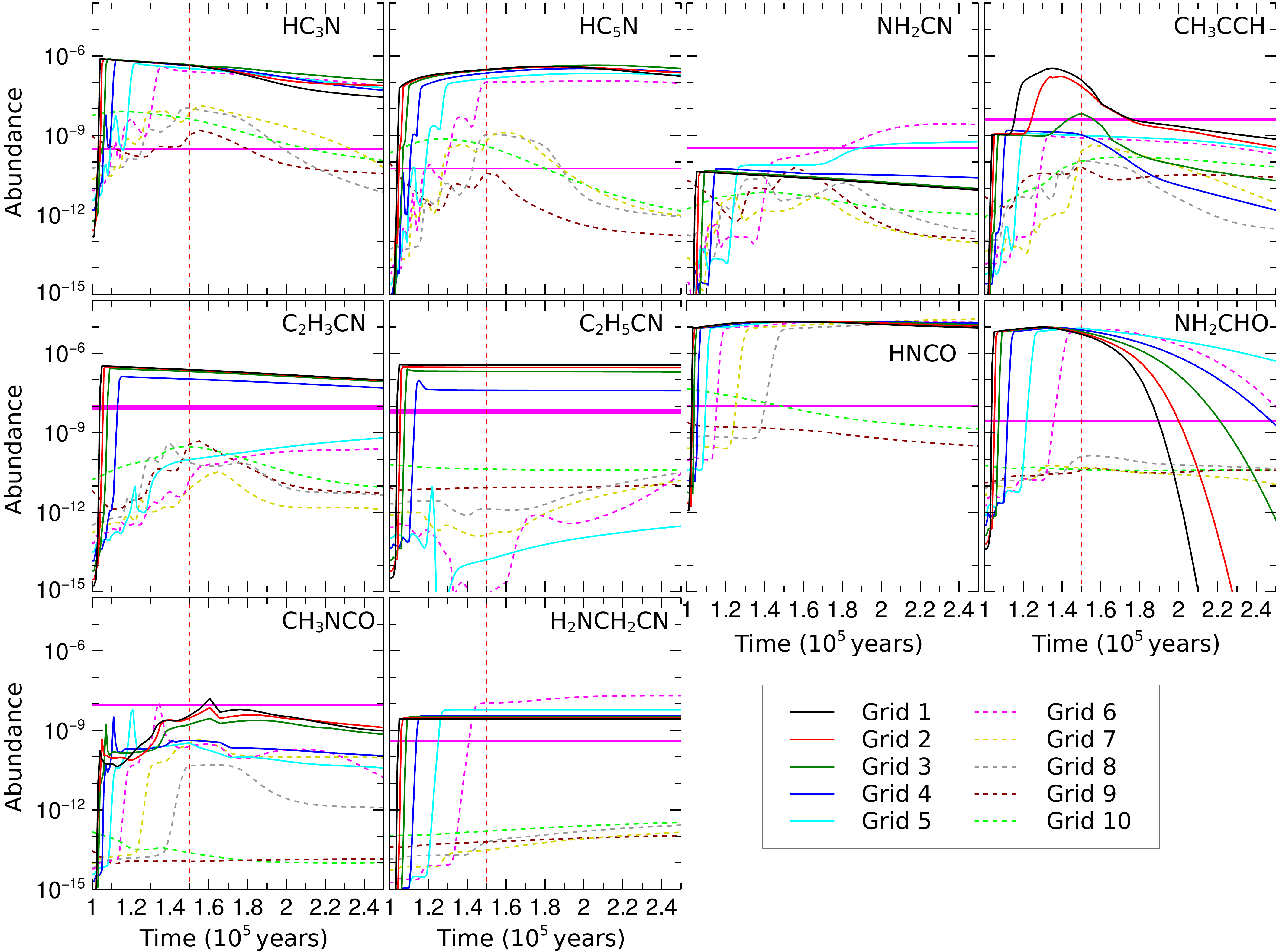}
\caption{Same as Figure \ref{fig:abun-evo-II-15K}, but for initial dust and gas temperature of 20 K.}
\label{fig:abun-evo-II-20K}
\end{figure*}

\begin{figure*}[t]
\includegraphics[width=.96\textwidth]{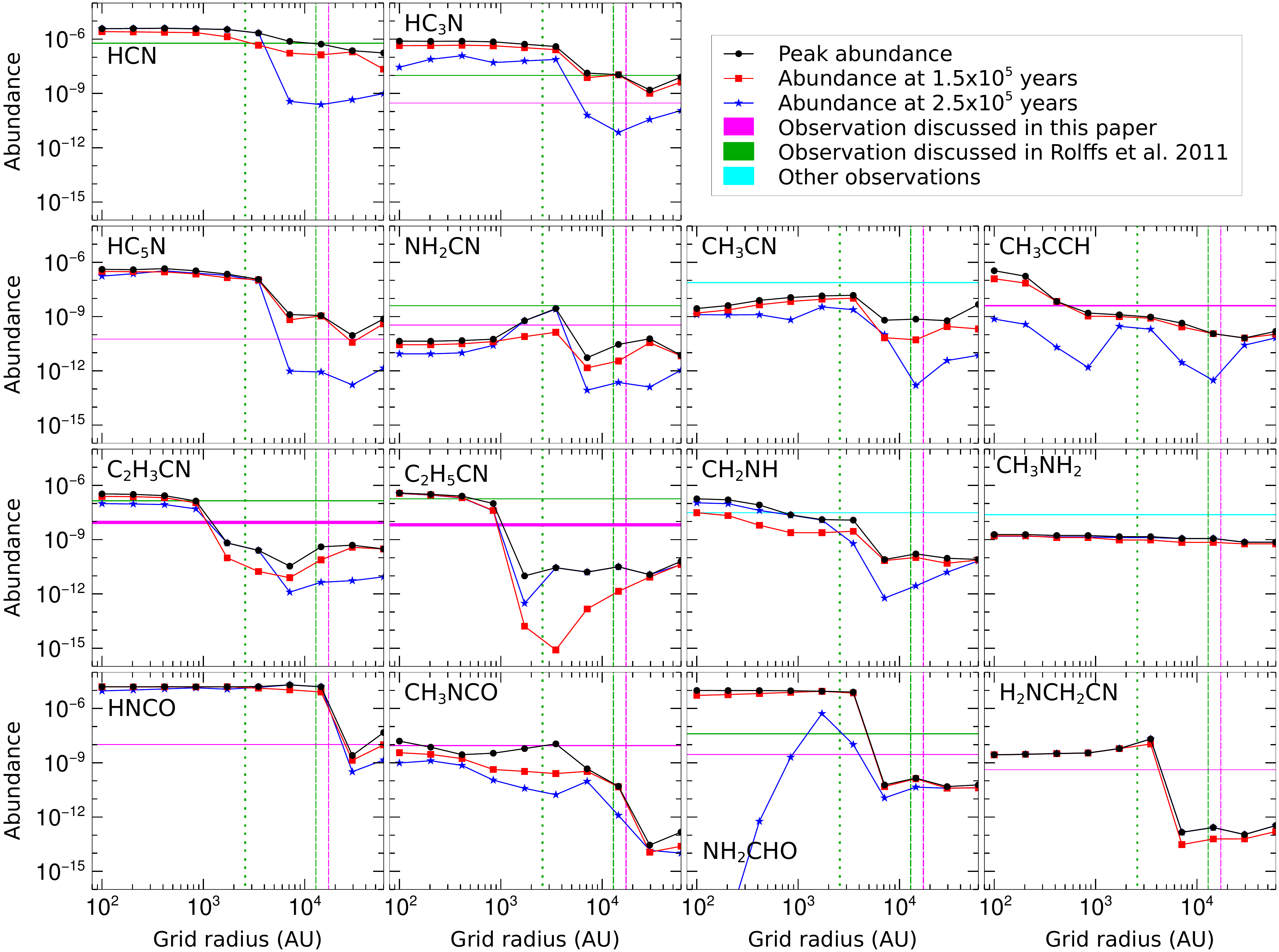}
\caption{Same as Figure \ref{fig:peak-model-II-15K}, but for initial dust temperature of 20 K.}
\label{fig:peak-model-II-20K}
\end{figure*}

%\clearpage
% \software{CASA 4.7.2 \citep{mcmu07}}
\bibliographystyle{aa}
\bibliography{Nitrile_G10}
\end{document}